\documentclass[12pt,a4paper]{article}
\usepackage{amsmath}
\usepackage{bbm}
\usepackage{lscape}
\usepackage{amsfonts}
\usepackage{amssymb}
\usepackage{adjustbox}
\usepackage{eurosym}
\usepackage{hhline}
\usepackage{breqn}
\usepackage{soul}
\usepackage{color}
\usepackage{csquotes}
\usepackage{chngcntr}
\usepackage{changepage}
\usepackage{graphicx}
\usepackage{epstopdf}
\usepackage{floatrow}
\usepackage{float}
\floatstyle{plaintop}
\restylefloat{table}
\usepackage{dsfont}
\usepackage{courier}
\usepackage{natbib}
\usepackage{setspace}
\usepackage{rotating}
\usepackage{enumitem}
\usepackage[titletoc,title]{appendix}
\usepackage{caption}
\date{}

\newtheorem{rk}{Remark}

\let\OLDthebibliography\thebibliography
\renewcommand\thebibliography[1]{
  \OLDthebibliography{#1}
  \setlength{\parskip}{0pt}
  \setlength{\itemsep}{0pt plus 0.3ex}
}
\def\spacingset#1{\renewcommand{\baselinestretch}%
{#1}\small\normalsize} \spacingset{1}
\begin{document}

\title{Efficient Estimation in Extreme Value Regression Models of Hedge Fund Tail Risks} 

\author{J. Hambuckers$^{1,\dagger}$, M. Kratz$^{2}$ and A. Usseglio-Carleve$^{3}$\\[1ex]
\small $^1$ University of Li\`{e}ge - HEC Li\`{e}ge, Belgium. $\dagger$ Corresponding author. Email: jhambuckers@uliege.be\\
\small $^2$ ESSEC Business School, CREAR, France. Email: kratz@essec.edu\\
\small $^3$ Avignon Universit\'e, France. Email: antoine.usseglio-carleve@univ-avignon.fr\\
} 
\maketitle

\begin{abstract}
\noindent We introduce a method to estimate simultaneously the tail and the threshold parameters of an extreme value regression model. This standard model finds its use in finance to assess the effect of market variables on extreme loss distributions of investment vehicles such as hedge funds. However, a major limitation is the need to select \textit{ex ante} a threshold below which data are discarded, leading to estimation inefficiencies. To solve these issues, we extend the tail regression model to non-tail observations with an auxiliary splicing density, enabling the threshold to be selected automatically. We then apply an artificial censoring mechanism of the likelihood contributions in the bulk of the data to decrease specification issues at the estimation stage. We illustrate the superiority of our approach for inference over classical peaks-over-threshold methods in a simulation study. Empirically, we investigate the determinants of hedge fund tail risks over time, using pooled returns of 1,484 hedge funds. We find a significant link between tail risks and factors such as equity momentum, financial stability index, and credit spreads. Moreover, sorting funds along exposure to our tail risk measure discriminates between high and low alpha funds, supporting the existence of a fear premium.
\end{abstract}
\vspace*{.5cm}
\noindent\textbf{Keywords}: Extreme value theory, generalized Pareto regression, censored maximum likelihood.\\ 
\vspace*{0.5cm}
\noindent {\emph MSC 2020}: 
62G32,  
62H12, 
62M10, 
62P05.   

\newpage

\section{Introduction}\label{sec:intro}

Extreme Value Regression (EVR), pioneered by \cite{davisonsmith1990}, finds its use in the analysis of factors affecting the likelihood of extreme events. In EVR models, one assumes that the tail distribution of a variable of interest is well approximated by a generalized Pareto distribution (GPD), a direct result of extreme value theory \citep{balkema,pickands}. In addition, owing to heterogeneity in the data (e.g., related to time or contextual factors), the scale and shape parameters of the GPD are assumed to be functions of covariates supposed to influence the distribution of extreme events. From a practical standpoint, the estimation of this model relies on the peaks-over-threshold (POT) approach: the analyst first selects a high threshold value. Then, they discard all data smaller than this threshold, and the parameters are estimated using solely the observations larger than the threshold, termed extreme values. The underlying idea is to select a threshold high enough to ensure that the GPD approximation is good, but also small enough to maintain a reasonable sample size. However, the regression structure in the tail poses several challenges to selecting these thresholds properly and to obtaining reasonably accurate estimates. Our main methodological contribution is therefore to propose an improved estimation method that also removes most of the subjectivity routinely applied to the choice of the threshold.

\begin{figure}[htbp]
    \centering
    \begin{tabular}{cc}
      \includegraphics[scale=.5]{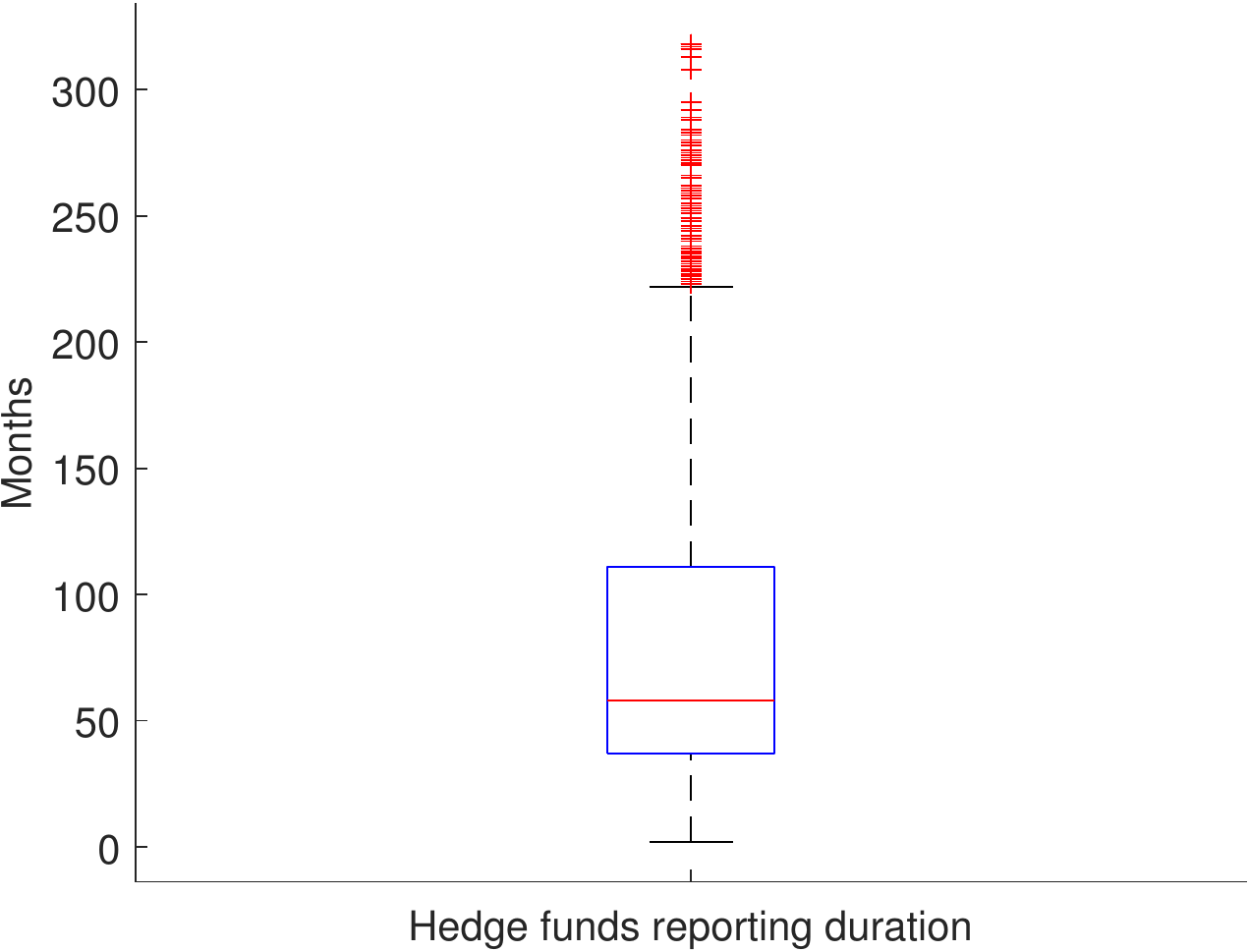}    & \includegraphics[scale = .5]{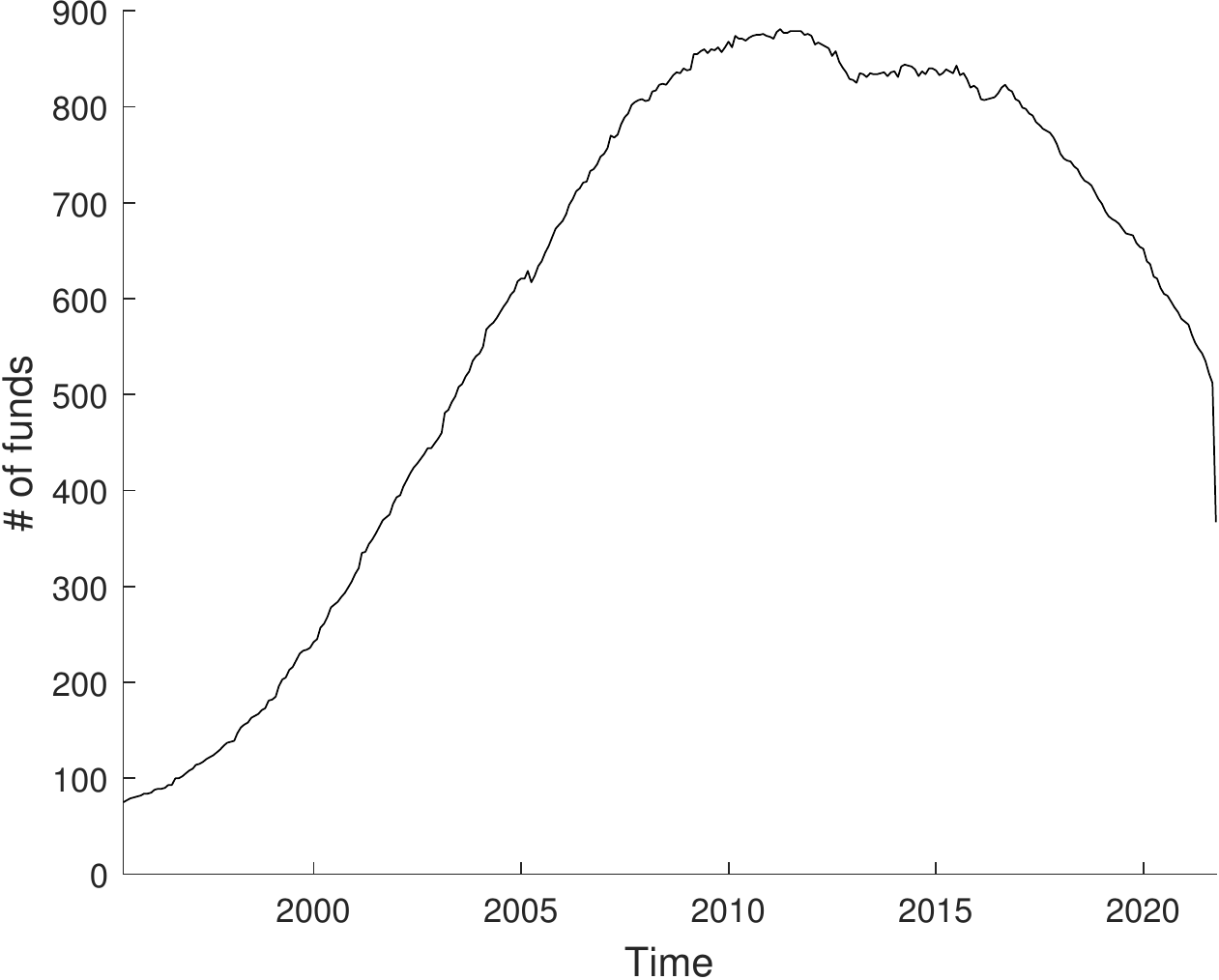} \\
    \end{tabular}
    \caption{\footnotesize Left: Total reporting duration of \textit{Long/Short Equity} hedge funds registered in the EH database over the period 01/1995-12/2021. Right: monthly number of hedge funds' losses with a \textit{Long/Short Equity} investment style and a total reporting duration of at least 60 months.}
    \label{fig:intro}
\end{figure}

EVR models have been successfully applied to the study of environmental and climate data \citep[see, e.g.,][]{chavez2005,opitz2018}, insurance problems \citep{beirlant2003,biagini2021}, and financial risk modeling \citep{chavez2016,castro2018,bee2019}\footnote{See also \cite{einmahl2023} for an excellent review on applications of extreme value analysis of heterogeneous data in finance and economics.}. Particularly in finance, EVR models are useful to study the link between the state of the financial markets and the likelihood of extreme losses of economic entities. As a motivational application, we consider the problem of estimating the conditional tail distributions of a large cross-section of hedge funds, following the idea of \cite{kelly2014}. Hedge funds are investment vehicles relying on sophisticated trading strategies and complex financial products to generate an economic profit \citep{getmansky2015}. Identifying financial conditions that influence their extreme downside risks is important to anticipate threats to financial stability \citep{billio2012}, but also to evaluate their performance, since hedge funds often pursue strategies akin to “selling earthquake insurance” \citep{stulz2007}. Measuring their propensity to suffer from extreme losses is therefore essential. However, this task is particularly complicated by the short history and unbalanced nature of available data, two characteristics that prevent us from applying time series methods or standard extreme value theory, as used for stocks \citep[see, e.g., ][]{kelly2014,huang2012}. Figure~\ref{fig:intro} (left panel) illustrates this issue: we display the distribution of the reporting duration in a panel of 3,100 funds with a \textit{Long/Short Equity} investment strategy, between January 1995 and December 2021. The median reporting duration is only 58 months. In addition, the cross-section of funds at each point in time is also limited: in Figure~\ref{fig:intro} (right panel), we display the number of funds reporting for a given month, given that the fund remains in the database for at least 60 months (a typical minimum duration used in empirical studies). The median reporting duration is around 680, but sometimes as low as 75. 

To overcome this issue, \cite{mhalla2022} and \cite{dupuis2022} advocate pooling all of the funds' returns and using an EVR model relying on financial factors to control for heterogeneity across funds and time\footnote{Similarly, \cite{kelly2014} compute a nonparametric tail risk measure of the market by applying the same pooling principle, although in a static way, to the cross-section of stock returns.}. A critical aspect of EVR models, though, is the need to choose a threshold for the POT approach. Common practices involve using either a global threshold (e.g., an empirical quantile), local thresholds (e.g., empirical quantiles computed over sub-periods), or regression thresholds (e.g., conditional quantiles obtained from quantile regression) to select the sample of extreme values. However, as we illustrate in our simulation study in Section~\ref{sec:simu}, the POT approach discards a significant portion of the data, causing a major loss of efficiency at the estimation stage. Hedge funds data being particularly scarce, there is therefore a strong need to find alternatives that exploit the available information more effectively. In addition, several other difficulties with threshold selection in the context of regression have been documented in the literature: a threshold selected in a preliminary step generates additional estimation uncertainty in the other parameters that is hard to account for in inferential procedures \citep{he2022}. Furthermore, it does not fulfill the threshold stability property \citep{eastoe,decarvalho2021}, with the consequence that conclusions regarding the effects of covariates on the tail distribution might radically differ with a change in threshold choice. In light of these concerns, an automatic selection procedure acknowledging the link between threshold and covariates is therefore needed, as advocated early on by \cite{beirlant2003}. 

So far, the literature on this problem is limited with regard to regression effects. To the best of our knowledge, only \cite{decarvalho2021} explicitly introduces an automatic threshold selection procedure for EVR models in a Bayesian context. On the other hand, automatic threshold selection has been discussed extensively in the non-regression setting (see, e.g., \citealp{bader2018}, \citealp{dacorogna2023} and references therein) and inspires the present work. Furthermore, the present work is complementary to the recent contributions of \cite{decarvalho2021} and \cite{naveau2016} on unconditional threshold selection, and extends the framework of \cite{debbabi} and \cite{debbabi2014} with a regression structure. 

To solve the threshold selection problem, we use the principle of distributional regression \citep{rigby2005,hothorn2014,kneib2021} to formulate an auxiliary splicing distribution model that extends the EVR model below the supposed extreme value threshold. Splicing distributions are common tools in insurance to fit claim-size data \citep[see][for a review]{reynkens2017}, and comprise combining a distribution for the body of the data (e.g. Gaussian) with a distribution for the tail (e.g. GPD). To ensure that the resulting distribution is proper, constraints on the derivatives of the distributions are imposed at their junction point. Inspired by this setting, we develop a conditional version of a splicing distribution, in which the tail parameters are expressed as functions of candidate covariates. As a corollary of the definition of the parameters in our model, the junction point between the body and the right tail is automatically defined as \textit{conditional} on the covariates intervening in these parameters, and can be interpreted as a conditional threshold in an EVR model. We then use this formulation of the problem to estimate the regression parameters in the tail, without resorting to a preliminary threshold choice. As a baseline model, we rely on a three-component distribution model advocated in \cite{debbabi}, combining Gaussian (G) and GPD distributions, bridged with an exponential (E) distribution. This model is referred to as G-E-GPD and describes in a general way average and extreme behaviors via the use of the central limit theorem (CLT) and EVT.

Although splicing distributions are attractive thanks to their flexibility, estimating their parameters is challenging in practice: standard maximum likelihood estimation often fails because of the difficulty of choosing simultaneously a correctly-specified parametric model for both the bulk and the tail of the data. Therefore, we introduce a robust\footnote{The term ``robust" refers to an estimation procedure that suffers from limited bias when the data suffer from arbitrary contamination with respect to our distributional assumptions.} estimation strategy for this model based on a censored likelihood principle advocated by \cite{diks2011} and \cite{aeberhard2021}. Our estimator belongs to the class of M-estimators, and lowers the impact of a misspecified body or left tail of the distribution on the estimated tail parameters. In this estimation procedure, we replace the likelihood contributions of the data in these regions by a conditional probability of belonging to the said regions. With this approach, we keep the ``best of both worlds": on the one hand, we use the asymptotically justified GPD model without the need to know the exact distribution of the tail. On the other hand, we exploit data belonging to the body of the distribution to choose an appropriate threshold. The result is a more efficient estimation of the regression parameters in the tail compared to classical POT-EVR approaches. The use of censored likelihood methods and M-estimators has been previously advocated in the study of extreme values, although in unconditional and multivariate contexts, and not in a regression setting. See, e.g., \cite{vandewalle2007}, \cite{huser2014} and \cite{einmahl2018}.

We discuss the theoretical properties of this estimator, and provide a data-driven method to select the robustness parameter determining the degree of censoring. We illustrate the good behavior of this approach in practice with realistic simulation studies. In particular, we demonstrate that our splicing approach is much more efficient at estimating regression effects than classical POT-based methods, even under a general misspecification of the G-E-GPD model.

We then illustrate this approach by studying the conditional tail distribution of hedge funds, using a database of around 189,000 observations covering 1,484 funds over a period of 26 years, briefly sketched above. Motivated by considerations outlined in \cite{agarwal2017} and \cite{kelly2014}, we address the questions as to whether hedge funds tail risks fluctuate with the state of the market; and whether tail risk is an important predictor of hedge funds' performance. We answer these two questions positively and contrast the results of the proposed approach with those obtained with classical POT-EVR models, as well as with the semi-parametric approach of \cite{kelly2014}. We demonstrate the significant superiority of our measure, both in terms of interpretability and explanatory power. We find that hedge funds' tail risk is positively related to favorable funding conditions and a booming stock market: a decrease in spreads, market volatility, and financial stability indicators, as well as an increase in time-series equity momentum are associated with a significant increase in tail risk. Then, sorting funds according to the correlation between their returns and one-month lagged values of our tail risk measures, we find that low-correlation funds exhibit significantly higher average alpha than high-correlation funds. This result delivers additional empirical evidence to the ``fear premium" hypothesis investigated in \cite{gao2018}.

The rest of the paper is organized as follows: in Section~\ref{sec:methodo}, we introduce our model and estimation strategy. In Section~\ref{sec:theory}, we derive several theoretical results associated with our model. In Section~\ref{sec:simu}, we conduct an extensive simulation study. In Section~\ref{sec:empirical}, we conduct our empirical analysis and conclude in Section~\ref{sec:conclusion}.
Additional results and discussions are given in the Appendix.

\section{Methodology}\label{sec:methodo}
Similarly to \cite{kelly2014}, we assume that the conditional distribution of the stationary loss process $Y_{it}$ at time $t\in\{1,\ldots,T\}$  for a given entity $i$ (e.g. a hedge fund) belongs to the maximum domain of attraction of a Fr\'{e}chet distribution. That is, the upper tail of the conditional loss distribution above a threshold $u_{it}$ is well approximated by a GPD:
\begin{equation}\label{eq:pot}
\mathbb{P}\left(Y_{it}\leq y_{it}|Y_{it}>u_{it}\right) \underset{u_{it}\rightarrow y_{it}^{F}}{\longrightarrow} 1-\left(1+\dfrac{\xi_{it}(y_{it}-u_{it})}{\sigma_{it}}\right)^{-1-1/\xi_{it}},
\end{equation}
where $\xi_{it}=\xi(\mathbf{x}_{it})$ and $\sigma_{it}=\sigma(\mathbf{x}_{it})$, the conditional shape and scale parameter of the tail distribution, respectively, are both assumed to be strictly positive \citep{pickands,balkema}. $\mathbf{x}_{it}$ is a d-dimensional vector of covariates observed over time, and $y_{it}^{F}$ denotes the upper end point of $Y_{it}$. See also \cite{coles2001,chavez2005} and \cite{embrechtsbook} for theoretical details; and \cite{chavez2016,hambuckers2018} and \cite{mhalla2022} for examples of applications in the field of finance. Here, contrary to \cite{kelly2014} who adopt a cross-sectional view, we specify $\xi_{it}$ and $\sigma_{it}$ as explicit functions of $\mathbf{x}_{it}$. As in classical GLM, we connect $\mathbf{x}_{it}$ to the parameters of the GPD with a log-link function, specifying
\begin{eqnarray}
\log(\xi_{it})=\mathbf{x}_{it}^{T}\pmb{\beta}^{\xi},\\
\log(\sigma_{it})=\mathbf{x}_{it}^{T}\pmb{\beta}^{\sigma},
\end{eqnarray}
where $\pmb{\beta}^{\xi}$ and $\pmb{\beta}^{\sigma}$ are the vectors of regression coefficients, including the constants. Our primary interest is in $\xi_{it}$, which is used as a tail risk measure, and in $\pmb{\beta}^{\xi}$, which captures the marginal effects of changes in covariates on the tail risk. Since $\pmb{\beta}^{\xi}$ and $\pmb{\beta}^{\sigma}$ are constant across time and entities, estimates are obtained by pooling all observations across these two dimensions, then by applying the POT-EVR approach, and estimating the parameter with maximum likelihood procedures. Notoriously, the choice of $u_{it}$ is known to be often arbitrary and the POT-EVR approach to be inefficient, since it discards a large portion of the data. In the following, we detail our procedure to improve on these issues. 

\subsection{Splicing auxiliary regression model}

To bypass the need to select \textit{ex ante} $u_{it}$, we propose extending the EVR model below $u_{it}$, thus defining a flexible auxiliary regression model for the full range of the data. Hence, instead of relying solely on tail data, we exploit the informational content of data in the body of the distribution. Of course, this extended modeling comes at a price: while the POT-EVR approach is immune to misspecification issues if $u_{it}$ is large enough, it is not the case for the auxiliary regression model below the threshold. Nevertheless, contrary to tail events that are difficult to model, reasonable assumptions such as exponentially-decreasing density functions are known to provide decent approximations of the true density in the body of the data. In addition, in the next subsection, we detail an improved maximum likelihood estimation strategy of the auxiliary regression model that reduces estimation bias in case of misspecification. 
Our approach can be seen as a compromise between an asymptotically correct EVR model based on few data, and an over-parametrized model for the full range of the data\footnote{See also \cite{beirlant2004b} for an earlier discussion.}.

We therefore assume that the \textit{full} conditional density of $Y_{it}$ can be approximated by a \textit{splicing regression model} of the G-E-GPD type, denoted by 
$$
f(y_{it};\pmb{\theta},\mathbf{x}_{it})=f(y_{it};\mu_{0},s(\mathbf{x}_{it}),\xi(\mathbf{x}_{it}),\sigma(\mathbf{x}_{it})),
$$ 
and associated with a vector of $d$ predictors denoted $\mathbf{x}_{it}$, for $i=1,\cdots,I$ and $t=1,\cdots,T$. We assume that the same set of predictors drives the different distribution parameters, but this assumption can be easily relaxed. $\xi(\mathbf{x}_{it})$ and $\sigma(\mathbf{x}_{it})$ are those of the GPD in~\eqref{eq:pot}. We acknowledge that financial data usually exhibit heteroscedasticity via the specification of the variance parameter $s(\mathbf{x}_{it})$ as a function of the covariates, while $\mu_{0}$ is a location parameter assumed to be constant. This last hypothesis reflects that our data have either a constant mean or that the conditional mean has been removed in a preliminary step. Dropping the explicit reference to $\mathbf{x}_{it}$, the pdf of the G-E-GPD model is then defined by:
\begin{equation}\label{eq:density}
f(y_{it};\pmb{\theta},\mathbf{x}_{it}) = \left\{
                \begin{array}{lcl}
                  \gamma_{1,it}\,\phi(y_{it};\mu_{0},s_{it}^{2}) & \text{if} & y_{it}\leq u_{it}^{*} \\
                  \gamma_{2,it} \,e(y_{it};\lambda_{it})	&\text{if} & u_{it}^{*}\leq y_{it}\leq u_{it} \\
                  \gamma_{3,it} \,h(y_{it}-u_{it};\xi_{it},\sigma_{it}) & \text{if} & u_{it}\leq y_{it} \\
                \end{array}
              \right.
\end{equation}
where $\gamma_{j,it}$, $j=1,2,3$, are the respective weights of the components,
$\phi(\cdot;\mu,s^2)$ refers to the Gaussian pdf with mean $\mu$ and variance $s^2$, $e(\cdot;\lambda)$ to the exponential pdf with parameter $\lambda>0$ defined by $e(y;\lambda)=\lambda\,e^{-\lambda\,y}$, 
and $h(z;\cdot)$ denotes the pdf of the GPD given by
$$
h(z;\xi,\sigma)=\left\{
                \begin{array}{lcl}
                  (1+\xi\,z/\sigma)^{-1-1/\xi}/\sigma & \text{if} &\xi\neq 0\\
                 e^{-z/\sigma}/\sigma & \text{if} & \xi = 0 \\                  
                \end{array}
              \right.$$
where $\sigma>0$, $z\in\mathbb{R}_+$ if $\xi\geq 0$ and $z\in [0,-\sigma/\xi]$ if $\xi<0$. Hence, in this model, an observation $y_{it}$ will be considered extreme if $y_{it}\geq u_{it}$, and its exceedance above $u_{it}$ assumed to follow a GPD with tail parameter $\xi_{it}$. The parameter $u_{it}$ plays the role of the conditional threshold used in the POT approach. Note also the role of the exponential bridge: it allows for a smooth transition between the two main components of the density. Furthermore, it allows for a greater ability to determine automatically the threshold above which observations are considered as extremes, since it provides an additional degree of freedom (via $u_{it}^{*}$) to describe the distribution of the data (see \citealp{debbabi} for more details). From~\eqref{eq:density}, we can therefore devise a likelihood-based estimation procedure for $\xi_{it}$ that uses all the data at hand. Finally, notice also that the proposed structure circumvents some of the conceptual issues associated with the absence of threshold stability property in EVR models \citep{eastoe}: by estimating directly $\xi(\mathbf{x}_{it})$ and $\sigma(\mathbf{x}_{it})$ without selecting \textit{ex ante} a threshold, it becomes irrelevant to investigate the effect of a change in threshold on the estimated regression model. 

To reduce the dimensionality of the problem, we assume that $f$ is continuous and differentiable (${\cal C}^1$) at the two junction points between the components. Moreover, since $f$ is a pdf, we have that 
$$\int\limits_{\mathcal{R}}f(y;\pmb{\theta},\mathbf{x})dy = 1.$$
From these equalities, we reduce the vector of free distribution parameters to  $[\mu_{0}\,,s_{it}\,, \sigma_{it},\, \xi_{it}]$, the other parameters satisfying
\begin{equation}\label{relations-param}
{\small\left\{
  \begin{array}{ll}
u_{it}^{*}=\mu_{0}+\lambda_{it} \,s_{it}^2 \,;\qquad	\lambda_{it}=(1+\xi_{it})/\sigma_{it};  &
\gamma_{1,it}=\gamma_{2,it} \,  \displaystyle \frac{e(u_{it}^{*}\,;\lambda_{it})}{\phi(u_{1,i}\;;\mu_{0},s_{it}^2)}; 
\\
\gamma_{2,it}=\Big[ \xi_{it} \, e^{-\lambda_{it}\,u_{it}} +  \left(1+\lambda_{it}\,\displaystyle \frac{ \Phi(u_{it}^{*}\,;\mu_{0},s_{it}^2)}{\phi(u_{it}^{*}\,;\mu_{0}, s_{it}^2)} \right) e^{-\lambda_{it}\,u_{it}^{*}}\Big]^{-1}\!\!; &
\gamma_{3,it}=\sigma_{it}\,\gamma_{2,it} \, e(u_{it};\lambda_{it}).
  \end{array}
\right.}
\end{equation}
See \cite{debbabi} for details. In addition, to enforce $\displaystyle u_{it}^{*}\leq u_{it}$, $\forall i,t$,  we rely on Theorem 3.4.5 in \cite{embrechtsbook}, pg. 160 (eq. 3.46), and define the threshold $u_{it}$ as
\begin{equation}\label{eq:threshold}
u_{it}=u_{it}^{*}+\sigma_{it}/\xi_{it}.
\end{equation} 
Indeed, the model is formally identified for $u_{it}=c\sigma_{it}/\xi_{it}\geq u_{it}^{*}$ with $c$ being any positive constant (see \citealp{embrechtsbook}). Eq.~\eqref{eq:threshold} is therefore similar to setting $c$ as $u_{it}^{*}\xi_{it}/\sigma_{it}+1$. Notice that, for $\sigma_{it} \approx 0$, $u_{it}^{*}\approx u_{it}\rightarrow +\infty$, and the model is equivalent to the Gaussian location-scale regression model. 

Thus, as soon as we allow at least $\xi_{it}$ to be driven by covariates, we also implicitly define a model in which the threshold parameter $u_{it}$ depends on the covariates, a feature in line with the hypotheses of EVR models. As a corollary, estimating model~\eqref{eq:density} automatically sets the threshold $u_{it}$ at a level for which the distribution of the data is supposed to be well approximated by a GPD, lifting the need to estimate this conditional threshold \textit{ex ante} and in isolation. To ensure sufficient flexibility and account for heteroscedasticity, we allow $s_{it}$ to be a function of the covariates as well, specifying
\begin{equation}
\log(s_{it})=\mathbf{x}_{it}^{T}\pmb{\beta}^{s},
\end{equation}
and 
$\pmb{\theta}=\left(\mu_{0},\pmb{\beta}^{s},\pmb{\beta}^{\sigma},\pmb{\beta}^{\xi}\right)\in \Theta \subset \mathbb{R}^{3d+4} $ is the vector of all parameters to be estimated. Notice that we can easily assume that different sets of covariates enter the equations of each distribution parameter, or that some of them are left constant. 

\subsection{Estimation under misspecification}

A recurrent challenge with splicing models involves defining parametric models valid for the complete support of the data. In particular, if the chosen distribution for the bulk of the data is too far from the true data generating process, we quickly encounter numerical issues with likelihood-based estimation procedures. This is a problem to which EVR is rather immune, since the GPD approximates most tail distributions encountered in practice well. To retain simultaneously the computational ease of a parametric likelihood-based estimation procedure and the robustness of EVR to specification problems, we propose using a robust weighted maximum likelihood estimator (WMLE) instead of the classical MLE to estimate $\pmb{\theta}$, as suggested by \cite{hu2002} and \cite{wang2005}. The general idea of this estimator involves controlling for potential specification problems in the body of the data (and in particular in the left tail), by lowering the importance of these observations in the likelihood function. To do so, we use a weighting scheme based on an \textit{artificial censoring procedure}, reducing our problem to the maximization of a \textit{censored likelihood function}, a proper scoring function introduced in \cite{cuesta2008} and \cite{diks2011}.
First, let us start by defining the maximum likelihood estimator:
\begin{equation}
\hat{\pmb{\theta}} = \arg\max_{\pmb{\theta}\in\Theta}\sum\limits_{i=1}^{I}\sum\limits_{t=t_{i,1}}^{t_{i,n_{i}}}\log(f(y_{it};\pmb{\theta},\mathbf{x}_{it})),
\end{equation} where $y_{it}$ denotes the observation at time $t$ for entity $i$ reporting over $n_{i}$ periods, with $t_{i,1}$ the first period for which entity $i$ reports in the database, and $t_{i,n_{i}}$ the last period. Remember that since our application of interest relates to hedge funds, hardly any entities (i.e. funds) report over the complete period under study. We are therefore dealing with an unbalanced panel. Our WMLE is then given by:
\begin{equation}\label{eq:wmle}
\begin{aligned}
\hat{\pmb{\theta}}^{w}(\tau) = & \arg\max_{\pmb{\theta}\in\Theta}\sum\limits_{i=1}^{I}\sum\limits_{t=t_{i,1}}^{t_{i,n_{i}}}\mathbbm{1}(y_{it}\geq q(\tau))\log(f(y_{it};\pmb{\theta},\mathbf{x}_{it}))\\
&+\mathbbm{1}(y_{it}<q(\tau))\log(F(q(\tau);\pmb{\theta},\mathbf{x}_{it})),
\end{aligned}
\end{equation} where $q(\tau)$ is a censoring threshold. $F(q(\tau);\pmb{\theta},\mathbf{x}_{it})$ is the cumulative distribution function of $y_{it}$ at point $q(\tau)$. The idea behind this artificial censoring mechanism is to give less weight to observations $y_{it}<q(\tau)$ in the left tail and the body of the distribution, where our distributional assumptions are probably wrong and of less importance. On the contrary, in the right tail, since our model relies on EVT, we have an asymptotically justified model holding under fairly general conditions and which should not be problematic in the estimation procedure. The principle of estimator~\eqref{eq:wmle} is also known as \textit{minimum scoring rule inference} \citep{dawid2016} and guarantees unbiased estimation under limited assumptions. The theoretical properties of \eqref{eq:wmle} are discussed in Section~\ref{sec:theory}. For the censoring threshold $q(\tau)$, we propose using either an unconditional threshold at level $\tau$, or observation-specific thresholds, such as conditional quantiles. In the latter case, our estimator of $\pmb{\theta}$ becomes a conditional weighted maximum likelihood estimator (CWMLE) and is given by: 
\begin{equation}
\begin{aligned}
\hat{\pmb{\theta}}^{cw}(\tau) =& \arg\max_{\pmb{\theta}\in\Theta}\sum\limits_{i=1}^{I}\sum\limits_{t=t_{i,1}}^{t_{i,n_{i}}}\mathbbm{1}(y_{it}\geq q_{it}(\tau))\log(f(y_{it};\pmb{\theta},\mathbf{x}_{it}))\\
&+\mathbbm{1}(y_{it}<q_{it}(\tau))\log(F(q_{it}(\tau);\pmb{\theta},\mathbf{x}_{it})),
\end{aligned}
\end{equation}
where $q_{it}(\tau)$ denotes the \textit{conditional} quantile of $y_{it}$ at level $\tau$, estimated, e.g., via quantile regression \citep{koenker1978} and defined as
$$\mathbb{P}(y_{it}\leq q_{it}(\tau)|\mathbf{X}=\mathbf{x}_{it})=\tau.$$
In Section~\ref{sec:simu}, we compare the use of an observation-specific threshold to the simpler approach of using a global empirical (unconditional) quantile at level $\tau$. To choose $\tau$ (termed the robustness parameters), we describe a data-driven approach in Section~\ref{subsec:tau_empirical}. Final estimation is conducted with numerical optimization routines. As starting values, we use the following quantities throughout the paper for the constant parameters:
\begin{itemize}
\itemsep-0.5em 
\item $\mu_{0}$: the mean of the sample from which we removed the top 20\% of observations.
\item $\beta^{s}_{0}$: the log of the mean absolute deviation of the sample, from which we removed the top 20\% of observations.
\item $\beta^{\xi}_{0}$: the log of the unconditional MLE computed on the top 5\% of the data.
\item $\beta^{\sigma}_{0}$: the log of the unconditional MLE computed on the top 5\% of the data.
\end{itemize}

All initial estimates of the regression parameters are set to small values close to $0$, i.e. $0.001$.

\subsection{Theoretical properties and inference}\label{sec:theory}

In order to make inference on the parameters vector $\pmb{\theta}$, and propose confidence intervals for each of its components, we focus in this section on the asymptotic properties of the estimators introduced in the previous section. It is worth noticing that most of these properties may be derived from the existing literature. Indeed, let us recall that our WMLE is defined as:
$$ \hat{\pmb{\theta}}^{w}(\tau) = \arg\max_{\pmb{\theta} \in \Theta} M_n(\pmb{\theta})= \arg\max_{\pmb{\theta} \in \Theta } \frac{1}{n} \sum\limits_{i=1}^{n} m_{\pmb{\theta}}(y_i,\mathbf{x}_{i}), $$  
where $\pmb{\theta}=\left(\mu_{0},\pmb{\beta}^{s},\pmb{\beta}^{\sigma},\pmb{\beta}^{\xi}\right)\in \Theta \subset \mathbb{R}^{3d+4}$, $n=\sum \limits_{i=1}^I n_i$ and
$$ m_{\pmb{\theta}}(y,\mathbf{x})= \mathbbm{1}(y \geq q(\tau))\log(f(y;\pmb{\theta},\mathbf{x}))+\mathbbm{1}(y<q(\tau))\log(F(q(\tau);\pmb{\theta},\mathbf{x})).$$  
By considering $q(\tau)$ as known (in practice, we plug in an empirical quantile of $(y_{it})_{i,t}$), this estimator is an M-estimator, as defined in \cite{van2000asymptotic}. The asymptotic normality of this estimator may thus be derived from the literature on M-estimation, as follows. Introducing the vector-valued function $\psi_{\pmb{\theta}}(.,.)=\frac{\partial}{\partial \pmb{\theta}} m_{\pmb{\theta}}(.,.)$, then, under the following assumptions:
\begin{enumerate}[label=(\Roman*)]
    \item $\psi_.(y,\mathbf{x})$ fulfills a locally Lipschitz-type condition, i.e. for $\pmb{\theta}_1$ and $\pmb{\theta}_2$ in the neighborhood of $\pmb{\theta}_0$,  
    $$ || \psi_{\pmb{\theta}_1}(y,\mathbf{x}) -  \psi_{\pmb{\theta}_2}(y,\mathbf{x}) || \leq \overline{\psi}(y,\mathbf{x}) \,|| \pmb{\theta}_1 - \pmb{\theta}_2 ||, $$
    where $\overline{\psi}$ is a measurable function such that $\mathbb{E}[\overline{\psi}(y,\mathbf{x})^2]<\infty$,
    \item $\mathbb{E}\left[ || \psi_{\pmb{\theta}_0}(y,\mathbf{x}) ||^2 \right]<\infty$ and the map $\pmb{\theta} \rightarrow \mathbb{E}\left[ || \psi_{\pmb{\theta}_0}(y,\mathbf{x}) || \right]$ is differentiable at $\pmb{\theta}_0$ with nonsingular derivative matrix $\pmb{V}_{\pmb{\theta}_0} $,
    \item $\hat{\pmb{\theta}}^{w}(\tau) \overset{\mathbb{P}}{\rightarrow} \pmb{\theta}_0$\; and \;$\displaystyle \frac{1}{n} \sum\limits_{i=1}^{n} \psi_{\hat{\pmb{\theta}}^{w}(\tau)}(y_i,\mathbf{x}_{i})= o_{\mathbb{P}} \left( n^{-1/2} \right)$,
\end{enumerate}
we have, as $n\to \infty$,
\begin{equation}
\sqrt{n} \left( \hat{\pmb{\theta}}^{w}(\tau) - \pmb{\theta}_0  \right) \rightarrow \mathcal{N} \left( \pmb{0}, \pmb{V}_{\pmb{\theta}_0}^{-1} \mathbb{E} \left[ \psi_{\pmb{\theta}_0}(y,\mathbf{x}) \psi_{\pmb{\theta}_0}(y,\mathbf{x})^{\top} \right] \left( \pmb{V}_{\pmb{\theta}_0}^{-1} \right)^{\top} \right).\label{eq:CI}
\end{equation}

\noindent This asymptotic normality may thus be used to construct confidence regions for our estimators, under the assumption that Conditions (I), (II), and (III) are fulfilled. These conditions are difficult to check theoretically for our model, which is why we turn to their empirical counterparts. Nevertheless, in Section~\ref{sec:theoretical_suite} of the Appendix, we illustrate this study in a simplified setting, considering a Pareto distribution as a representative example of the maximum domain of attraction of a Fr\'echet distribution. 
\newpage
\begin{rk}~
    \begin{itemize}
    \item[(i)] The term $\mathbb{E} \left[ \psi_{\pmb{\theta}_0}(y,\mathbf{x}) \psi_{\pmb{\theta}_0}(y,\mathbf{x})^{\top} \right]$ in the asymptotic variance above may be estimated by its empirical counterpart 
    $$ 
    \frac{1}{n} \sum\limits_{i=1}^{I}\sum\limits_{t=t_{i,1}}^{t_{i,n_{i}}} \psi_{\hat{\pmb{\theta}}^{w}(\tau)}(y_{it},\mathbf{x}_{it}) \psi_{\hat{\pmb{\theta}}^{w}(\tau)}(y_{it},\mathbf{x}_{it})^{\top}, 
    $$
    where $\psi$ is obtained numerically. The derivative matrix $ \pmb{V}_{\pmb{\theta}_0}$ can also be approximated numerically.  
    %
    \item[(ii)] Condition (III) requires the consistency of the estimator $\hat{\pmb{\theta}}^{w}(\tau)$ of $\pmb{\theta}_0$. Conditions can be introduced to ensure this consistency, namely:
    \begin{enumerate}[label=\roman*)]
        \item $\underset{\pmb{\theta} \in \Theta}{\sup} \left| \left| \frac{1}{n} \sum\limits_{i=1}^{n} \psi_{\pmb{\theta}}(y_i,\mathbf{x}_{i}) - \mathbb{E} \left[ \psi_{\pmb{\theta}}(y,\mathbf{x}) \right]   \right| \right| \overset{\mathbb{P}}{\rightarrow} 0$, as $n\to\infty$;
        \item For all $\varepsilon>0$, 
        $\displaystyle
        \underset{||\pmb{\theta}-\pmb{\theta}_0|| \geq \varepsilon}{\text{inf}} \left| \left| \mathbb{E} \left[ \psi_{\pmb{\theta}}(y,\mathbf{x}) \right] \right| \right| > 0 = \left| \left| \mathbb{E} \left[ \psi_{\pmb{\theta}_0}(y,\mathbf{x}) \right] \right| \right|$.
    \end{enumerate}
    These conditions may also be replaced by similar conditions on the empirical criterion function $M_n(\pmb{\theta})$: 
    $$
    \sup_{\pmb{\theta} \in \Theta}  \left| M_n(\pmb{\theta})- \mathbb{E}[m_{\pmb{\theta}}(y,\mathbf{x})] \right| \overset{\mathbb{P}}{\rightarrow} 0 \quad \text{and}\quad \sup_{\pmb{\theta} : ||\pmb{\theta}-\pmb{\theta}_0|| \geq \varepsilon} \mathbb{E}[m_{\pmb{\theta}}(y,\mathbf{x})] < \mathbb{E}[m_{\pmb{\theta}_0}(y,\mathbf{x})].
    $$ 
    \end{itemize} 
\end{rk}

\subsection{Data-driven choice of $\tau$}\label{subsec:tau_empirical}

Our estimators $\hat{\pmb{\theta}}^{w}(\tau)$ and $\hat{\pmb{\theta}}^{cw}(\tau)$ depend on the robustness parameter $\tau$. This quantity determines the intensity of the censoring mechanism: since $q(\tau)$ and $q_{it}(\tau)$ are defined as quantiles and conditional quantiles, respectively, a large $\tau$ implies that a large fraction of the observations is censored in the likelihood function. If the data exhibit large deviations from the G-E-GPD model, a large $\tau$ may be thus required to overcome misspecification issues. However, there is simultaneously a need to set $\tau$ as small as possible to avoid losing valuable information from data points belonging to a region of the distribution that is specified correctly. 

With the ambition to respect the features of the data, we introduce in this section a data-driven selection method, relying on the modified Anderson-Darling (AD$^{m}$) statistic. This quantity assesses the goodness-of-fit of a parametric distribution and gives a special weight to extreme values in the upper tail of a distribution \citep{babu2016}. The use of this statistic continues the approach advocated by \cite{davisonsmith1990}, \cite{heo2013}, \cite{choulakian2001}, and \cite{bader2018} (among others) to select an appropriate threshold in classical EVT.

Denoting by $\pmb{\theta}^{w}(\tau)$ or $\pmb{\theta}^{cw}(\tau)$ the (C)WMLE of the splicing model obtained with a specific $\tau$, we compute the corresponding pseudo-residuals obtained from the probability integral transform (PIT), namely
$$\hat{\epsilon}_{it}(\tau)=\Phi^{-1}(F(y_{it};\hat{\pmb{\theta}}^{w}(\tau),\mathbf{x}_{it})),$$
with $\Phi^{-1}$ being the quantile function of the standardized normal. Under a correct specification and a good estimation of the splicing model, we expect these residuals to be approximately normally distributed \citep{dunn1996}, in particular in the tail. The AD$^{m}$ statistic is then given by
\begin{equation}\label{eq:mad}
    AD^{m}(\tau) = n\int\limits_{-\infty}^{+\infty}\dfrac{\left(\Phi(x)-\tilde{F}_{n}(x,\tau)\right)^{2}}{1-\Phi(x)}d\Phi(x)
    \end{equation}
where $\tilde{F}_{n}(\cdot,\tau)$ denotes the empirical distribution function of the re-indexed pseudo-residuals $\hat{\epsilon}_{k}(\tau)$, for $k=1,\ldots,n$ and $n=\sum_{i=1}^{I}n_{i}$. Eq.~\eqref{eq:mad} measures departures from the expected Gaussian distribution. The sample version of Eq.~\eqref{eq:mad} is given in \cite{heo2013} (equation (9)). Computing $AD^{m}(\tau)$ for various values of $\tau$,
we select $\tau^{opt}$, such that 
$$
\tau^{opt}=\arg\min_{\tau} AD^{m}(\tau).
$$ 
From a practical standpoint, both the simulation study and the empirical analysis presented in the next sections suggest that choosing $\tau \in \left[0.10,0.30\right]$ improves numerical stability and delivers improved estimations. In the Appendix, we provide further evidence that choosing $\tau\approx 0.20$ is a good trade-off between accuracy and computational simplicity for a unimodal distribution. 

\section{Simulation evidence}\label{sec:simu}

\noindent In this section, we assess the finite sample properties of the proposed methodology under realistic data-generating processes (DGP). Our objective is to quantify the gains in terms of bias and variance of our splicing approach over standard POT approaches. The DGP are detailed in the next subsections. We consider three settings: (i) correct specification, (ii) contaminated body, and (iii) misspecified model. In the first setting, the data are exactly distributed as the splicing model, whereas in the second and third settings the splicing model suffers from misspecifications in the body of the distribution. In addition, in the third setting, the splicing model is only asymptotically correct in the upper tail.  

To select $\tau$, we use the procedure described in Subsection~\ref{subsec:tau_empirical}, based on the $AD^{m}$ statistic. The selection is made over a grid of 20 values equally spaced between $0.05$ and $0.5$. We compare the quality of the estimated parameters when the weight function is taken either as the empirical quantile of the data at level $\tau$, or as estimated conditional quantiles obtained from parametric quantile regression at level $\tau$. 

As performance measure, we report the root mean squared error (RMSE) of the estimated parameters $\hat{\pmb{\beta}}^{\xi}$, given as (we denote generically the $j$-th element of the vector $\pmb{\beta}^{\xi}$ by $\beta^{\xi}_{j}$, $j=0,\ldots,d$, with $\beta^{\xi}_{0}$ referring to the constant term):
\begin{eqnarray}\label{eq:rmse_beta}
RMSE_{j}=\sqrt{\dfrac{1}{B}\sum\limits_{b=1}^{B}\left(\dfrac{\hat{\beta}^{\xi,(b)}_{j}-\beta^{\xi}_{j}}{|\beta^{\xi}_{j}|}\right)^{2}},
\end{eqnarray}
where $\hat{\beta}^{\xi,(b)}_{j}$ denotes an estimate computed on sample $b$, for $b=1,\ldots,B$. As a second criterion, we compute the empirical coverage probability of the 95\% confidence intervals for $\beta^{\xi}_{j}$, $j=0,\ldots,d$, obtained from Eq.~\eqref{eq:CI}. We also report the average length of these confidence intervals. We compare the results obtained with the proposed WMLE with those of the MLE and classical POT-EVR methods, with different threshold selection methods:
\begin{itemize}
\itemsep0em 
    \item A conditional quantile regression approach obtained from a quadratic polynomial in the covariate. We use as thresholds the estimated conditional quantiles at levels 90\%, 95\% or 99\%. A quadratic polynomial is used here to give the POT approach more flexibility, since our DGPs imply that the true conditional quantile is not linear in the covariate.
    \item An unconditional quantile approach. We simply use as thresholds the empirical quantiles at levels 90\%, 95\%, or 99\%. 
\end{itemize}
We then estimate the POT-EVR model on the selected extremes with maximum likelihood procedures. We assume that $\xi_{it}$, $\sigma_{it}$, and $s_{it}$ are driven by the same covariates. Throughout the simulation study, we set the number of samples $B$ to $200$ for each DGP. To mimic the fact that we usually work with panel data, we simulate realizations of the DGPs, at each time point $t=1,\ldots,T$ with $T=250$, as if we had $I=40$ reporting entities over time. As such, observations at a given time point $t$ have the same conditional density and $n=TI$ (the total sample size) is equal to $10,000$. Although appearing large, this sample size is much smaller than the total sample size in our empirical application, in which we use around 189,000 month-fund observations. It also gives us the opportunity to test the POT approaches in a setting with a sufficiently large number of extreme observations.

\subsection{Data-generating processes}\label{subsec:simu_corspec}

In this section, we describe the three DGPs used to simulate the data. First, we consider a model for which the splicing regression model is perfectly specified. Then, moving to more realistic DGPs, we introduce different model misspecifications. Throughout all DGPs, we assume that one ($d=1$) covariate drives $\xi_{it}$ (the tail index), $\sigma_{it}$ (the scale parameters), and $s_{it}$ (the volatility parameter of the body), reflecting the fact that financial data display heteroscedasticity and tail heterogeneity. The covariate is simulated from an AR(1) process, to mimic the fact that our explanatory variables (e.g. the VIX) are collected over time and usually exhibit autocorrelation:

$$x_{t}=.2+.5x_{t-1}+\epsilon_{t}, \text{ with } \epsilon_{t}\sim N(0,0.1).$$
For simplicity, we assume the covariate to be common to all entities at a given point in time, and no entity-specific covariates. We choose $\beta^{\xi}=\left[\log(0.2),\,1\right]$, $\beta^{s}=\left[\log(0.045),\,-.5\right]$ and $\beta^{\sigma}=\left[\log(0.08),\,.2\right]$. We set $\mu_{0}=0$. These values were chosen in line with estimates obtained in our empirical analysis.

\paragraph{Correct specification (DGP I)}
First, assume that the response variable $y_{it}$ is simulated from the splicing regression model~\eqref{eq:density}, making the model perfectly specified and expected to be estimated with classical likelihood-based inference well. We thus expect the censoring threshold $\tau$ to be close to 0. For this DGP, we expect the MLE of the splicing regression model to be the best estimator. Furthermore, we expect the POT estimator to be much less efficient, since it uses only a fraction of the data.
\paragraph{Contaminated body (DGP II)}
To consider a more realistic situation, we introduce in DGP II a misspecification in the body of the splicing distribution. To do so, we start by generating the data following the same splicing model as in DGP I. Then, we apply the following transformations to the simulated data:
\begin{itemize}
\itemsep-0.5em  
    \item We identify the 10\% smallest observations $y_{it}$ in the sample, and we denote by $t^{*}$ the empirical quantile at level 10\%. 
    \item These observations are then replaced by $y_{it}^{*}=t^{*}-(H^{-1}(z_{it}^{*},\sigma^{*},\xi^{*})),$ with $z_{it}^{*}$ being a random number simulated from a uniform distribution, $H^{-1}$ being the inverse of the GPD, $\sigma^{*}=0.08$ and $\xi^{*}=.1$. 
\end{itemize}

The data are therefore ``doubly GP'' distributed and heavily asymmetric, with the lower tail being heavier than the Gaussian distribution, a feature common in financial data. Nevertheless, the upper tail stays correctly specified and follows a GP regression model. Thus, the MLE is misspecified 
 (although only a small fraction of the data are contaminated), and we should observe an increase in its mean squared error. On the contrary, the POT estimator remains correctly specified for thresholds large enough, and we do not expect to observe major differences with respect to DGP I.

\paragraph{$T$-locations scale distribution (DGP III)}
Finally, we turn to the case in which the underlying model is only asymptotically valid in the tail. We consider a $t$-location scale distribution, characterized by 3 parameters: a mean parameter $\mu_{0}$, a scale parameter $s(x_{t})$, and a shape parameter $\nu(x_{t})$. As for the classical $t$-distribution, the tail index is given by $\nu(x_{t})$, such that the equivalent shape parameter of the limiting GP regression model is simply $\xi(x_{t})=\dfrac{1}{\nu(x_{t})}$ \citep{mcneil2015}. Consequently, specifying $$\nu(x_{t})=\exp(\beta_{0}^{*}+\beta_{1}^{*}x_{t}),$$ the true limiting GP regression model has a shape parameter given by
\begin{equation}
\xi(x_{t}) = \exp(-\beta_{0}^{*}-\beta_{1}^{*}x_{t}) =\exp(\beta_{0}+\beta_{1}x_{t}),
\end{equation}
with $\beta_{0}=-\beta_{0}^{*}$ and $\beta_{1}=-\beta_{1}^{*}$. We choose $\beta_{0}^{*}=-\log(0.2)$ and $\beta_{1}^{*}=-1$, such that the GP regression model has a dynamic of the shape parameter similar to DGP I and DGP II. 
Here, both the MLE and the POT estimators will suffer from an additional bias compared to DGP I and DGP II. 
\subsection{Description of the simulated data}

\begin{figure}[htbp]
    \centering
    \begin{tabular}{ccc}
\includegraphics[scale=.32]{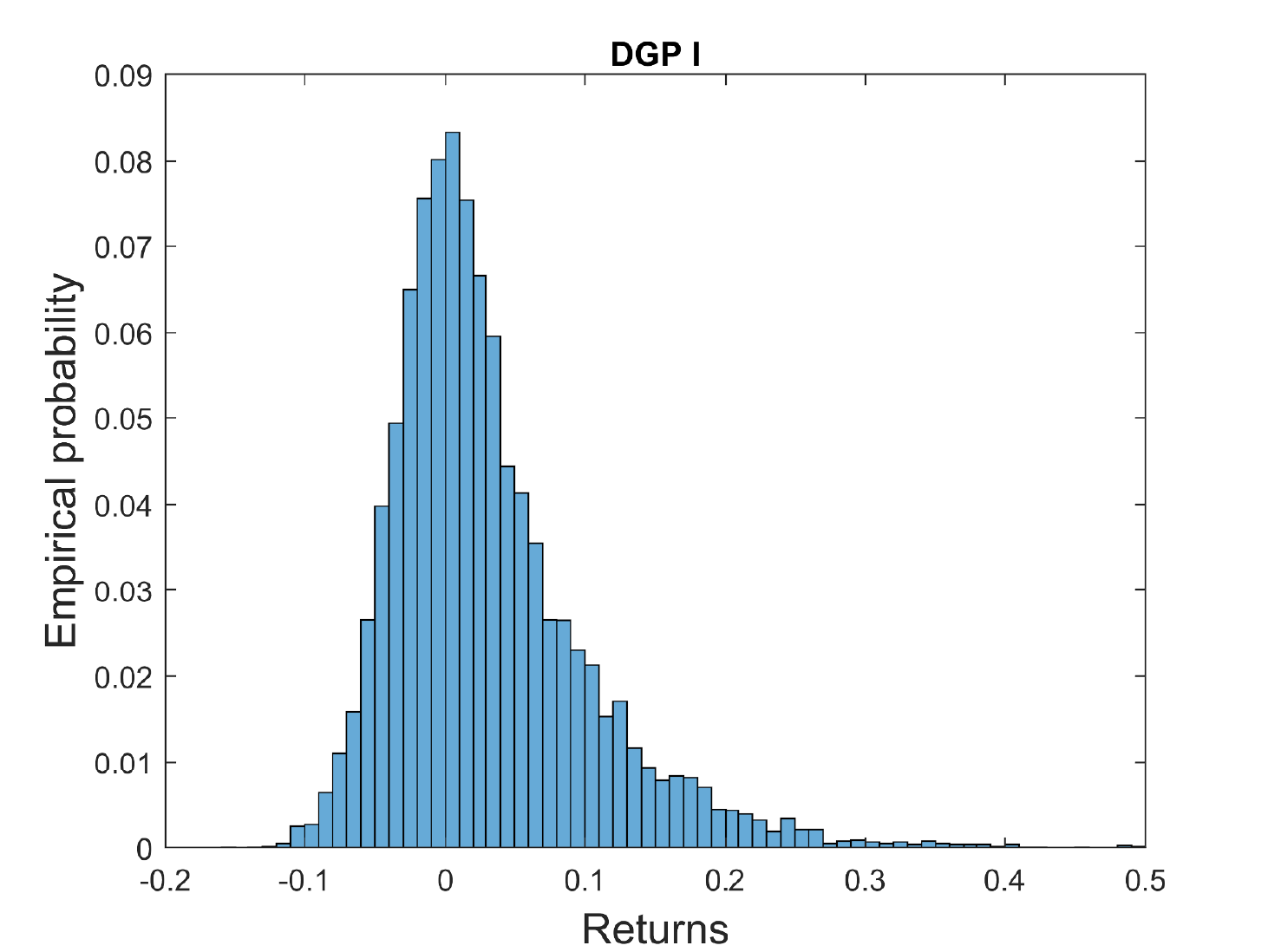} & \includegraphics[scale=.32]{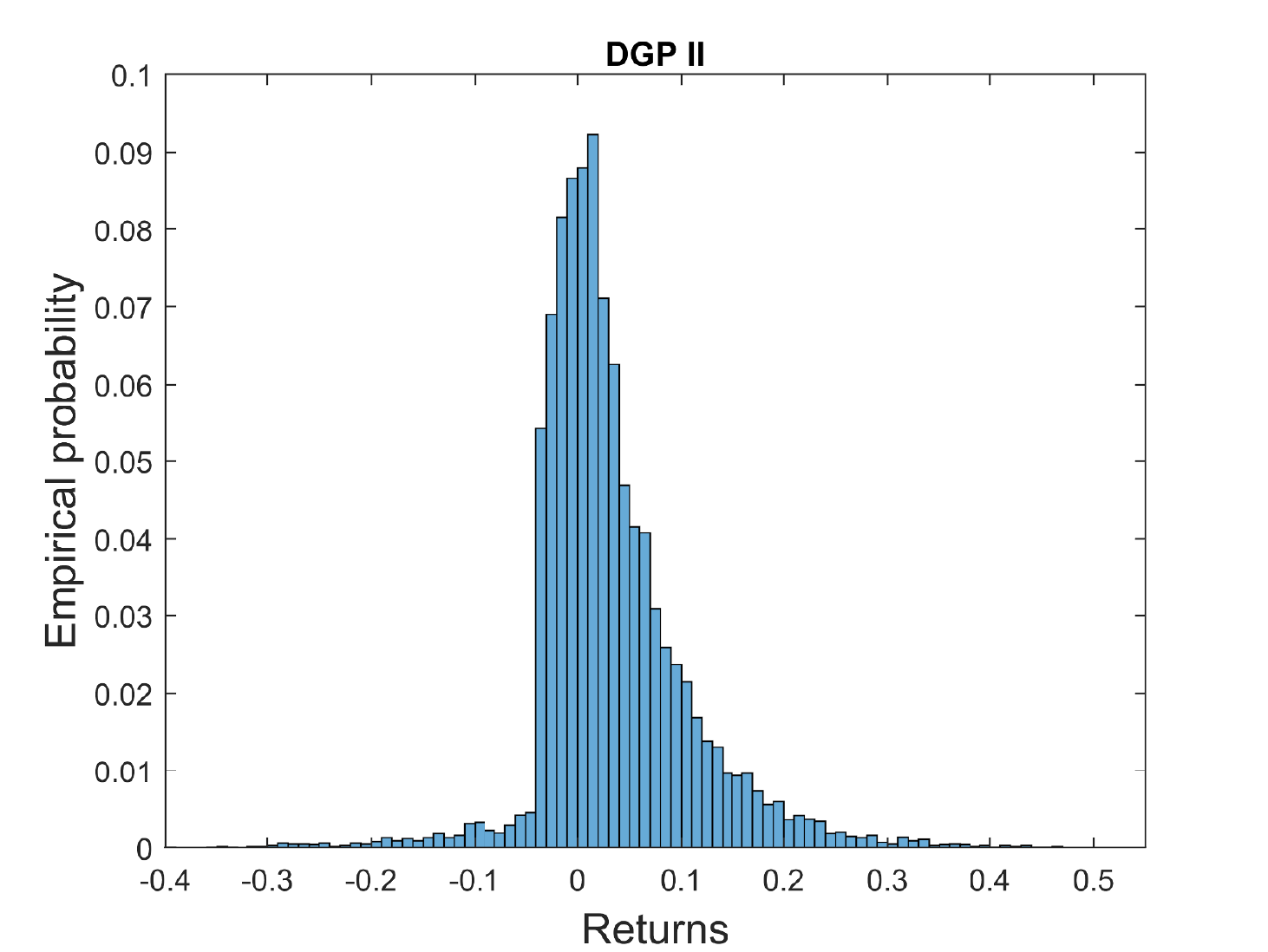} &\includegraphics[scale=.32]{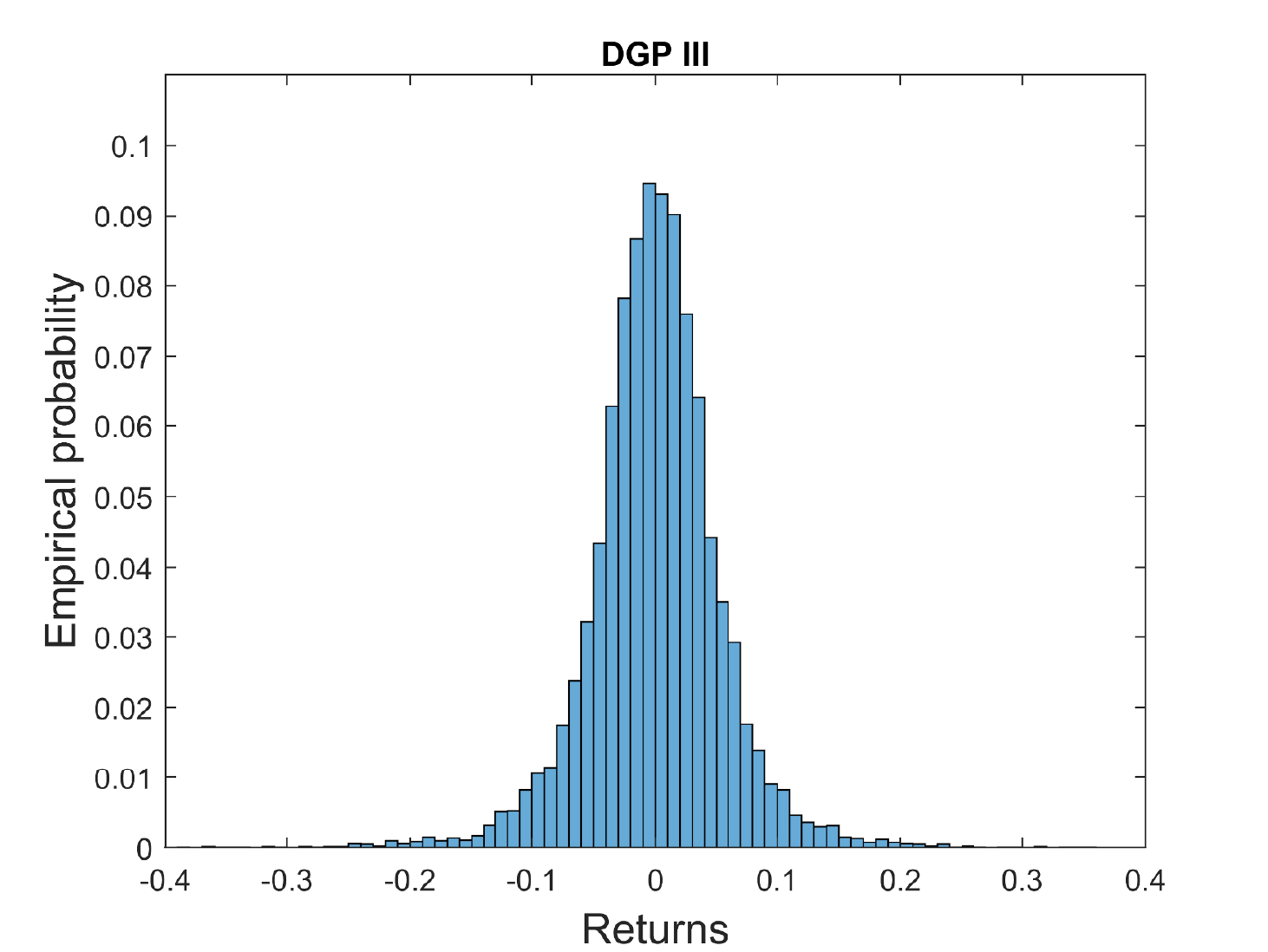}\\
(i) DGP I & (ii) DGP II & (iii) DGP III
\end{tabular}
    \caption{\footnotesize Simulated data from the different DGPs.}
    \label{fig:hist_dgp}
\end{figure}

We start by briefly describing the features of the simulated data. In Figure~\ref{fig:hist_dgp}, we display the histogram of the pooled data for each DGP. For DGP I, the data exhibit a heavy right tail and positive skewness. In this setup, the true conditional threshold, above which our random variable follows the extreme value regression model, is given by Eq.~\eqref{eq:threshold}, and is a nonlinear function of $x_{t}$. 
On average, around 0.9\% of the simulated data are larger than their corresponding $u_{it}$. For DGP II, we introduce a misspecification in the left tail, which is now heavier compared to DGP I. Similar fractions of observations are larger than $u_{it}$, since the models are identical in the right tail. Finally, for DGP III, both tails are heavier than the normal as well, and the right tail is not exactly GP-distributed. 

\subsection{Results}

\begin{figure}[htbp]
    \centering
    \begin{tabular}{cc}
    \includegraphics[scale = .45]{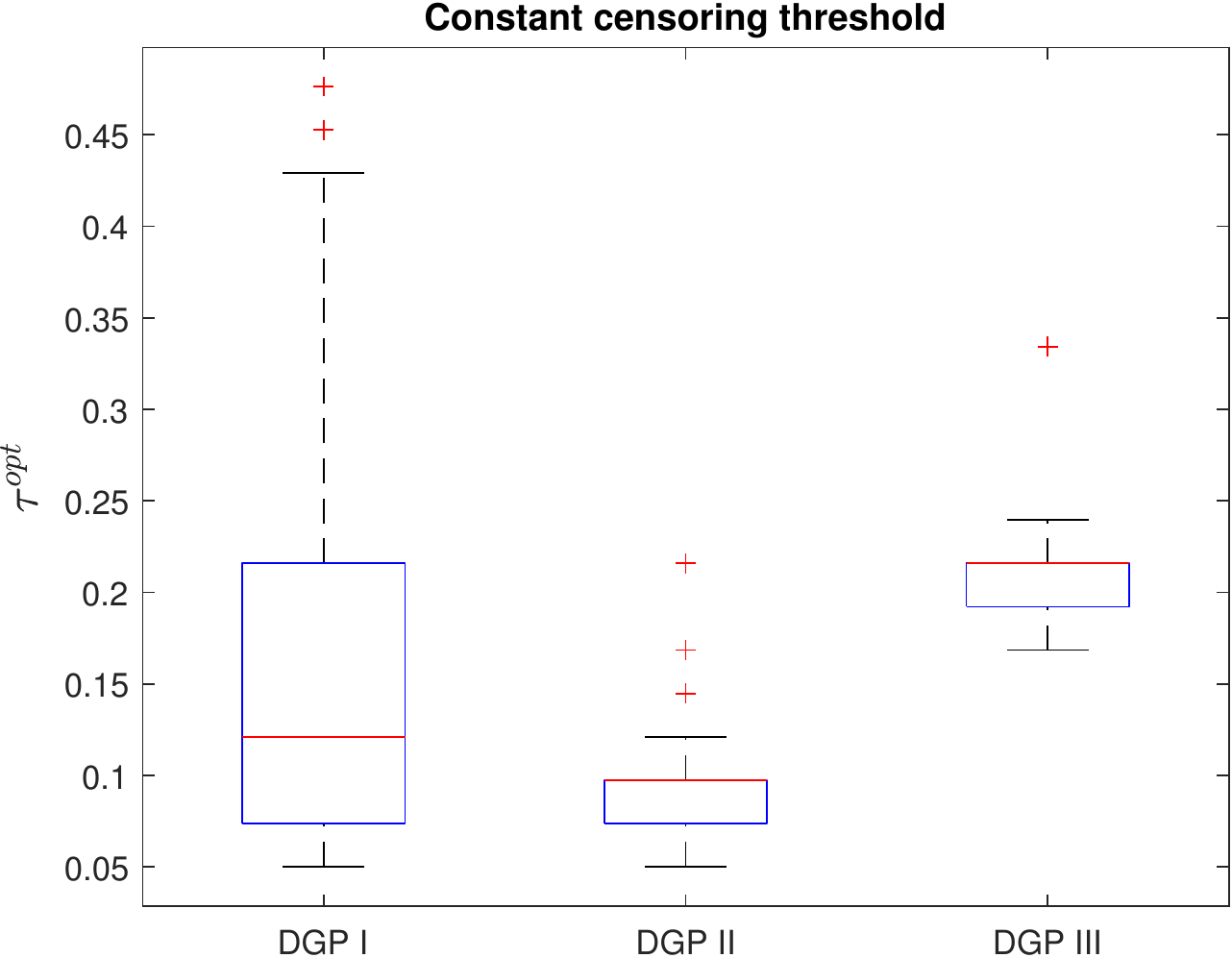} &     \includegraphics[scale = .45]{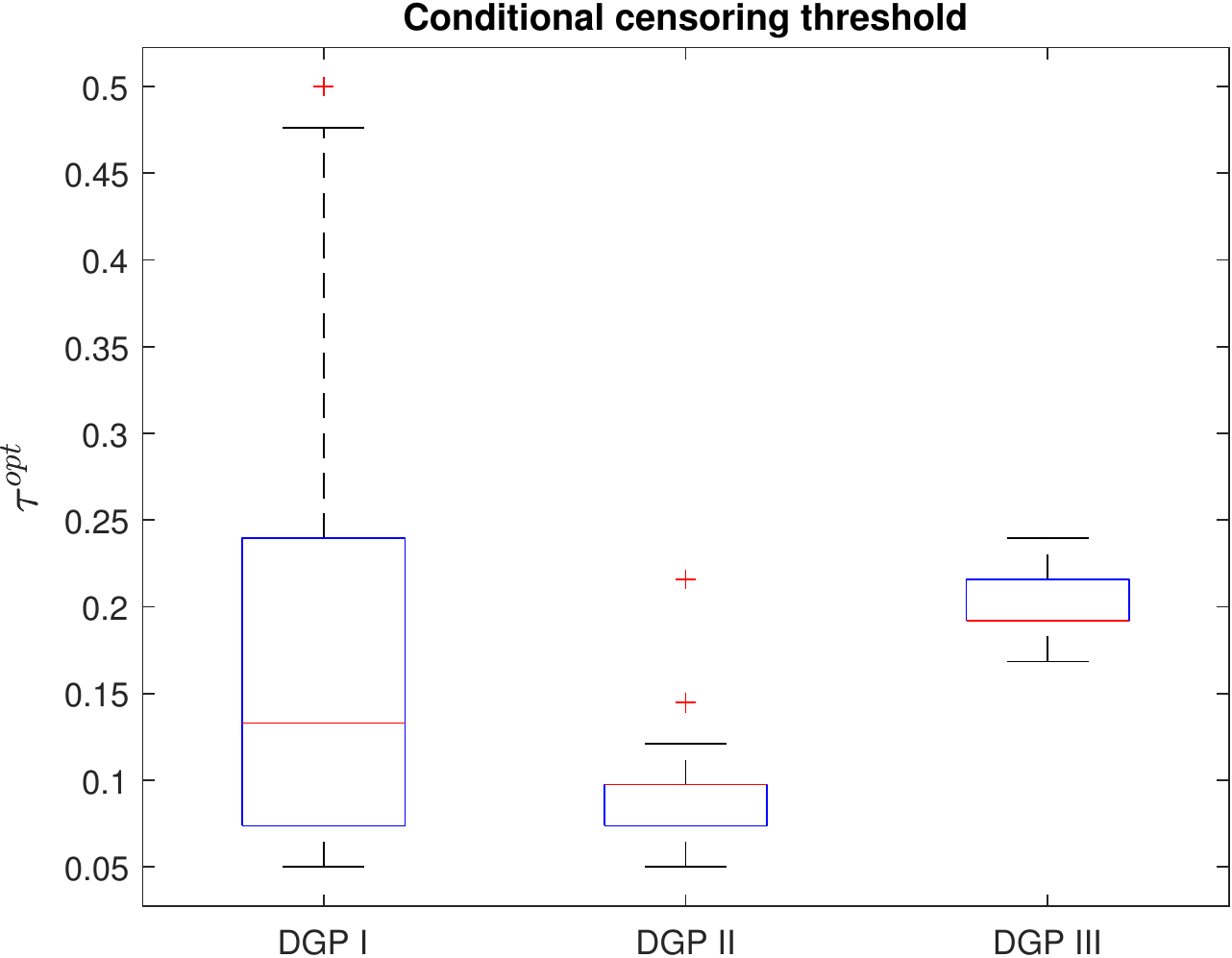}\\
    (i) Constant censoring threshold & (ii) Conditional censoring threshold
    \end{tabular}
    \caption{\footnotesize Selected censoring threshold $\tau^{opt}$, over 200 samples, with the censoring threshold being either (i) the empirical quantile ($q(\tau)$) or (ii) the estimated conditional quantile  at level $\tau$ ($q_{it}(\tau)$).}
    \label{fig:tau}
\end{figure}

We use now the methods described in Section~\ref{sec:methodo} to infer on the parameters of the GP regression model. We report first information on the selected $\tau$ (Figure~\ref{fig:tau}) under the different DGPs. 

\begin{figure}[H]
    \centering
    \begin{tabular}{c}
     \includegraphics[scale = .55]{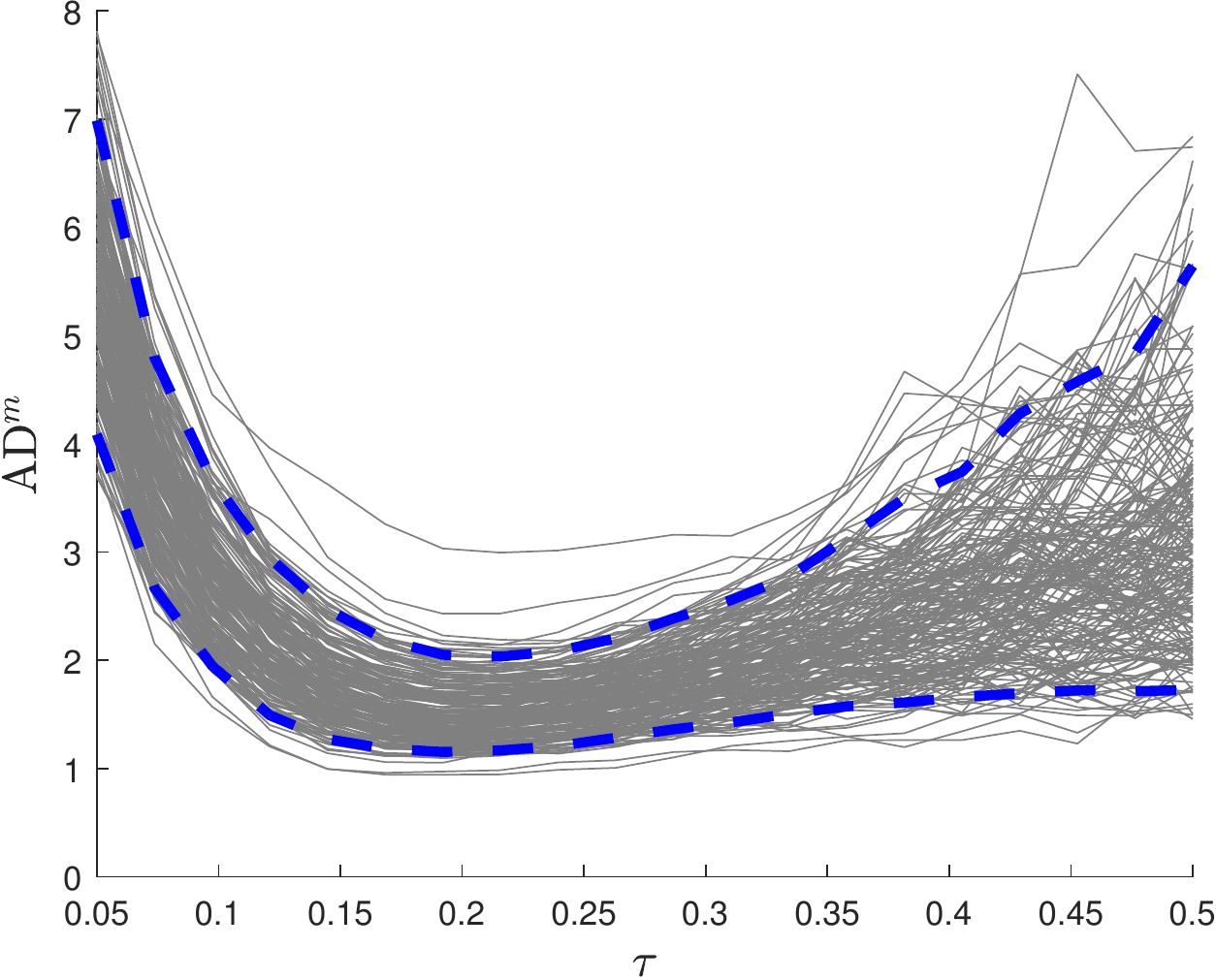} \\
       \end{tabular}
    \caption{\footnotesize Representative AD$^{m}$ statistics (grey), as functions of $\tau$ and using conditional censoring thresholds $q_{it}(\tau)$, for DGP III. Dashed blue: 5\% and 95\% empirical quantiles of AD$^{m}$ for a given $\tau$ level, across simulated samples.}
    \label{fig:mad_curve}
\end{figure}

For DGP I (correct specification), the selected $\tau$ values vary mostly between $0.05$ and $0.2$. This result supports the idea that a better fit in the right tail is obtained by decreasing, in the estimation procedure, the weights of observations in the left tail of the distribution. However, inspecting the sequences of the obtained AD$^{m}$ statistics for various $\tau$, we do not observe a clear minimum and only small differences for AD$^{m}$ computed with different $\tau$ values. This result is in line with the fact that, under a correct specification of the model, our censored estimate of $\xi$ is unbiased for any $0\leq\tau<1$. For DGP II, we observe values of $\tau$ systematically close to 0.1. Hence, the selection procedure succeeds in identifying the contamination threshold (here corresponding to the unconditional quantile at level 10\% of the data)\footnote{Repetitions of the simulation experiment with different contamination rates can be found in the Appendix.}. Finally, turning towards DGP III, the approach based on AD$^{m}$ delivers consistent selections across samples, with a selected $\tau^{opt}$ always around $0.2$, suggesting that a larger fraction of the data deviates from the G-E-GPD model. The dispersion of $\tau^{opt}$ is much lower in DGP III compared to DGP I. In Figure~\ref{fig:mad_curve}, we display typical AD$^{m}$ values obtained with $q_{it}(\tau)$, for DGP III. We observe a clear minimum for each sample. 

\begin{figure}[H]
    \centering
    \begin{tabular}{ccc}
    \includegraphics[scale = .38]{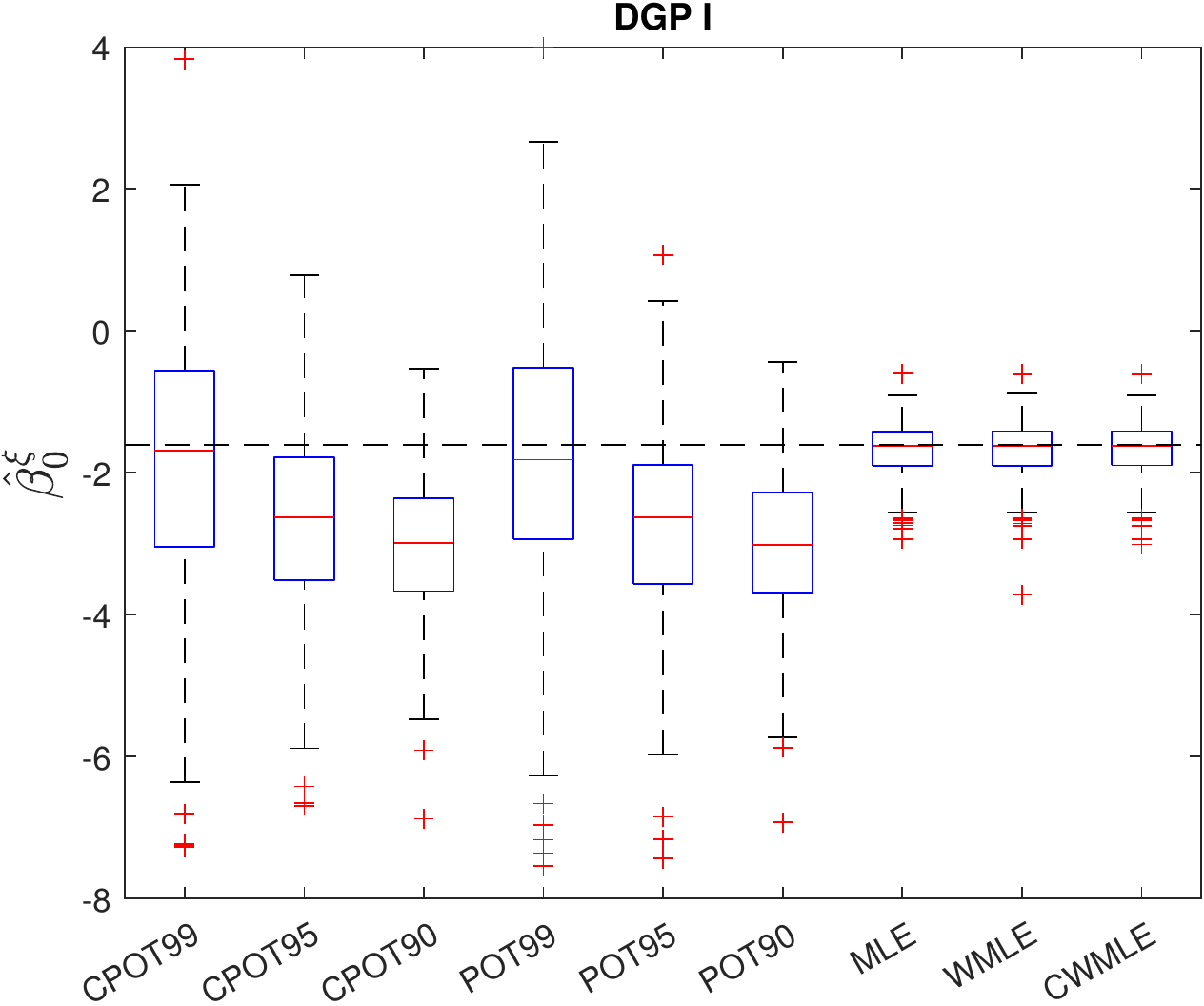}  & \includegraphics[scale = .38]{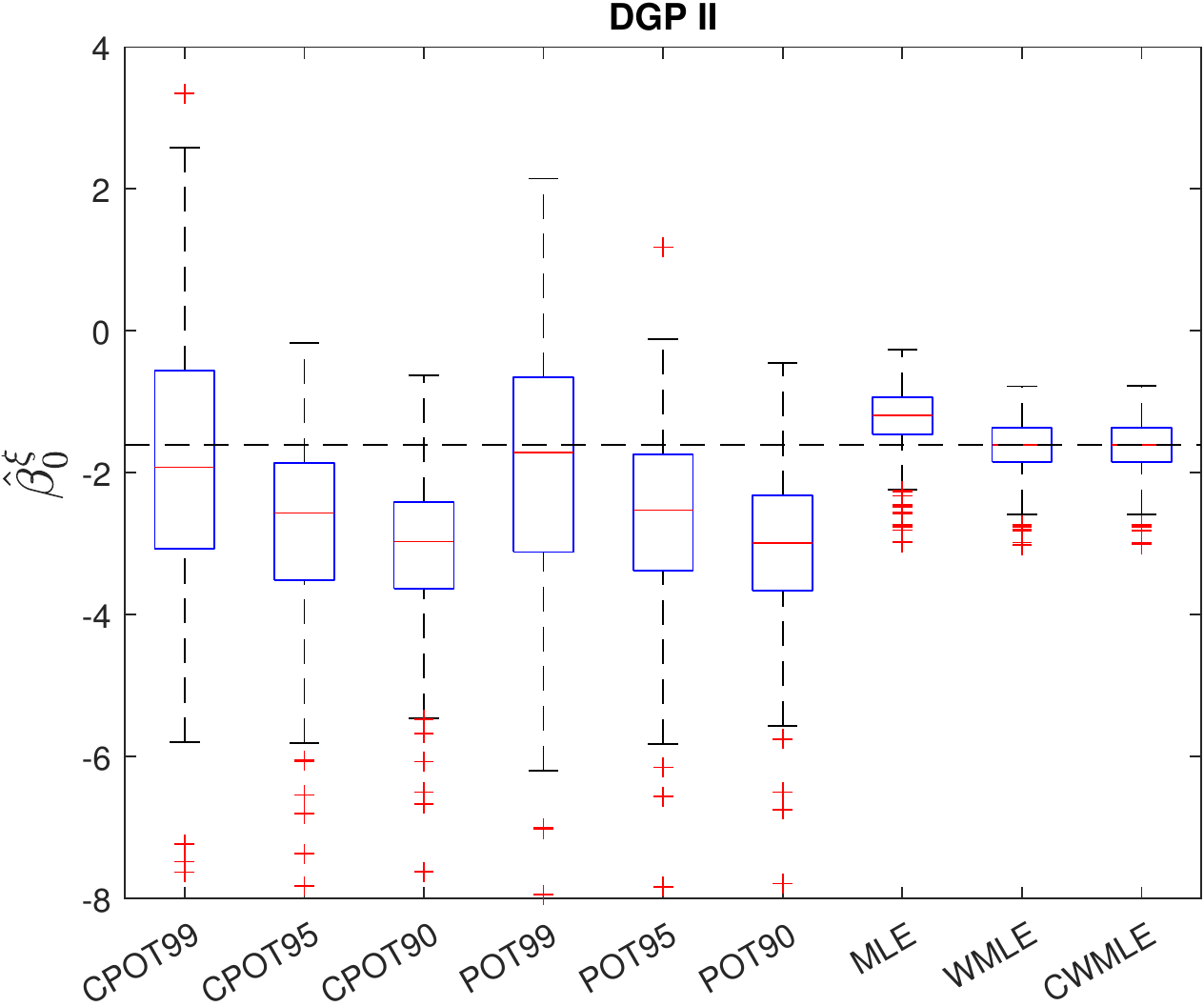} & \includegraphics[scale = .38]{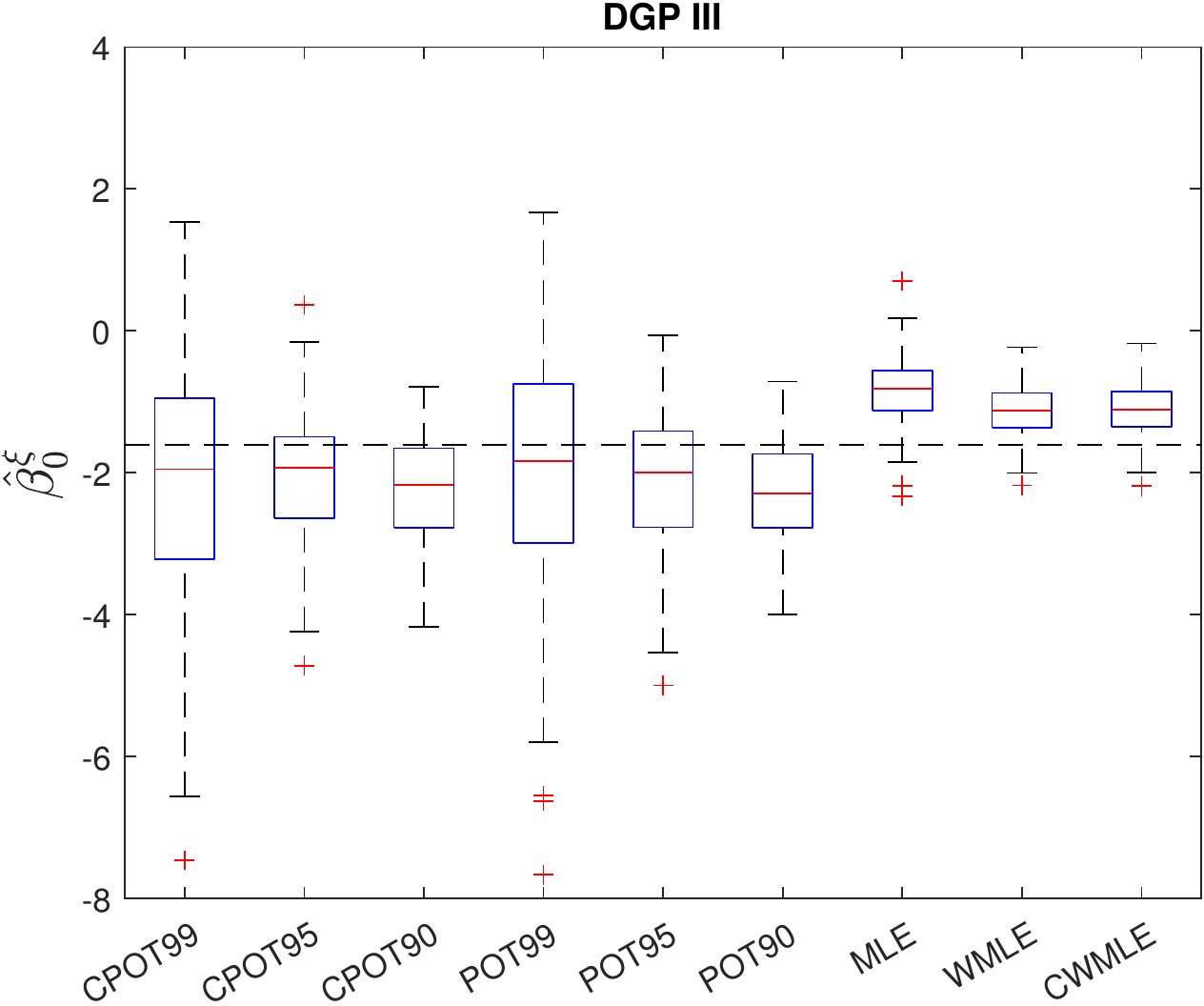} \\
    
    \includegraphics[scale = .38]{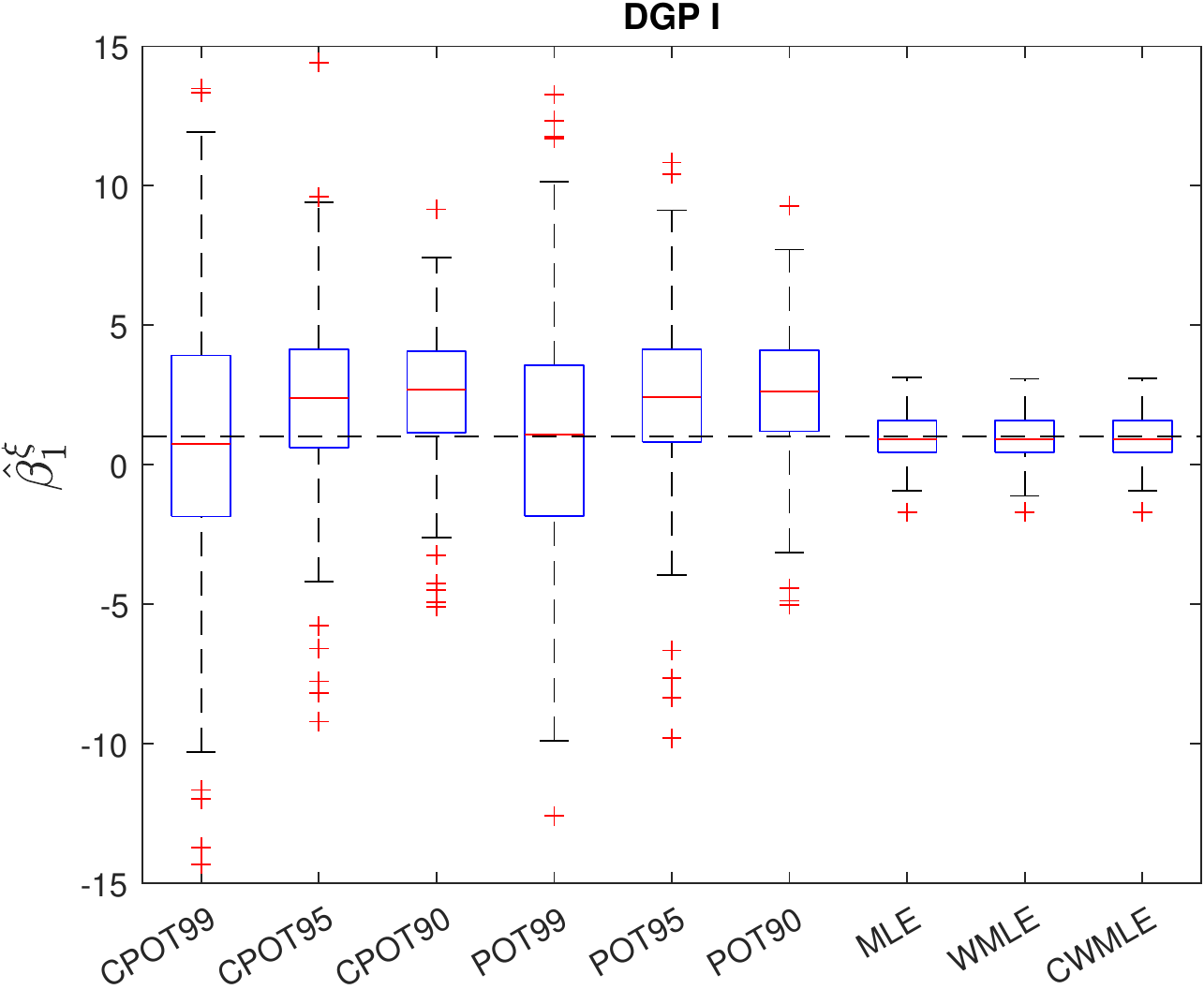} &
     \includegraphics[scale = .38]{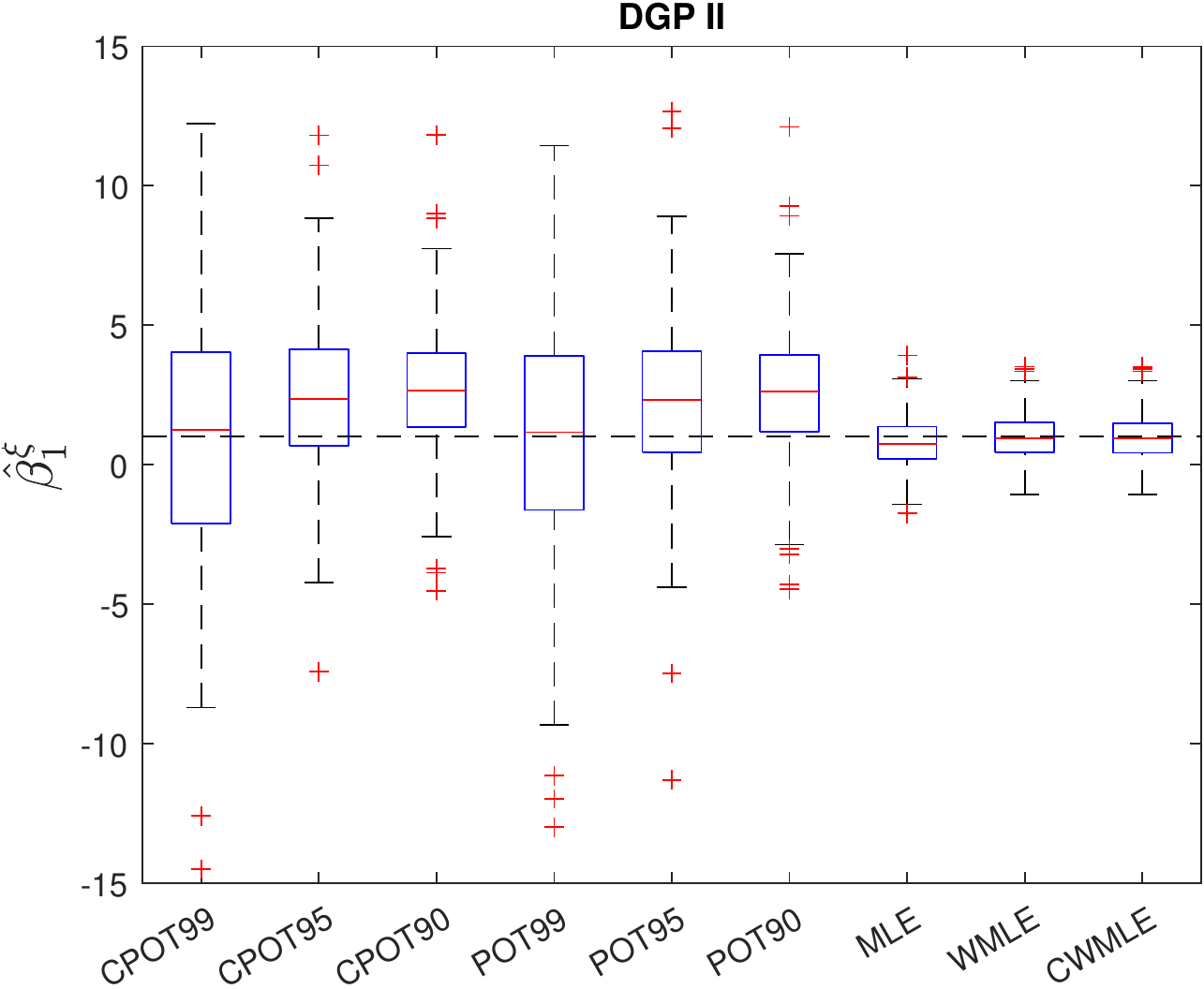} 
     &              \includegraphics[scale = .38]{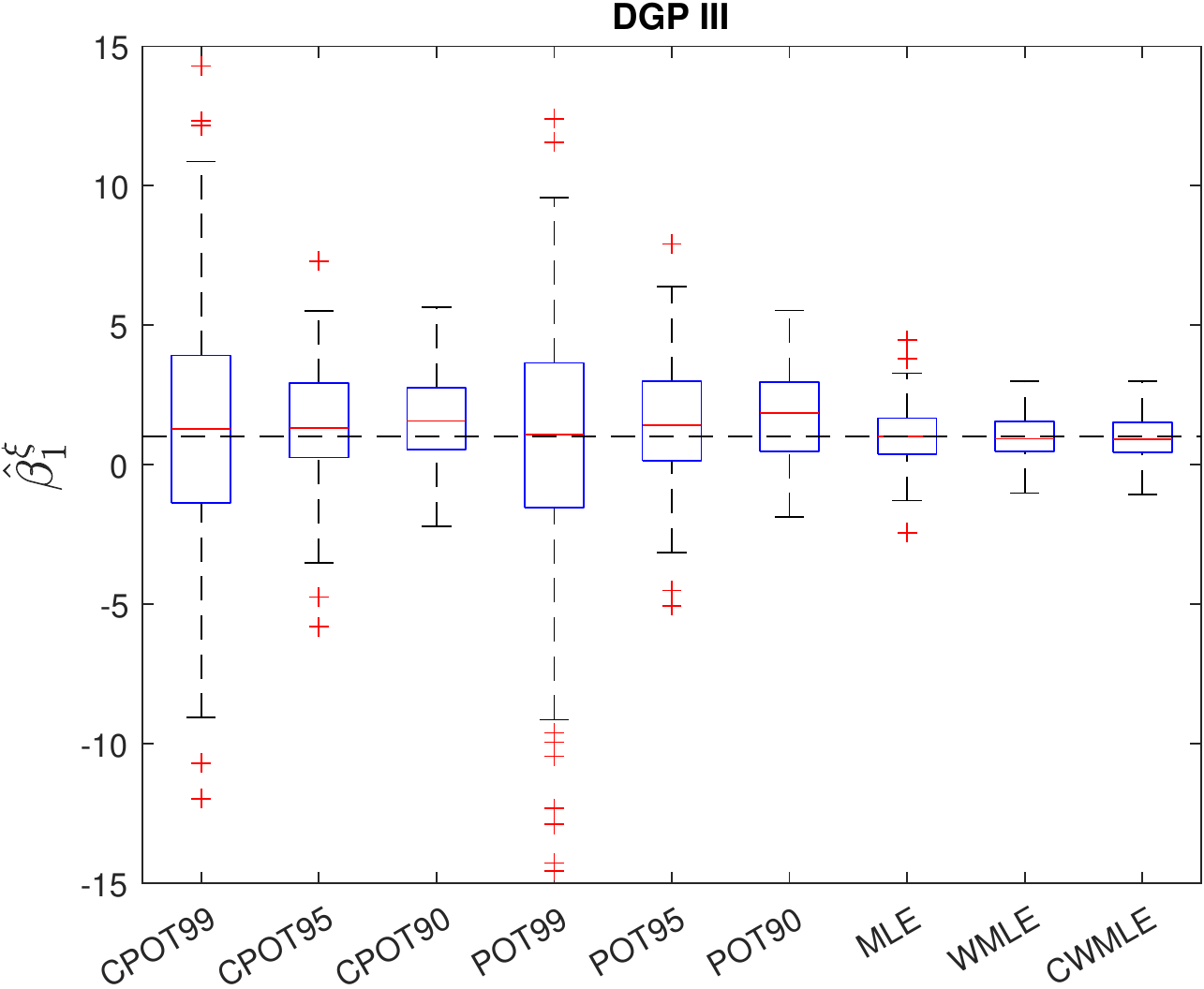}
    \end{tabular}
        \caption{\footnotesize Estimated parameters $\hat{\beta}_{0}^{\xi}$ (top)  and $\hat{\beta}_{1}^{\xi}$ (bottom) obtained with the proposed method and various POT approaches. Dashed: values of the true parameters. Columns \texttt{CPOT99}, \texttt{CPOT95} and \texttt{CPOT90} display the results obtained with the (conditional) POT approach, using quantile regression at the levels 99\%, 95\%, and 90\% for the threshold. Columns \texttt{POT99}, \texttt{POT95} and \texttt{POT90} refer to the results of POT obtained with an unconditional empirical quantile at the same levels. Columns \texttt{WMLE} and \texttt{CWMLE} denote the results obtained with the censored likelihood procedure, using either a global censoring threshold, or a censoring threshold based on quantile regression. \texttt{MLE} relates to classical MLE. Details on the POT computations can be found in the Appendix.}
        \label{fig:estimates}
    \end{figure}

\begin{figure}[htbp]
    \centering
    \begin{tabular}{ccc}
        \includegraphics[scale=.37]{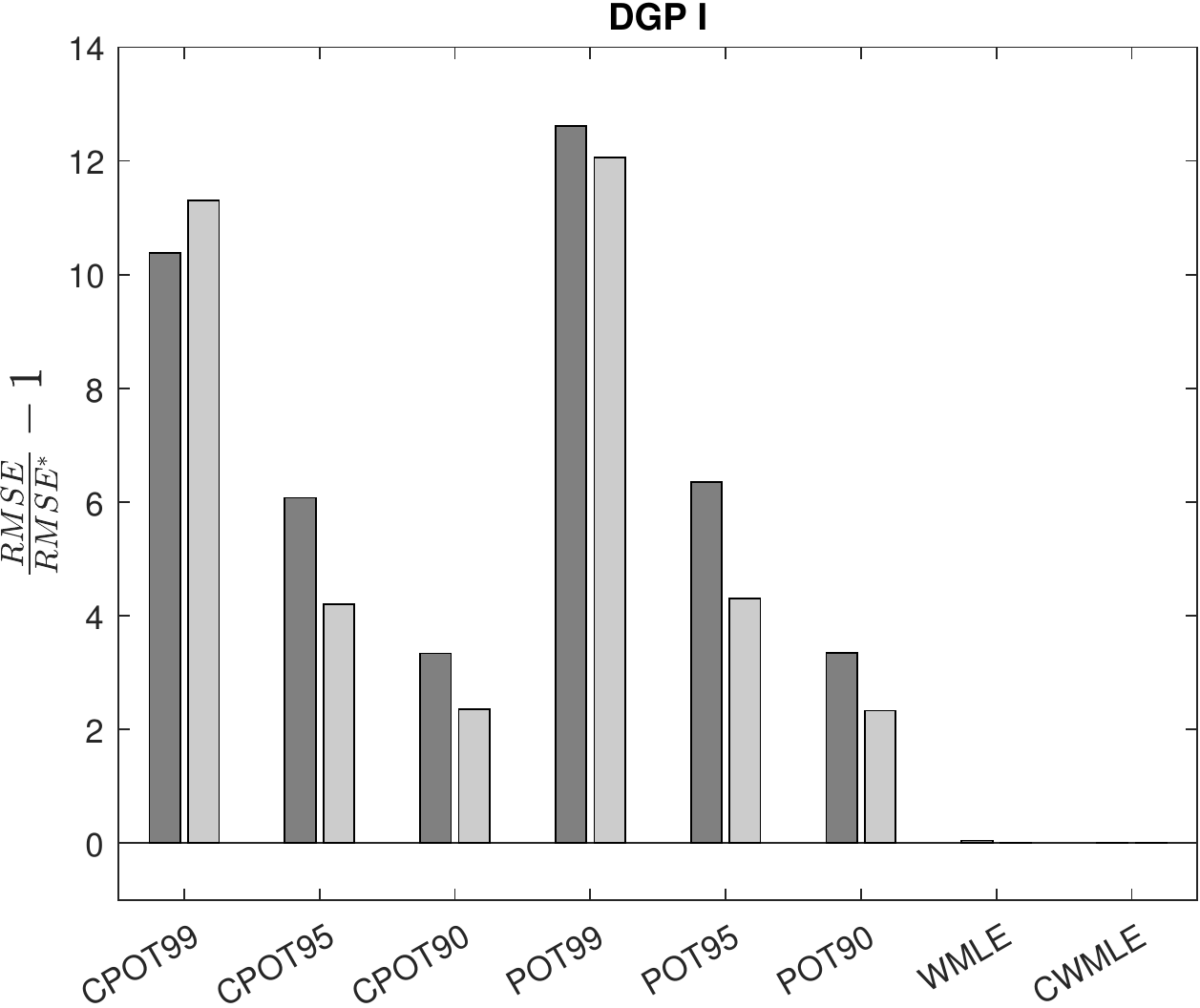} &\includegraphics[scale=.37]{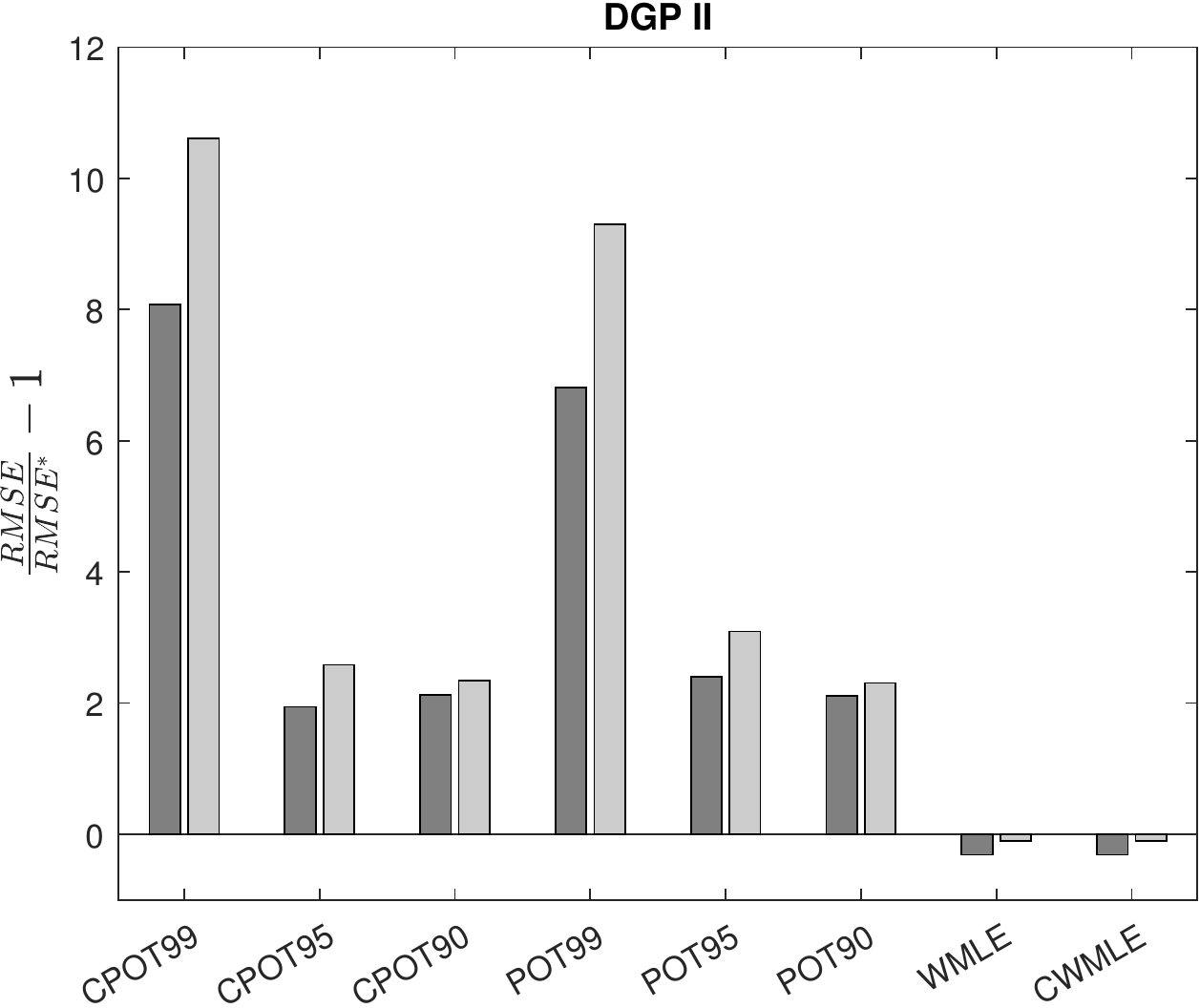} &\includegraphics[scale=.37]{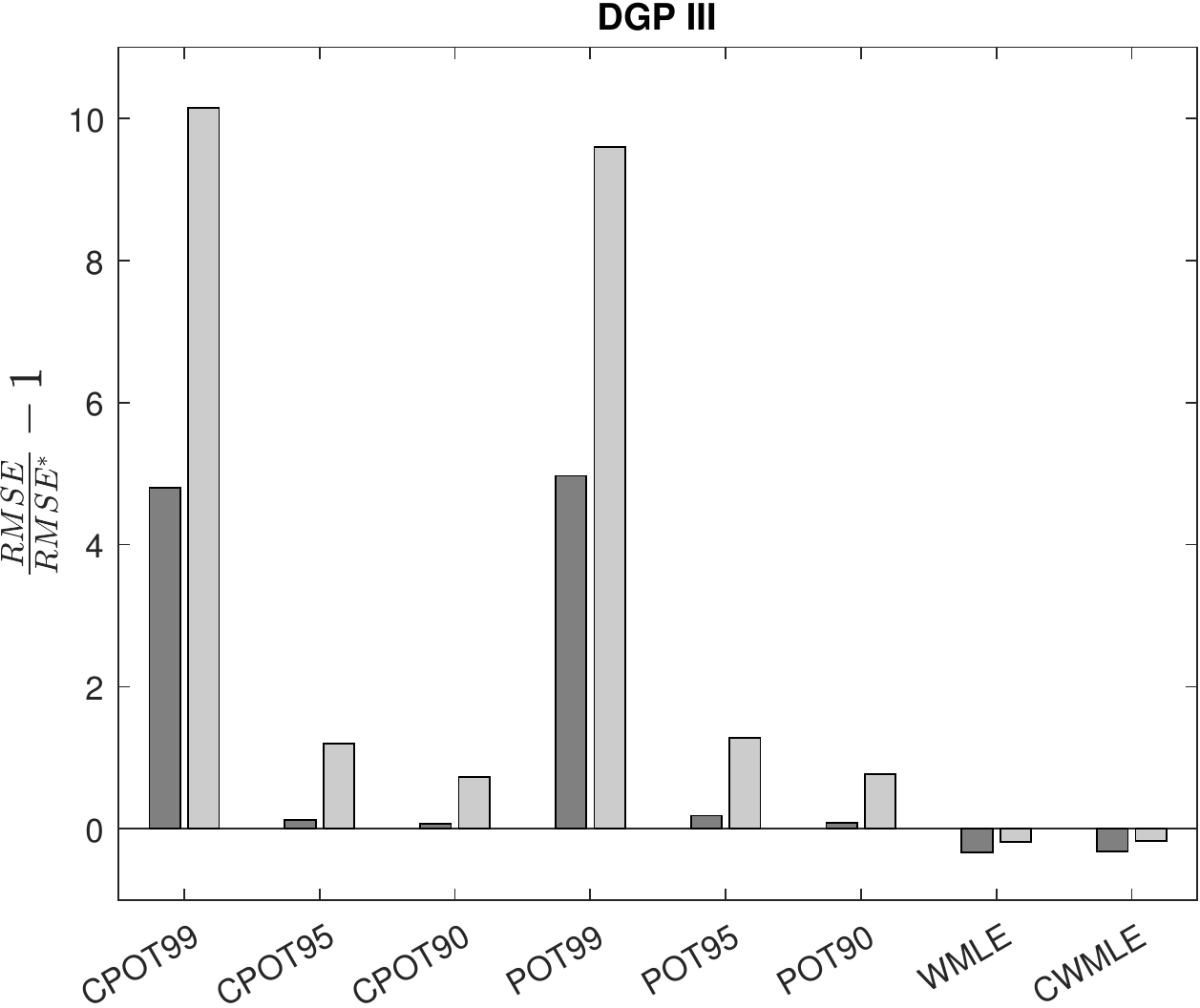}  
    \end{tabular}
    \caption{\footnotesize $RMSE/RMSE^{*}-1$ for $\hat{\beta}_{0}^{\xi}$ (dark grey) and $\hat{\beta}_{1}^{\xi}$ (light grey), for DGP I to III.}
    \label{fig:rmse}
\end{figure}

We now compare the results of the proposed estimation method with those of the POT alternatives. In Figure~\ref{fig:estimates}, we first display the estimated parameters $\hat{\beta}_{0}^{\xi}$ and $\hat{\beta}_{1}^{\xi}$ obtained from 200 simulated samples. For DGP I, unsurprisingly, our approach (columns \texttt{WMLE} and \texttt{CWMLE}) delivers unbiased and precise estimates, comparable to the MLE (column  \texttt{MLE}).
On the contrary, the POT approaches exhibit much more variability, and display a significant bias when the chosen threshold is too small (e.g. columns \texttt{CPOT90} and \texttt{POT90}, relying on conditional and unconditional quantiles at level 90\%). Similar results are observed for DGP II and III, despite the misspecification of the splicing model: \texttt{WMLE} and \texttt{CWMLE} deliver better results than the POT. In addition, for these DGPs, our approach is superior to the classical MLE, which exhibits a larger bias for $\hat{\beta}_{0}^{\xi}$. In the Appendix, Section~\ref{sec:empirical_suite}, we report detailed estimates of the bias. 
In Figure~\ref{fig:rmse}, we compare the RMSE associated to $\hat{\beta}_{0}^{\xi}$ and $\hat{\beta}_{1}^{\xi}$. For clarity, we display the quantities $RMSE/RMSE^{*}-1$, with the superscript "$*$" referring to the errors obtained with the MLE. As such, negative values indicate a lower error rate than the MLE. In terms of RMSE on the regression coefficients, the splicing approach is vastly superior to the POT approaches. For DGP I, the censored likelihood estimators (columns \texttt{WMLE} and \texttt{CWMLE}) are comparable to the MLE (with a difference of around 1\%). Moreover, for DGPs II and III, the censored likelihood estimators are much better than the MLE, with RMSE reductions ranging between 10\% and 35\%. The POT approaches perform relatively better for this DGP, especially for the constant term, but the RMSE is still much larger than those of the censored approaches.
We now turn to the coverage performance of confidence intervals obtained with the different methods. For the MLE and POT approaches, we use approximate asymptotic normality and the inverse of the Fisher information matrix obtained from the numerical Hessian. For the censored approaches, we use Eq.~\eqref{eq:CI}. In both cases, the variance-covariance (VCV) matrix is obtained from the gradient vector and Hessian matrix of the log-likelihood function computed with numerical procedures.
In Table~\ref{tab:coverage_ci}, we report the empirical coverage rates and median length of the 95\% confidence intervals constructed from the different estimators. 
\begin{table}[htbp]
    \centering
    \begin{scriptsize}
   \begin{tabular}{cccccccccccc}
    \hline
    \hline
        Crit. & DGP & Param. &\texttt{CPOT99} &\texttt{CPOT95} &\texttt{CPOT90} &\texttt{POT99} &\texttt{POT95} & \texttt{POT90} &\texttt{MLE} & \texttt{WMLE}&\texttt{CWMLE}\\
         \hline
         \hline
        Coverage & DGP I & $\beta_{0}^{\xi}$& 0.97 &    0.86 &   0.68 &  0.94&    0.87&   0.68 &    0.93 &   0.93 &  0.93\\
         & & $\beta_{1}^{\xi}$&  0.98 &    0.88&   0.85 &   0.98&    0.89 &    0.87 &    0.95 &    0.92 &    0.92\\
         \hline
         & DGP II& $\beta_{0}^{\xi}$& 0.94 &   0.90 &    0.71 &    0.94 &    0.90   & 0.74 &   0.69 &    0.89 &    0.90\\
        &  & $\beta_{1}^{\xi}$& 0.97 &   0.90 &    0.85 &  0.97 &   0.91 &    0.86 &    0.91 &  0.90 &   0.91\\
\hline
         & DGP III & $\beta_{0}^{\xi}$&  0.96 &    0.97 &   0.87 &   0.95 &    0.95 &    0.86 &    0.60 &    0.61 &    0.59\\
         & & $\beta_{1}^{\xi}$&  0.96 &    0.94 &    0.92 &    0.96 &    0.93 &    0.92 &   0.98 &    0.95 &    0.92\\
         \hline
         \hline
         Length & DGP I & $\beta_{0}^{\xi}$& 6.91 &   4.03 &    3.61 &    7.45 &    4.03&  3.61&  1.35 &    1.22&  1.21\\
         & & $\beta_{1}^{\xi}$& 16.33 &   8.91 &   7.67 &  16.00 &    9.02 &   7.74 &    2.94 &    2.62 &    2.62\\
        \hline
         & DGP II& $\beta_{0}^{\xi}$& 6.75 &    4.17 &    3.61 &   7.11 &    4.14&  3.63 &   1.36&  1.22 &    1.21\\
        &  & $\beta_{1}^{\xi}$& 16.30 &    9.23&  7.89& 16.67 &    9.19 &   7.82 &    3.07  &  2.59   & 2.61\\
\hline
         & DGP III & $\beta_{0}^{\xi}$& 6.88 &  3.39 &  2.62 &  7.16 &  3.44 &   2.64 &  1.80 & 1.25 &   1.25\\
         & & $\beta_{1}^{\xi}$& 15.93  &  7.75 &  5.84 &   16.13 &    7.82 &    5.97 &    4.21& 2.79 &   2.79 \\
        \hline
        \hline
    \end{tabular}
    \end{scriptsize}
    \caption{\footnotesize Empirical coverage and median length of the 95\% confidence intervals for the different estimators, over $B=200$ simulated samples.}
    \label{tab:coverage_ci}
\end{table}

\begin{figure}[H]
\centering
\begin{tabular}{ccc}
\includegraphics[scale = .38]{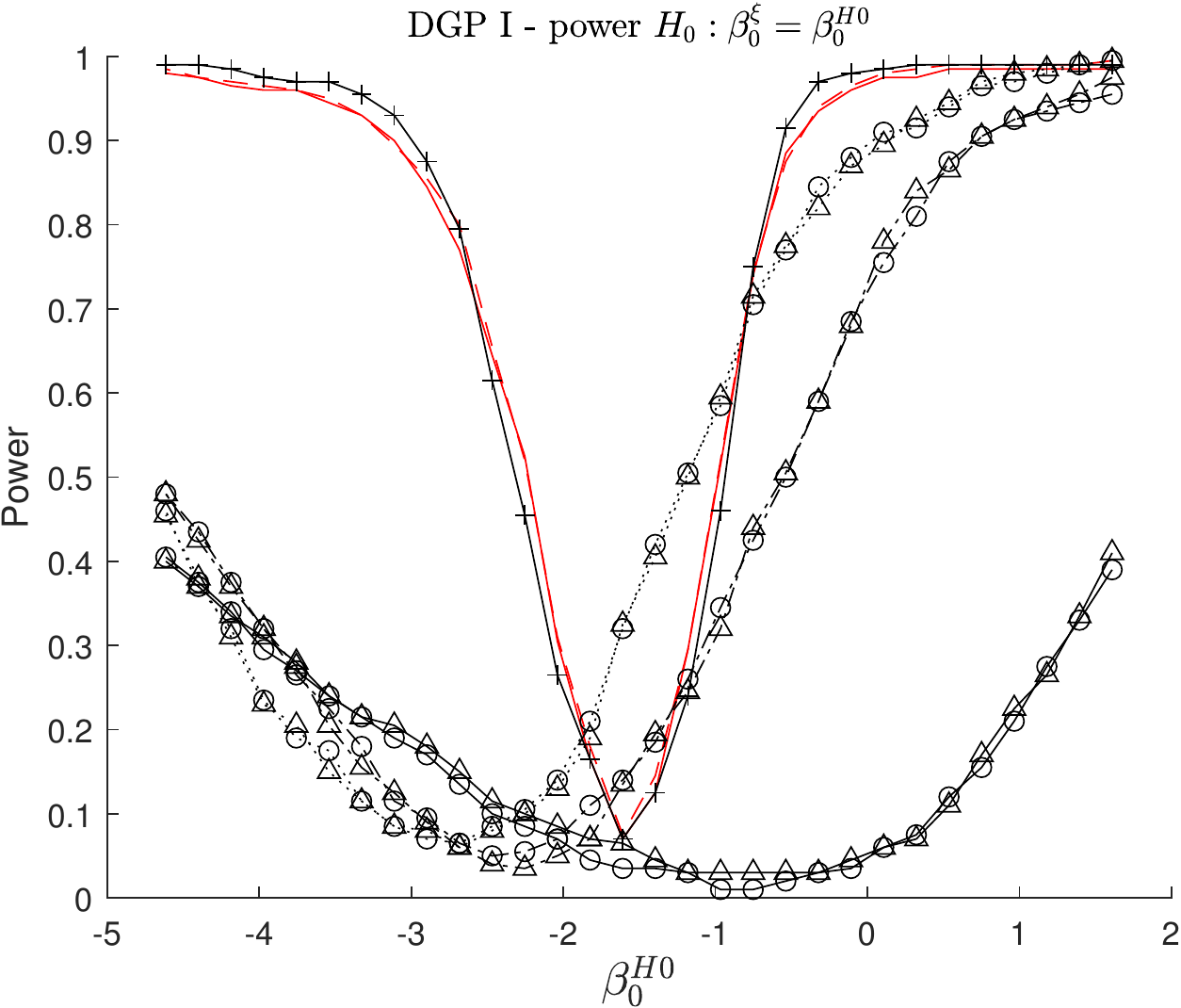}  &       \includegraphics[scale = .38]{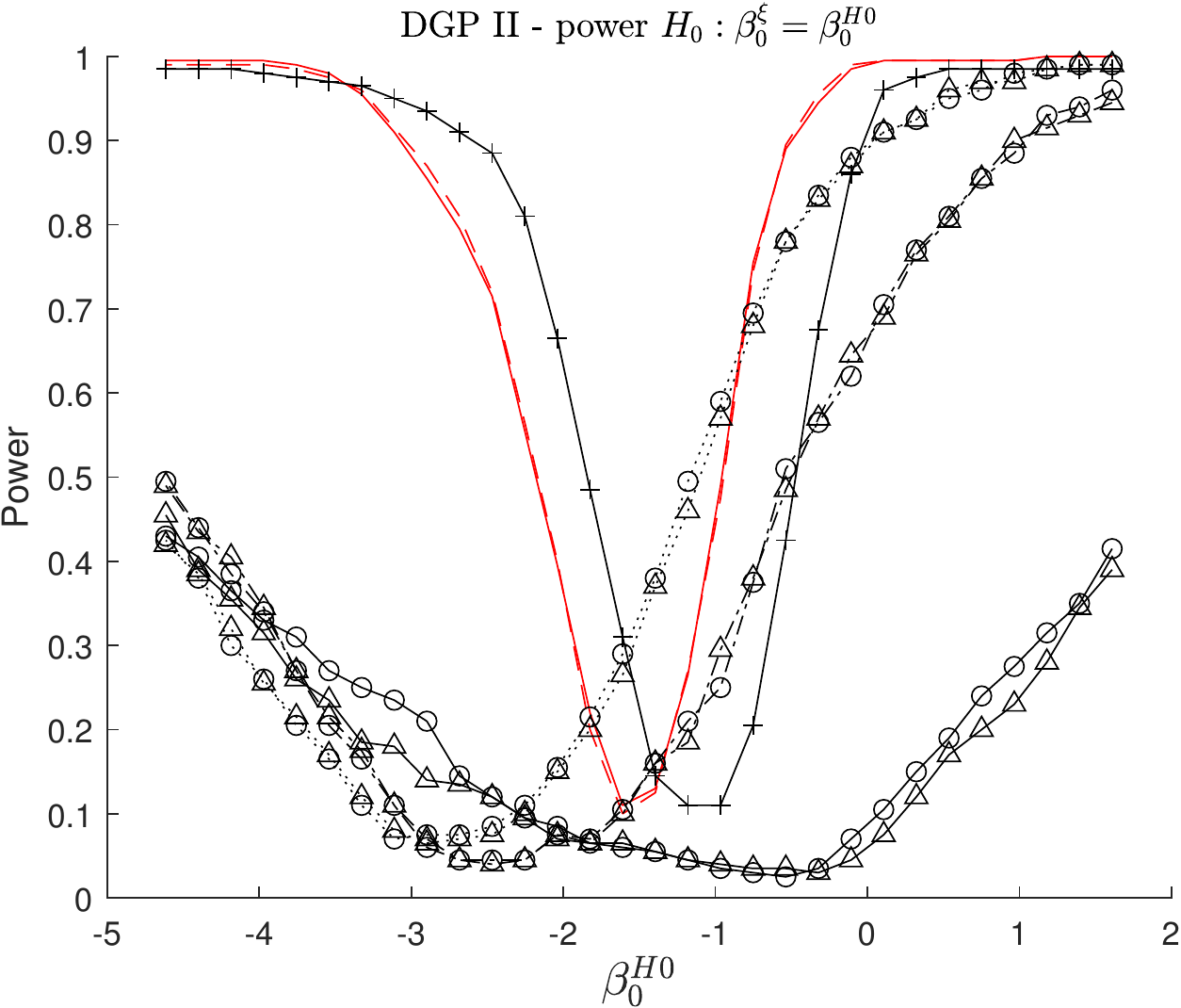} &       \includegraphics[scale = .38]{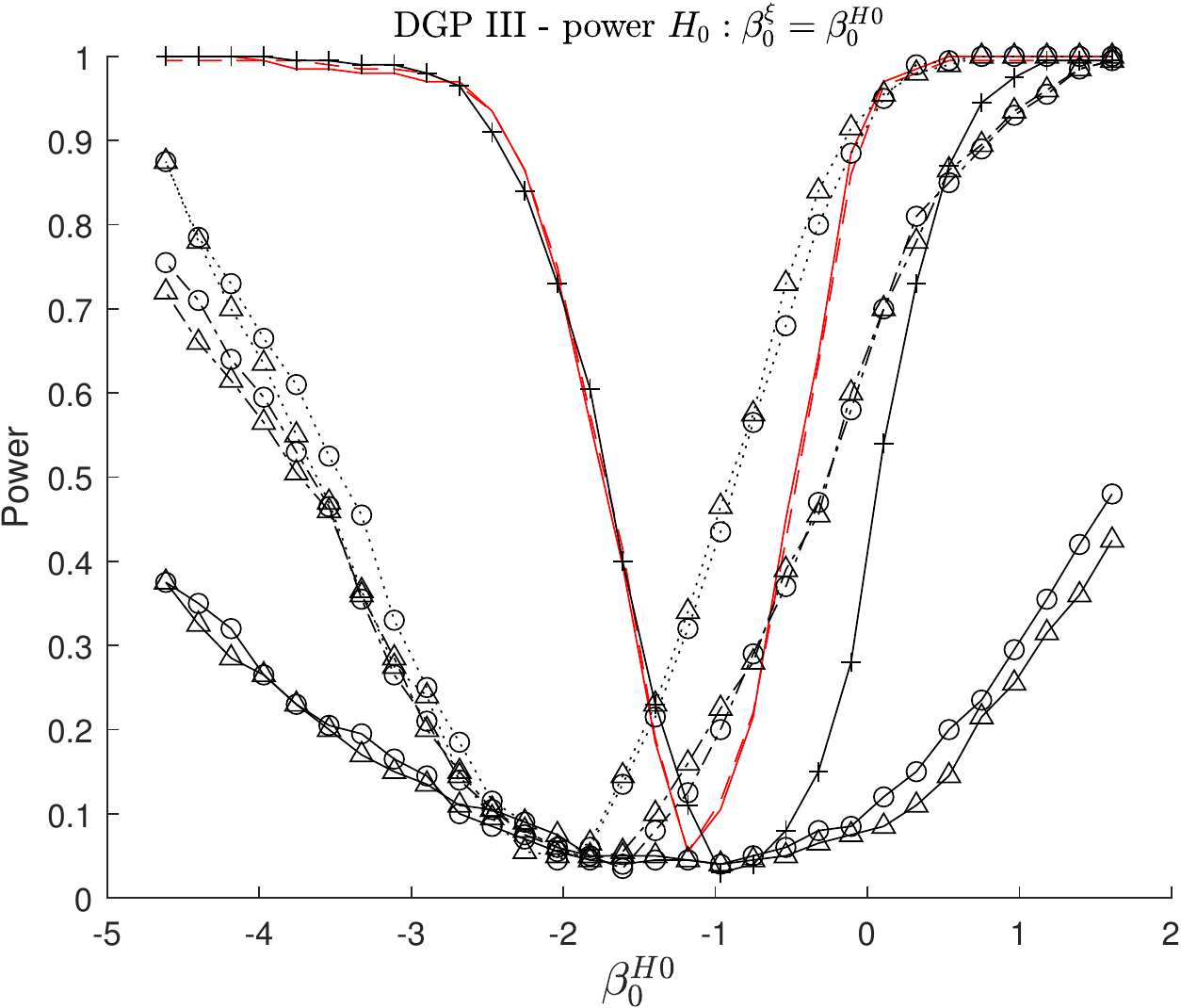} \\

\includegraphics[scale = .38]{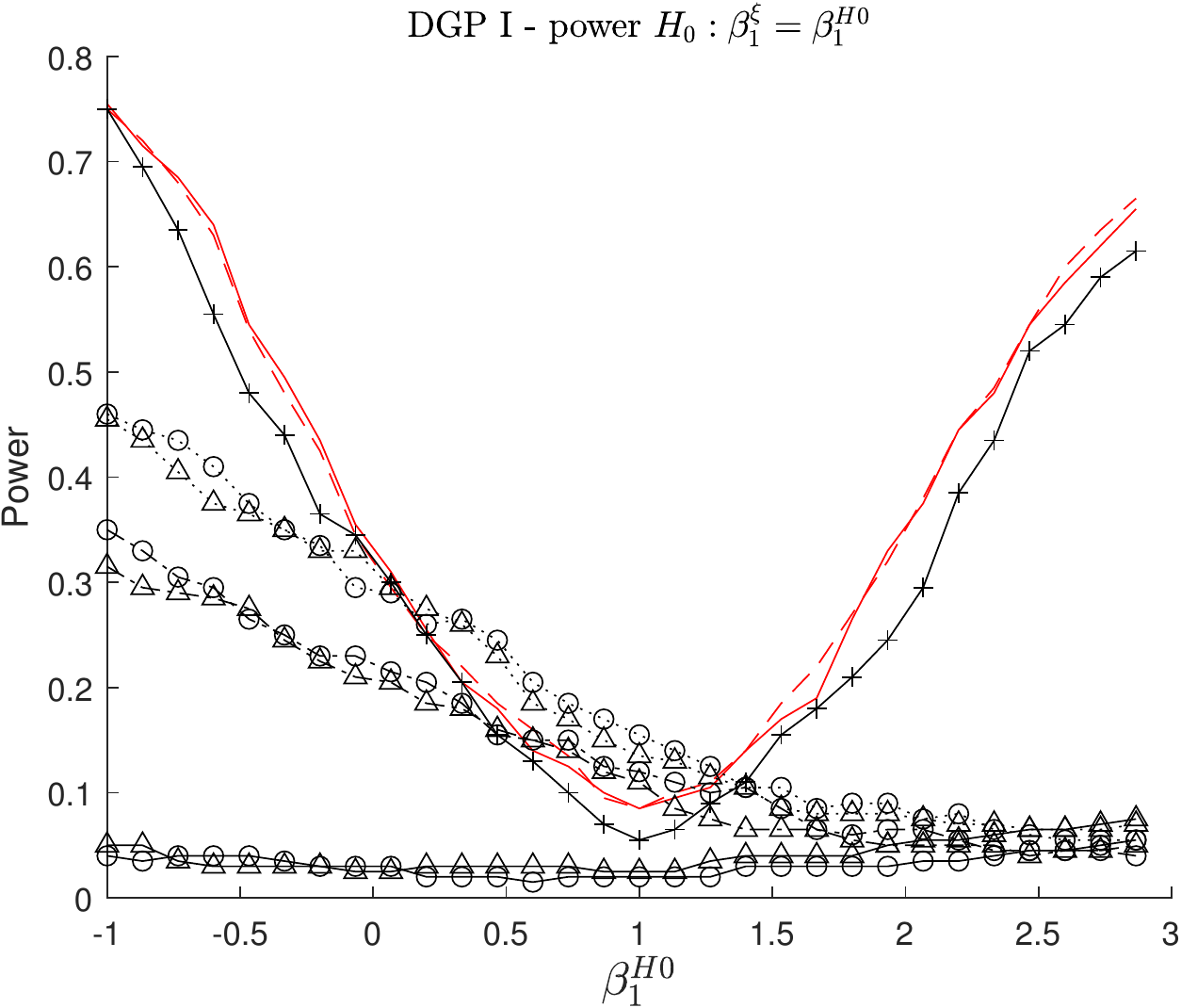}  &              \includegraphics[scale = .38]{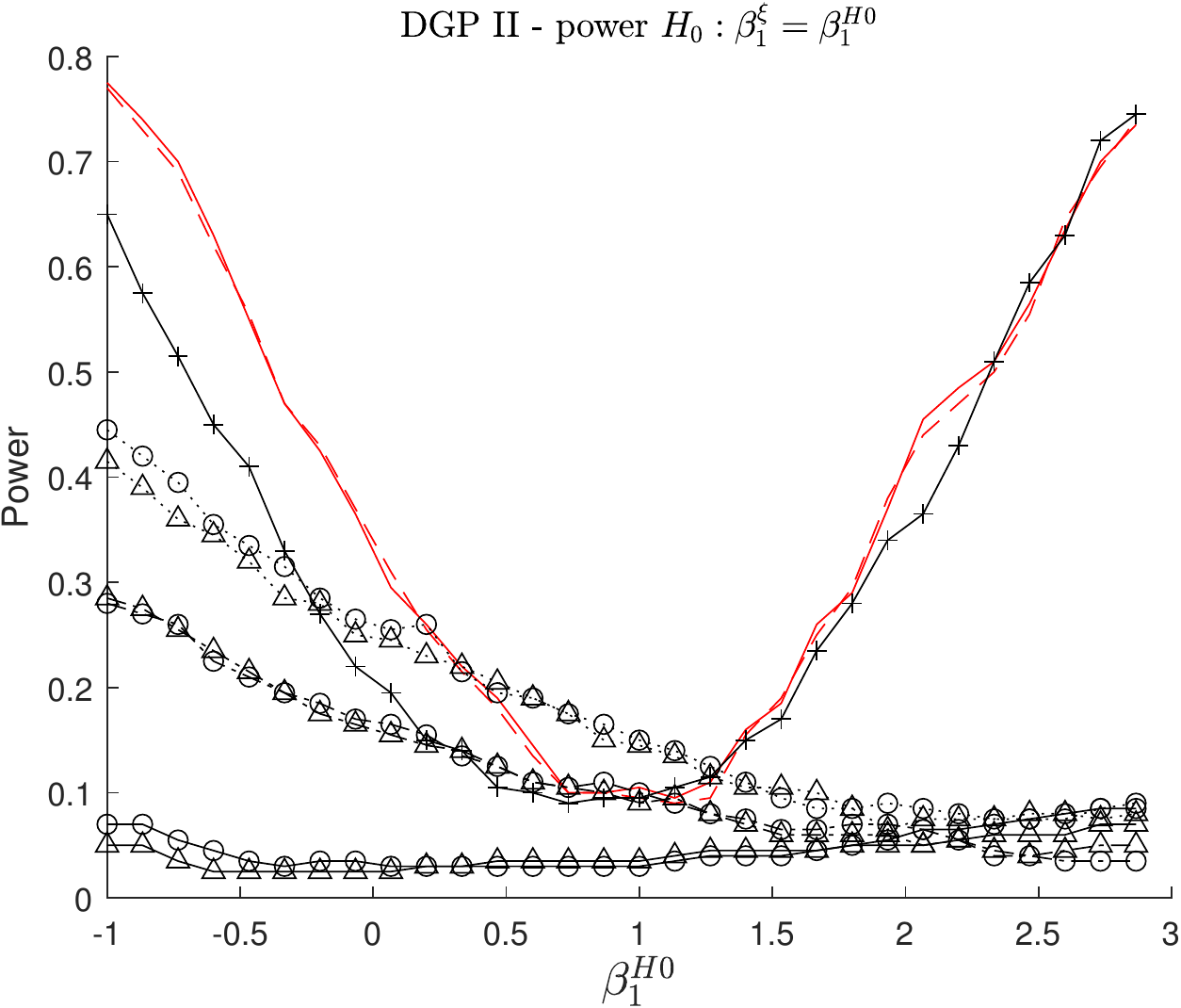}  &              \includegraphics[scale = .38]{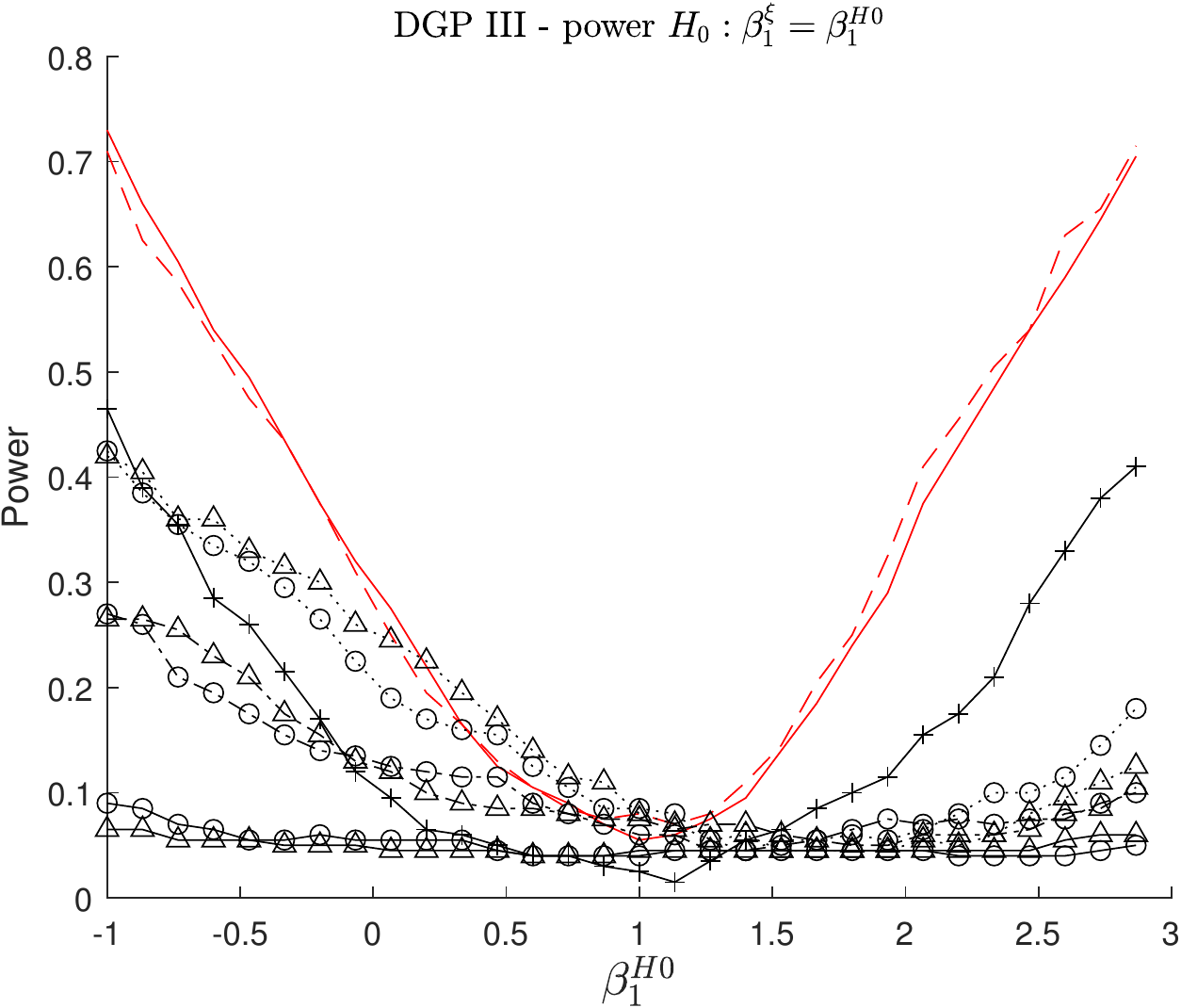}
\end{tabular}
    \caption{\footnotesize Power curves for the null hypotheses $H_{0}:\beta_{0}^{\xi}=\beta_{0}^{H0}$ (top) and $H_{0}:\beta_{1}^{\xi}=\beta_{1}^{H0}$ (bottom), for different values of $\beta_{0}^{H0}$ and $\beta_{1}^{H0}$. From left to right: DGP I to DGP III. For  $\beta_{0}^{H0}=\log(.2)$ and $\beta_{1}^{H0}=1$, we expect a rejection rate equal to the 5\% test level. Red solid and dashed: \texttt{WMLE} and \texttt{CWMLE}. $+$: \texttt{MLE}. $\triangle$: POT estimates based on unconditional thresholds. $o$: POT estimates based on conditional thresholds. Thresholds are taken as quantiles at the 99\% (black solid), 95\% (dashed dotted) and 90\% (dotted) levels in the POT approaches.}
    \label{fig:power_curves}
\end{figure}

We also display in Figure~\ref{fig:power_curves} the power curves obtained from testing $H_{0}:\beta_{0}^{\xi}=\beta_{0}^{H0}$ and $H_{0}:\beta_{1}^{\xi}=\beta_{1}^{H0}$ for various values of $\beta_{0}^{H0}$ and $\beta_{0}^{H0}$ with the different methods. While the coverage rate does not clearly discriminate between the methods, the median length of the confidence intervals and the power curves indicate a clear superiority of the splicing approaches: these methods exhibit respectable coverage rates for the regression coefficient (however below the nominal level, especially for the constant in DGP III), short lengths of the confidence intervals, and quickly increasing powers (dashed and solid red curves in Figure \ref{fig:power_curves}). The POT approaches, although exhibiting good coverage rates for high thresholds, fare poorly in terms of power and the length of their confidence intervals. The use of a smaller threshold leads to power improvements, but it comes at the cost of a larger bias and worse coverage rates. Comparison with the MLE clearly favors the censored estimators: as soon as the true DGP does not exactly correspond to the splicing regression models (DGPs II and III), the MLE suffers from significant biases, coverage rates well below the nominal level, and a loss of power (see, e.g., bottom right on Figure \ref{fig:power_curves}).

To conclude this section, we have demonstrated that the splicing approach combined with a censored likelihood estimation strategy clearly improves the bias-variance trade-off. In particular, when the true DGP is fully misspecified (i.e. DGP III), the bias and the variance of the estimated regression effect are relatively small, while the power and the coverage rate for the regression effect remain satisfactory. Additional simulation evidence can be found in the Appendix, Section~\ref{app:simu}.

\section{Empirical illustration: Hedge funds tail risks}\label{sec:empirical}

In this section, we investigate the determinants of hedge funds tail risks with the proposed approach. We use the tail risk definition given by \cite{kelly2014}, referring to the conditional tail index of the cross-sectional distribution of the returns, i.e. $\xi(\mathbf{x}_{t})$. This analysis is motivated by \cite{bali2007}, who highlight the lack of knowledge regarding hedge funds' risk profiles. Although it is well acknowledged that hedge funds exhibit tail risks (i.e. a relatively high probability of suffering large losses), there is no clear characterization of the link between economic risk factors and the conditional tail distribution of hedge funds. Previous studies relying on EVT to measure tail risks make use of multiple-step approaches, first inferring tail indices from the cross-section of returns, before conducting regression and portfolio performance analyses \citep{kelly2014,huang2012}. However, these measures are usually obtained with the POT, relying on a large and stable cross-section of assets (such as stocks) over time. This approach is not directly applicable to hedge funds: different time periods are characterized by an unequal reporting intensity of funds, making the use of month-by-month nonparametric tail estimates problematic or even impossible. Similarly, a lack of high-frequency data renders the approaches proposed by, e.g., \cite{bollerslev2015}, not reliably applicable to hedge funds.

\begin{figure}[!ht]
    \centering
    \begin{tabular}{cc}
\includegraphics[scale=.48]{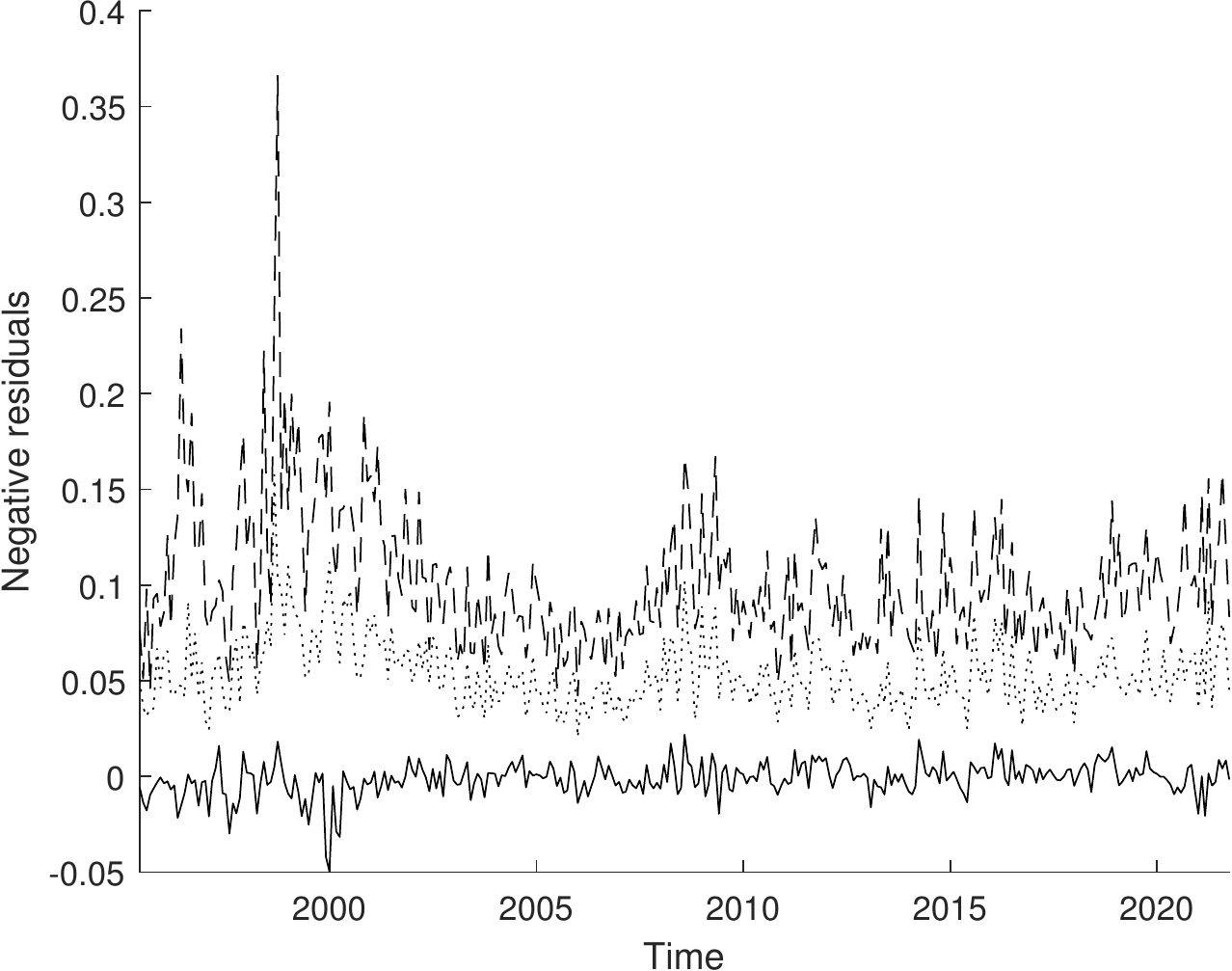} &\includegraphics[scale=.38]{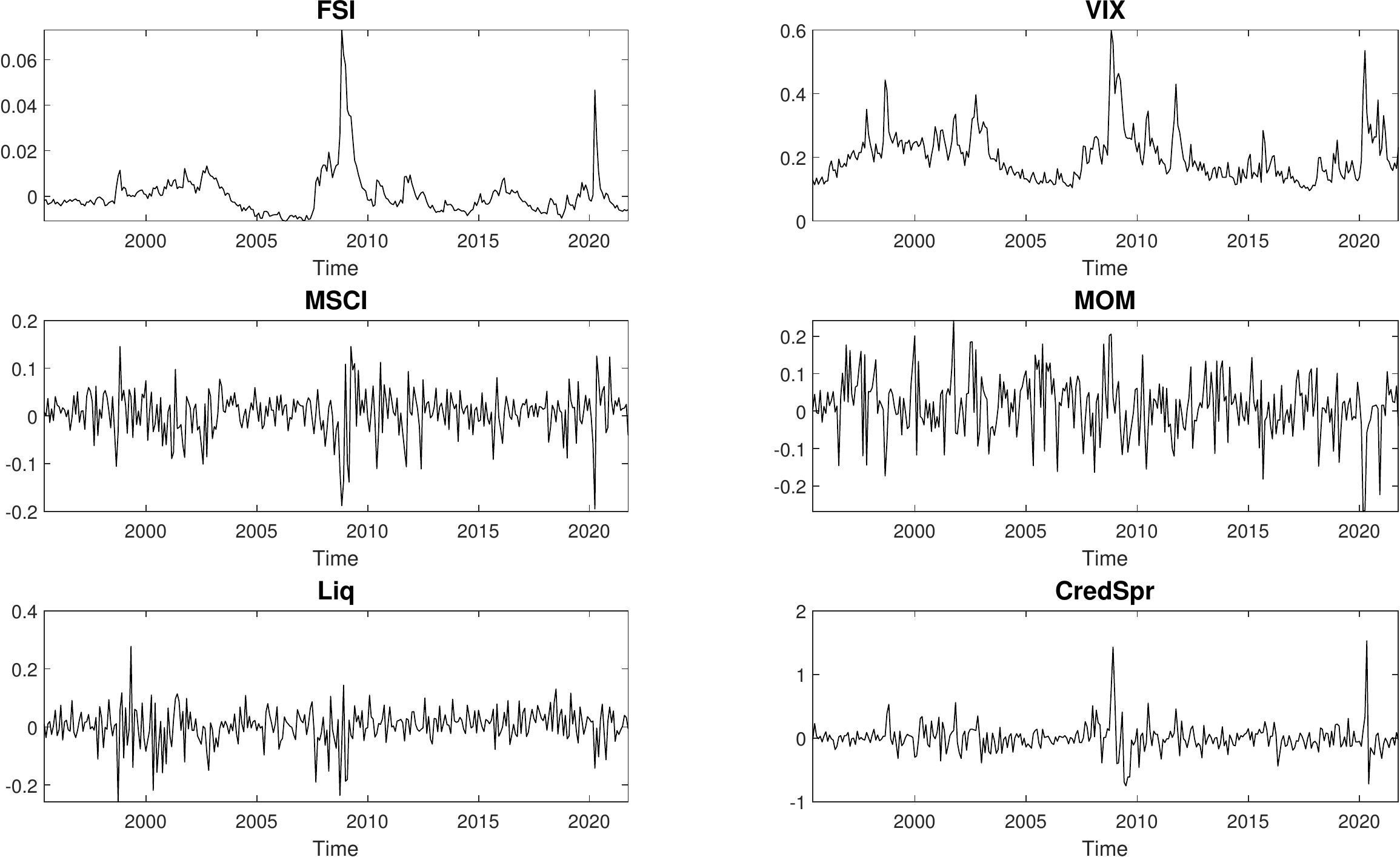}\\
(a) & (b)
    \end{tabular}
\caption{\footnotesize Panel (a): Mean (solid), 3$^{rd}$ quartile (dotted) and 99\% empirical quantile (dashed) of the cross-section of negative residuals $\hat{y}_{ti}$ of \cite{patton2013} model. Panel (b): Time series of the covariates.}
    \label{fig:descr_tail}
\end{figure}

We conduct our analysis using monthly gross-of-fee returns of \textit{Long/Short Equity} funds reporting in US dollars over the period 01/1995-09/2021 in the EurekaHedge database. To address backfill and survivorship biases, we removed the first 12 months reported by each fund, and solely kept funds with at least 60 months of uninterrupted reported history. In addition, we included both dead and live funds in our analysis. Our final sample consists of roughly 189,000 monthly returns spanning 1,484 funds. As a preliminary step, we remove time variations in the mean of the returns, following \cite{kelly2014} and \cite{huang2012}. To do so, we estimate the high-frequency asset pricing model of \cite{patton2013}, on a fund-by-fund basis\footnote{Details of this preliminary step are presented in the Appendix, Section~\ref{app:mean_filter}.}. We then use the negative residuals of this model (i.e. the residuals multiplied by $-1$) to estimate an extreme value regression model with the proposed splicing regression approach\footnote{To decrease specification issues related to this step, we also removed two residuals, being at least 16 times larger (in absolute value) than the sample standard deviation of the residuals across all funds.}. We denote by $r_{it}$ the observed return at time $t$ of fund $i=1,\ldots,I$, reporting during $n_{i}$ months. The total number of observations is thus $n=\sum\limits_{i=1}^{I}n_{i}$. The negative residual associated to $r_{it}$ is denoted $\hat{y}_{it}$ to stay consistent with the notation of Section~\ref{sec:methodo}. The tail risk analysis is then conducted on these pooled (negative) residuals. Residuals at a given point in time are therefore assumed to have their tail distribution driven by the same statistical model (see, e.g. \cite{kelly2014}, \cite{mhalla2022} and \cite{dupuis2022} for discussions on this pooling approach, and the Appendix, Section~\ref{sec:empirical_suite} for additional tests), while tail heterogeneity over time is assumed to be fully captured by the covariates. In Figure~\ref{fig:descr_tail}, panel (a), we display the mean, 3$^{rd}$ quartile and 99\% empirical quantile of the cross-section of negative residuals at each point in time. We observe significant variations over time, in particular at the onset of crisis periods.

As contemporaneous covariates for $\xi(\mathbf{x}_{t})$, we consider the following factors:
\begin{itemize}
\itemsep-0.5em 
    \item the market return (proxied by the returns of the MSCI World Index),
    \item the financial stability index (FSI) of the St. Louis Fed, aggregating interest rates, yield spreads, and volatility indicators,
    \item the (end-of-month) CBOE volatility index VIX,
    \item the credit spread factor of \cite{funghsieh},
    \item the liquidity factor of \cite{pastor2003},
    \item the global equity momentum factor of \cite{mosko2012},
    \item a crisis indicator, taking value 1 if the returns of the MSCI World index belongs to the worst 5\% returns over the considered period, 0 otherwise.
\end{itemize} 

The time series of the covariates are displayed in Figure~\ref{fig:descr_tail}, panel (b). These variables are suggested by \cite{kelly2021}, \cite{bali2014}, \cite{bollerslev2015}, \cite{agarwal2017}, and \cite{patton2013} as capturing tail risk over time. In particular, market returns can be seen as capturing systematic tail risk of hedge funds. The other factors span various dimensions of disturbances on the financial market, found to influence strongly the trading strategy adopted by hedge funds: uncertainty, liquidity, funding constraints, and benchmark pressures. We also let the variance parameter $s$ and the scale parameter $\sigma$ be functions of the VIX. Notice that, since our covariates are solely time-specific and not fund-specific, we have $\xi(\mathbf{x}_{it})=\xi(\mathbf{x}_{t})$, $\forall i=1,\cdots,I$. The predictor matrix of covariates has been standardized to have mean and variance of each column equal to 0 and 1, respectively.

\subsection{Main results}

\begin{figure}[!ht]
    \centering
    \includegraphics[scale=.5]{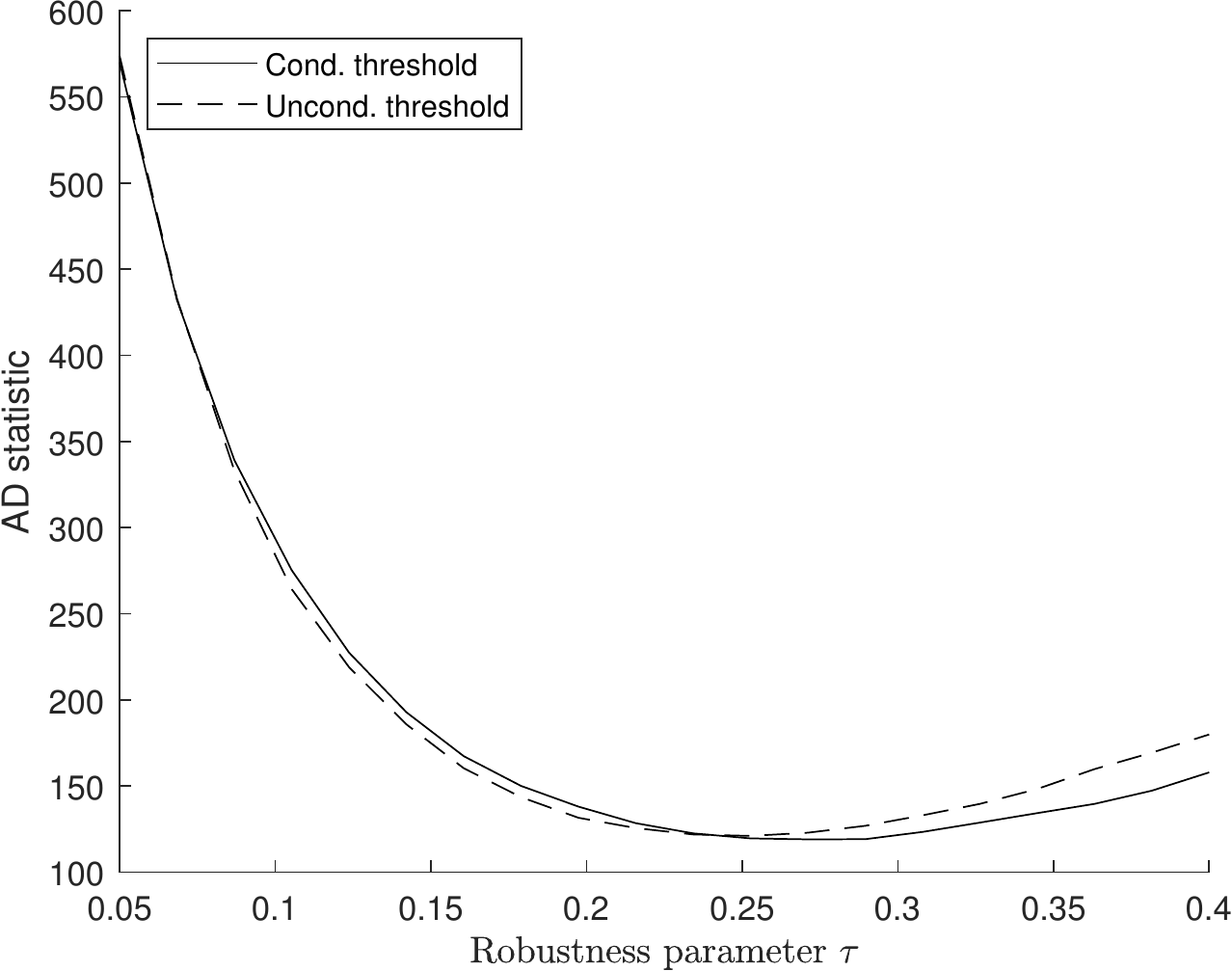}\includegraphics[scale=.5]{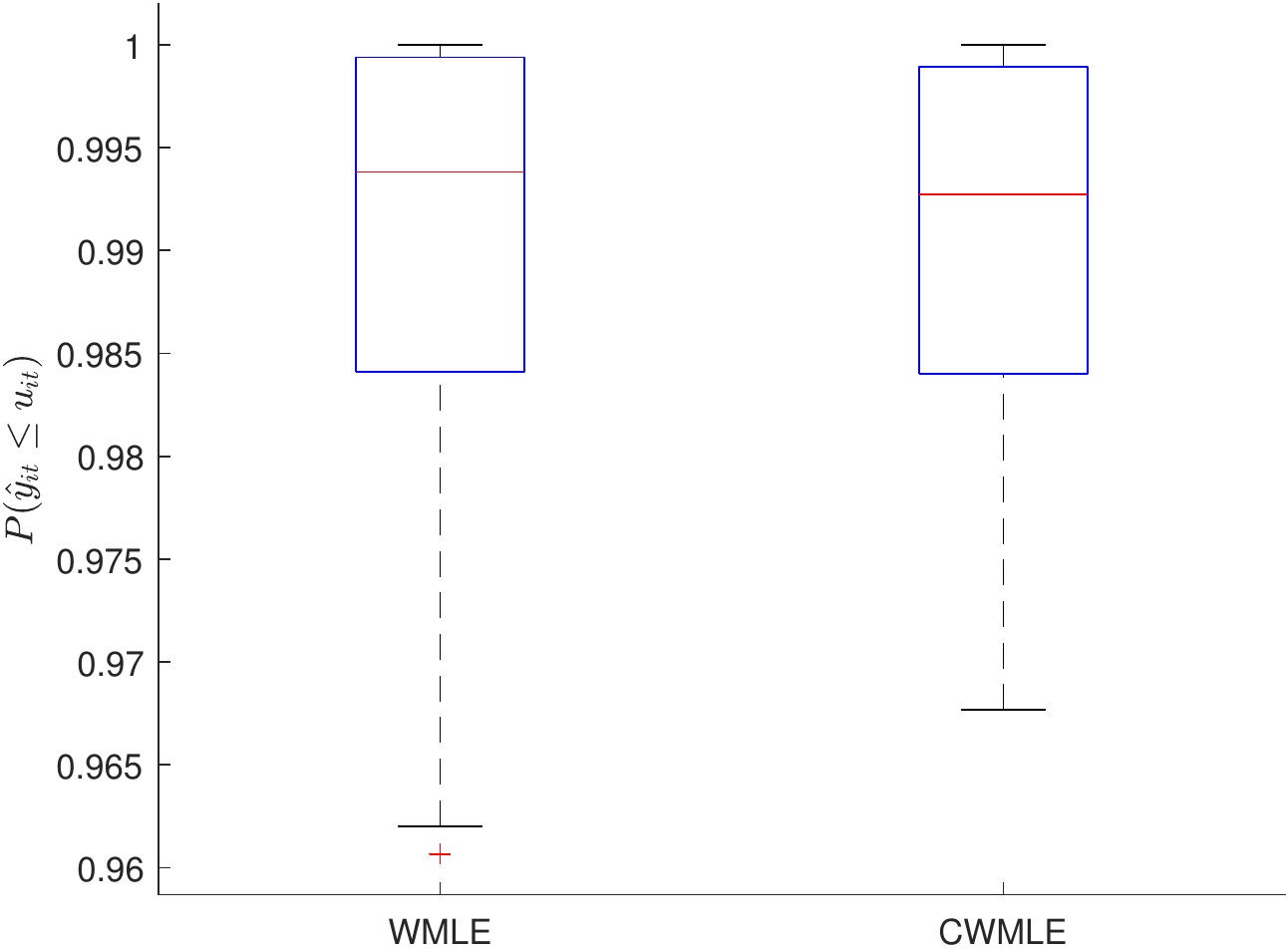}
    \caption{\footnotesize Left: $AD^{m}$ statistics as a function of $\tau$, using either an unconditional quantile (dashed) or a conditional quantile (solid) at level $\tau$ as censoring threshold. Right: boxplot of the distribution $\mathbbm{P}(\hat{y}_{ti}\leq u_{ti})$). }
    \label{fig:mad_curve_data}
\end{figure}

First, we use the proposed splicing approach to estimate the marginal effects of the covariates on $\xi(\mathbf{x}_{t})$ for the complete sample of funds. We consider both unconditional and conditional censoring thresholds. In light of the similar performance obtained in the simulation study, we discuss mostly the results obtained with the unconditional thresholds. In Figure~\ref{fig:mad_curve_data}, left panel, we display the $AD^{m}$ statistic as a function of $\tau$ for both approaches. We find an optimal value around $0.25$, a result consistent with our simulation study and indicative of a misspecification of the G-E-GPD model in the left tail. 
On the right panel of the same figure, we display the boxplot of the quantile levels corresponding to the implicit thresholds $u_{it}$ obtained with $\texttt{WMLE}$ and $\texttt{CWMLE}$  (i.e. $\mathbbm{P}(\hat{y}_{it}\leq \hat{u}_{it})$). This quantity indicates how far in the tail the GPD approximation starts under the G-E-GPD model, and how large the threshold used in the POT should be set. We find that conditional quantiles at levels between $0.985$ and $0.995$ (for an average around $0.99$) would be suitable thresholds for most combinations of the covariates. For some combinations, though, much lower threshold levels (as small as $0.96$) are selected. Overall, around 1\% of the observations in our sample are found to exceed these thresholds, a proportion of exceedances in accordance with those imposed in the simulation study.
\begin{table}[!ht]
    \centering
   \begin{footnotesize}
    \begin{tabular}{ccccccc}
    \hline
    \hline
    Covariate & $\texttt{WMLE}$ & $\texttt{CWMLE}$ & $\texttt{MLE}$ & $\texttt{POT95}$ & $\texttt{POT97.5}$\\
    \hline
    \hline
$\beta^{\xi}_{0}$   &  -1.32$^{***}$ &   -1.31$^{***}$ &   -0.52$^{***}$ &  -1.91$^{***}$ &  -1.94$^{***}$\\	
	 & 	$[-1.46,-1.19]$&$[-1.43,-1.19]$	&	$[-0.60,-0.43]$	&	$[-2.07,-1.74]$	&	$[-2.19,-1.70]$\\
  $\beta^{\xi}(\texttt{Crisis})$ &  0.03 &    0.04  &    0.24  &  -0.48  &   -0.56\\
  & $[-0.39,0.44]$ & $[-0.35,0.42]$ & $[-0.10,0.57]$ & $[-1.31,0.35]$ & $[-1.78,0.66]$\\
$\beta^{\xi}(\texttt{FSI})$		& -0.68$^{***}$ &   -0.67$^{***}$ &  -0.48$^{***}$  &   -0.14 &  -0.25	\\
	&	$[-0.79,-0.56]$ &	$[-0.79,-0.56]$	&	$[-0.64,-0.32]$ &		$[-0.35,0.08]$ & $[-0.57,0.07]$\\
$\beta^{\xi}(\texttt{VIX})$		&	-0.19$^{***}$ &   -0.17$^{***}$ &   -0.10  &   0.08 &   0.15\\
&		$[-0.32,-0.06]$ &	$[-0.29,-0.05]$	&	$[-0.20,0.00]$ &	$[-0.16,0.31]$	&	$[-0.19,  0.48]$ \\
$\beta^{\xi}(\Delta\texttt{MSCI})$		&	0.03 &    0.03& -0.01 &  -0.00 &    0.01	\\
	&	$[-0.02,0.09]$ &	$[-0.03,0.09]$	&	$[-0.07, 0.06]$	&	$[-0.15,0.14]$&	$[-0.20,0.21]$\\
$\beta^{\xi}(\texttt{MOM})$		& 0.21$^{***}$ &   0.20$^{***}$&    0.01 &   0.04&  -0.01\\
&	$[0.16,0.25]$ &$[0.16,0.25]$	&	$[-0.04,0.06]$	&	$[-0.09,0.17]$ & $[-0.19,0.18]$\\
$\beta^{\xi}(\texttt{Liq})$		&  -0.01 & -0.01 &    0.04  & -0.13$^{**}$ &   -0.10\\
&	$[-0.06,0.04]$ &$[-0.06,0.04]$	&	$[-0.01,0.09]$	&	$[-0.24,-0.01]$ & $[-0.26,0.07]$\\
$\beta^{\xi}(\texttt{CredSpr})$		& -0.47$^{***}$ &   -0.46$^{***}$ & -0.11$^{***}$ &  0.01& 0.09\\
		&	$[-0.54,-0.39]$ &$[-0.53,-0.39]$	&	$[-0.17,-0.04]$	&	$[-0.13,0.15]$ & $[-0.12,0.30]$\\
		\hline
		\hline
$\beta^{\sigma}_{0}$	&	-3.89$^{***}$& -3.89$^{***}$& -3.37$^{***}$  & -3.67$^{***}$ &   -3.55$^{***}$  		\\
&	$[-3.91,-3.88]$ 	&$[-3.90,-3.88]$ &	$[-3.38, -3.37]$	&	$[-3.70,-3.63]$	&	$[-3.60,-3.51]$ \\
$\beta^{\sigma}(\texttt{VIX})$		&0.08$^{***}$ &    0.10$^{***}$ & 0.13$^{***}$ & 0.06$^{***}$ &  0.05$^{***}$ 	\\
	&	$[0.07,0.09]$	&	$[0.09,0.11]$ &	$[0.12,0.13]$	&	$[0.03,0.09]$	&	$[0.01,0.09]$\\
$\beta^{s}_{0}$		&	-3.56$^{***}$ & -3.56$^{***}$ & -3.47$^{***}$ & - & - \\											
	&	$[-3.58,-3.55]$	&	$[-3.58,-3.55]$	&  $[-3.49,-3.45]$ & - & -\\										
$\beta^{s}(\texttt{VIX})$		&	0.02$^{***}$ &    0.02$^{***}$ &    0.03$^{**}$ & - & -	\\										
	& $[0.01,0.03]$ &$[0.02,0.03]$	&	$[0.00,0.05]$ & - & -		\\								
$\beta^{m}_{0}$	&	-0.002$^{***}$		&		-0.002$^{***}$		&-0.001$^{***}$ & - & -			\\						
&	$[-0.002,-0.002]$	&	$[-0.002,-0.002]$	&		$[-0.001,-0.001]$ & - & - \\	
\hline
\hline
$\tau^{opt}$ & 0.253 & 0.271 & - & - & -\\
\hline
\hline
    \end{tabular}
    \end{footnotesize}
    \caption{\footnotesize Estimated regression effects for the different estimation methods. Confidence intervals at the 95\% level are in brackets below the estimates. $\texttt{CWMLE}$ has been obtained using all variables as conditioning variables.$^{***}$ and $^{**}$ indicate coefficients significant at the 1\% and 5\% test levels, respectively.}
    \label{tab:coef_data}
\end{table}
In Table~\ref{tab:coef_data}, we report pointwise estimates of the regression coefficients and their confidence intervals, for both the G-E-GPD methods and the POT-based methods\footnote{We use the same set of covariates for both estimation methods for $\xi$ and $\sigma$, and an unconditional threshold in the POT. Results with regression thresholds are qualitatively alike and available upon demand.}.  Looking at the results for $\texttt{WMLE}$ (first column), we find the FSI ($\texttt{FSI}$), the credit spread factor ($\texttt{CredSpr}$), the VIX ($\texttt{VIX}$), and the momentum factors ($\texttt{MOM}$) to have regression coefficients significantly different from zero. A decrease in $\texttt{FSI}$, $\texttt{CredSpr}$, and $\texttt{VIX}$ are associated with an increase in tail risk, suggesting a propensity for the cross-section of hedge funds to be particularly exposed to tail risk in a context of cheap funding conditions and low uncertainty (the FSI loading positively on TED and OIS spreads). Similarly, an increase in time series equity momentum is associated with an increase in tail risk, consistent with the idea that tail risk is high in booming market conditions. Re-estimating the model using only these variables as predictors, we find very similar results in terms of pointwise estimates and confidence intervals (see the Appendix, Section~\ref{sec:empirical_suite}).

Focusing on the POT approaches, the results are harder to interpret: for $\beta^{\xi}$, no regression effects are found to be significantly different from zero, except $\texttt{Liq}$ for $\texttt{POT95}$. The G-E-GPD approach suggests thresholds in the range of 98\%-99.5\%, above which the GPD approximation should kick in. However, the use of the POT with such high thresholds would leave us with extremely few data, and lead to instabilities of the estimates. This result highlights therefore the efficiency gains obtained with the G-E-GPD approach.

\begin{figure}[htbp]
    \centering
      \includegraphics[scale=.36]{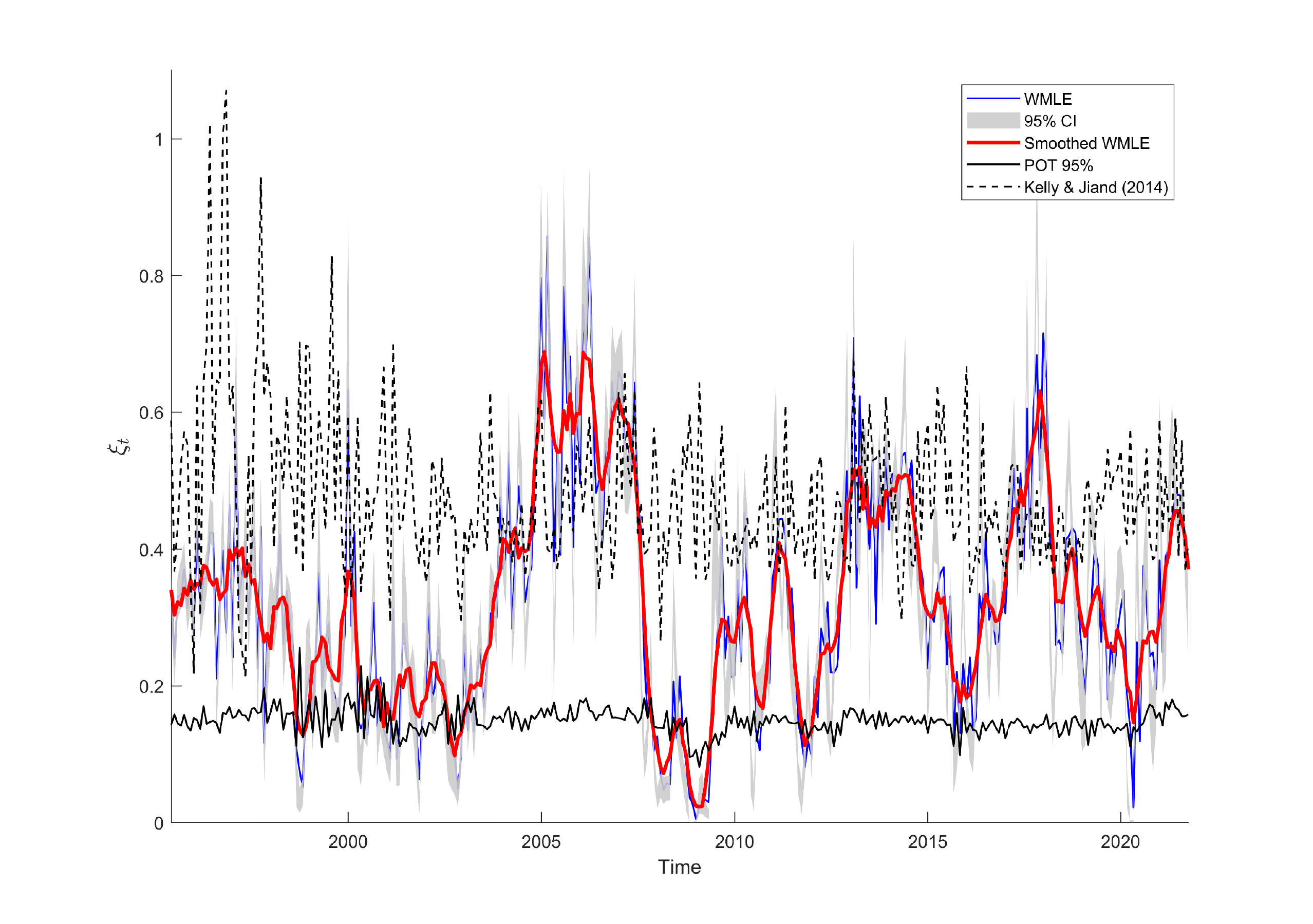}
    \caption{\footnotesize Time series of $\hat{\xi}(\mathbf{x}_{t})$ (solid blue). Red: smoothed estimate of $\texttt{WMLE}$ obtained from a moving average filter. Shades of grey: 95\% pointwise confidence intervals for $\xi(\mathbf{x}_{t})$. Solid black: POT estimates with 95\% empirical quantile as threshold. Dashed black: Hill estimator proposed by \cite{kelly2014} with the monthly 95\% empirical quantile as threshold.}
    \label{fig:ts_xi_t}
\end{figure}

Adopting a time-series view, we compute
$\hat{\xi}(\mathbf{x}_{t})$ for $t=1,\ldots,T$ with our $\texttt{WMLE}$. We display these quantities and the associated pointwise 95\% confidence intervals in Figure~\ref{fig:ts_xi_t}. The G-E-GPD approach (red solid, named HKU hereafter) suggests a much higher risk level of the funds' cross-section, and more variations over time, than the POT approaches (black solid) and the Hill estimator suggested by \cite{kelly2014} (dashed black, denoted KJ hereafter). In particular, we observe a tail risk increase before large systemic events, such as the global financial crisis in 2008 or the COVID crisis in 2020, followed by a rapid decline of the risk measure. These variations appear to reflect the dynamic nature of hedge fund investment strategies well, expected effectively to take more risk exposure in booming periods, and wind down their risks in times of market turmoils. On the contrary, the POT and KJ estimators do not seem to capture these differences over time. As for the time series characteristics of the different tail risk estimates, we find the HKU estimates to be highly persistent, with a monthly AR(1) coefficient of 0.727, while the POT and KJ estimators exhibit coefficients of 0.092 and 0.289, respectively. Hence, the HKU tail risk measure also seems more promising as a predictor of hedge fund returns \citep{kelly2014}. 

\subsection{Implications for hedge fund performance}

\cite{gao2018} and \cite{gao2019} suggest that exposure to tail risk is a good predictor of expected returns on several asset classes such as equities, but also hedge funds (see also \citealp{agarwal2017}). In particular, \cite{gao2018} conclude that over-performing hedge funds are those that exploit \textit{ex ante} market disaster concerns of investors while being less exposed to disaster risk, capturing a ``fear premium". 

To investigate whether our tail risk measure captures a similar mechanism, we hypothesize that funds following strategies generating future returns that are positively correlated to our tail risk measure are indicative of a lower abnormal return (or alpha), since it would signal an inability to limit exposure to disaster risk. On the contrary, negative exposure would indicate that times with a high tail risk are followed by smaller returns, indicative of selling disaster insurance and being negatively impacted by the materialization of these disasters. However, if hedge fund managers are more skilful at identifying overpriced fear premiums or periods that do not result in disaster shocks, then they could deliver a superior alpha \citep{gao2018}.

To test this assumption, we first estimate for each fund a conditional factor model for the excess return over the 3-month treasury bill rate $r_{t}^{f}$, $i=1,\ldots,I$:
\begin{equation}
r_{it}-r_{t}^{f}=\alpha^{*}_{i}+\pmb{\beta}^{*}_{i}\tilde{f}_{t}+\eta^{*}_{it},
\end{equation}
where $\alpha^{*}_{i}$ is the fund's unconditional alphas while $\pmb{\beta}^{*}_{i}$ is the vector of loadings for the vector of risk factors $\tilde{f}_{t}$. We want to test whether high exposure to tail risk identifies funds with low alphas. Tail risk exposure is obtained by regressing the lagged value $\hat{\xi}(\mathbf{x}_{t-1})$ on the excess returns at time $t$:
\begin{equation}
r_{it}-r_{t}^{f}=\gamma_{i0}+\gamma_{i}\hat{\xi}(\mathbf{x}_{t-1})+\delta_{it},
\end{equation}
where $\delta_{it}$ is centered on zero.  
If tail risk is positively correlated with a missing factor earning a positive risk premium, we should observe a negative cross-sectional correlation between $\gamma_{i}$ and $\alpha^{*}_{i}$ \citep{ardia2023}. All models are estimated by ordinary least squares.

As reference asset pricing models, we use the CAPM, the 7-factor model of \cite{funghsieh} (FH) and the 6-factor model of \cite{joenv2021} (JKKT)\footnote{This model combines market, value, and size factors with the time-series equity momentum factor of \cite{mosko2012}, the liquidity factor of \cite{pastor2003}, and the betting-against-beta factor of \cite{frazzini2014}.}. For the tail risk measure, we use our HKU estimator\footnote{For HKU, we re-estimate the G-E-GPD model using only the VIX, the FSI, the equity momentum, and the credit spread as covariates.}, the KJ estimator and the POT. We then split the funds into deciles, according to their estimated $\gamma_{i}$, and compute their average estimated $\alpha^{*}_{i}$. We limit ourselves to in-sample analysis, in a static estimation framework.

\begin{figure}[htbp]
    \centering
    \includegraphics[scale=.53]{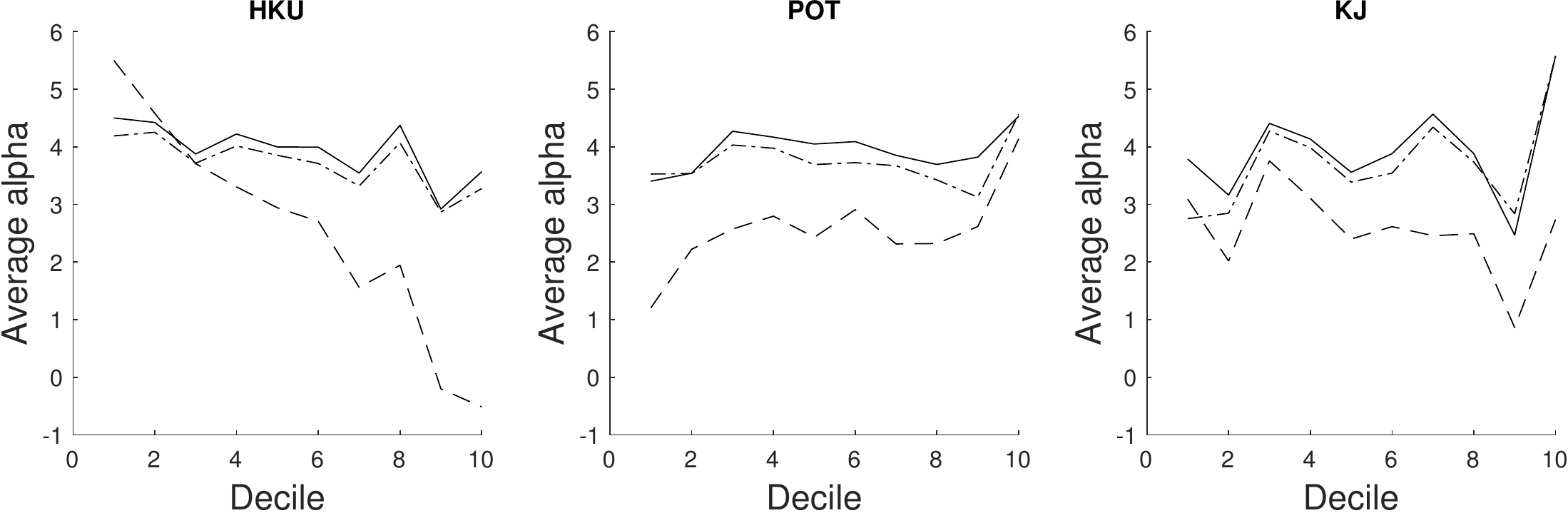}
    \caption{\footnotesize Average $\hat{\alpha}^{*}_{i}$ for deciles of funds sorted on their $\hat{\gamma}_{i}$, obtained with the three different tail risk measures. Solid: CAPM alpha. Dashed-dotted: FH alpha. Dashed: JKKT alpha.}
    \label{fig:portfolio}
\end{figure}

In Table~\ref{tab:ap_results}, we report first descriptive statistics of the estimated  annualized $\hat{\alpha}^{*}_{i}$ obtained with the classical asset pricing models. For the CAPM and the FH model, we find average $\hat{\alpha}^{*}_{i}$ values close to one another, as reported in \cite{ardia2023}, at 3.94\% and 3.73\% respectively. 
JKKT has a smaller average $\hat{\alpha}^{*}_{i}$, although still largely positive at 2.55\%, and a higher standard deviation. Computing the correlation between $\hat{\alpha}^{*}_{i}$ and $\hat{\gamma}_{i}$, we find it to be significantly negative for HKU at the 95\% confidence level for the three reference models. In particular, for JKKT, the correlation reaches $-25\%$. For POT and KJ, the correlation is found to be only mildly positive.
\begin{table}[!ht]
    \centering
    \begin{footnotesize}
    \begin{tabular}{cccccccccccc}
    \hline\hline
         Model &  CAPM  & FH  & JKKT  \\
        \hline\hline
        $\bar{\hat{\alpha}}^{*}_{i}$  & 3.94 & 3.73 & 2.55\\
        std($\hat{\alpha}^{*}_{i}$) & 6.70 & 6.51 & 7.34\\
        Adj. $R^{2}$ &  0.204 & 0.246 & 0.299 \\
        \hline
        HKU $\rho(\hat{\alpha}^{*}_{i},\hat{\gamma}_{i})$ & -0.063 & -0.063 & -0.250\\
          & $(-0.12,-0.02)$ & $(-0.12,-0.02)$ & $(-0.30,-0.20)$ \\
        POT $\rho(\hat{\alpha}^{*}_{i},\hat{\gamma}_{i})$ & 0.030 & 0.010 & 0.093\\
                & $(-0.02,0.08)$ & $(-0.04,0.06)$ & $(0.04,0.14)$ \\
        KJ $\rho(\hat{\alpha}^{*}_{i},\hat{\gamma}_{i})$ &  0.046 & 0.096 & -0.046\\
        & $(-0.01,0.10)$ & $(0.05,0.15)$ & $(-0.10,0.01)$ \\
        \hline\hline\
        Model & HKU & POT & KJ \\
        \hline\hline
$\bar{\hat{\gamma}}_{i}$ & 0.004 & 0.021 &-0.022 \\
std($\hat{\gamma}_{i}$) & 0.026& .288 &0.063 \\
$Q_{.10}(\hat{\gamma}_{i})$& -0.023 &  -0.276 & -0.093 \\ 
$Q_{.90}(\hat{\gamma}_{i})$& 0.033 &0.306 &0.039 \\
        \hline\hline
D1 $\bar{\hat{\alpha}}^{*}_{i}$ (CAPM)& 4.50  &  3.40   & 3.79\\
D10 $\bar{\hat{\alpha}}^{*}_{i}$ (CAPM) &3.57 &   4.53  &  5.58\\
D1-D10 & 0.93 &  -1.13  & -1.79\\
\hline
D1 $\bar{\hat{\alpha}}^{*}_{i}$ (FH) & 4.19   & 3.53   & 2.75\\
D10 $\bar{\hat{\alpha}}^{*}_{i}$ (FH) & 3.28  &  4.56  &  5.57\\
D1-D10 &0.91  & -1.04  & -2.82$^{***}$\\
\hline
D1 $\bar{\hat{\alpha}}^{*}_{i}$ (JKKT) &5.50   & 1.20  &  3.09\\
D10 $\bar{\hat{\alpha}}^{*}_{i}$ (JKKT) &-0.52  &  4.14   & 2.74\\
D1-D10 & 6.01$^{***}$ &  -2.93$^{**}$  &  0.35\\
        \hline\hline
    \end{tabular}          
    \end{footnotesize}
    \caption{\footnotesize Descriptive statistics for the cross-sectional distribution of $\hat{\alpha}^{*}_{i}$ (in \%, annualized) and $\hat{\gamma}_{i}$. Pearson's correlation coefficient is denoted by $\rho$, while $Q_{p}$ refers to the empirical quantile at level $p$. In the lower panel, we report the average $\hat{\alpha}^{*}_{i}$ in the first (D1) and last (D10) decile of funds sorted on their $\hat{\gamma}_{i}$, with the different asset pricing models. The line D1-D10 reports the difference. A significant difference from 0 at the 5\% and 1\% test level, using a simple Welch's t-test, is indicated by $**$ and $***$.}
    \label{tab:ap_results}
\end{table}
We now investigate the average $\hat{\alpha}^{*}_{i}$ per decile of funds, sorted on their $\hat{\gamma}_{i}$ (see Figure~\ref{fig:portfolio}). For CAPM and FH, no patterns are perceptible. However, for JKKT,  we observe a clear downward trend for HKU (Figure~\ref{fig:portfolio}, left panel): an increase in loading for HKU is associated with a decrease in $\hat{\alpha}^{*}_{i}$. The bottom decile displays an average annualized $\hat{\alpha}^{*}_{i}$ of 5.5\%, while the top decile exhibits only -0.52\%. For POT and KJ, we either observe a positive trend, in contradiction with the tested hypothesis, or no trend at all. 
Formally testing for differences between the average $\hat{\alpha}^{*}_{i}$ of the top and bottom deciles with Welch's t-test, we reject the null hypothesis of equal means for HKU at the 1\% test level. A similar result is obtained when repeating the test up to the 8$^{th}$ decile. These results support the hypothesis outlined at the beginning of this section, and suggest the existence of a fear premium exploited by hedge funds to generate abnormal returns. 
Thus, we can conclude from this analysis that our tail risk measure has the potential to improve on the evaluation of hedge fund performance (although a formal assessment of this ability is beyond the scope of the present paper and left to further research). Moreover, while results are clear-cut for HKU, we cannot reach a formal conclusion when using KJ and POT, probably owing to their high estimation uncertainty. These results highlight therefore the usefulness of an efficient estimation method for tail risk inference.

\section{Conclusion}\label{sec:conclusion}

Measuring tail risk and the regression effects of its determinants is a challenging task when studying hedge funds, due to the low reporting frequency in commercial databases. In particular, approaches based on the peaks-over-threshold (POT) discard an excessive portion of the data, rendering the estimation inefficient. Thus, our primary contribution is to introduce a method that estimates the tail and threshold regression models simultaneously. Consequently, we bypass the need to choose \textit{ex ante} a threshold, and we drastically reduce estimation uncertainty.

The proposed approach is based on an auxiliary splicing regression model for non-tail observations. To guard against estimation bias if this model is misspecified, we outline a 
censored maximum likelihood procedure that decreases the influence of non-tail observations on the estimator. In a simulation study, we demonstrate the superiority of this estimator over alternatives. 
  Then, applying the proposed methodology to a representative database of 189,000 hedge funds monthly returns, we identify the VIX, the FSI, the time-series equity momentum, and credit spreads as major factors driving the tail risk. 
Contrasting our results with the POT and the semi-parametric tail risk measure of \cite{kelly2014}, we find that our measure is less volatile, makes more sense economically, and conveys a greater predictive ability for abnormal excess returns of hedge funds, a result in line with the fear premium hypothesis of \cite{gao2018}. 

\section*{Acknowledgments}
The authors thank seminar participants of the Econometric Institute at the Erasmus University Rotterdam, HEC Lausanne Operation department and Humboldt-Universit\"{a}t zu Berlin statistics and econometrics groups for insightful comments, as well as P. H\"{u}bner for assistance in preparing the data. JH acknowledges the financial support of the National Bank of Belgium (project REFEX).
\nocite{}

\begin{singlespace}
{\scriptsize
\bibliography{preprintHKU.bib}}

\begin{thebibliography}{66}
\providecommand{\natexlab}[1]{#1}
\providecommand{\url}[1]{\texttt{#1}}
\expandafter\ifx\csname urlstyle\endcsname\relax
  \providecommand{\doi}[1]{doi: #1}\else
  \providecommand{\doi}{doi: \begingroup \urlstyle{rm}\Url}\fi

\bibitem[Aeberhard et~al.(2021)Aeberhard, Cantoni, Marra, and
  Radice]{aeberhard2021}
Aeberhard, W.H.; Cantoni, E.; Marra, G., and Radice, R.
\newblock Robust fitting for generalized additive models for location, scale
  and shape.
\newblock \emph{Statistics and Computing}, 31\penalty0 (11), 2021.

\bibitem[Agarwal et~al.(2017)Agarwal, Ruenzi, and Weigert]{agarwal2017}
Agarwal, V.; Ruenzi, S., and Weigert, F.
\newblock Tail risk in hedge funds: A unique view from portfolio holdings.
\newblock \emph{Journal of Financial Economics}, 125\penalty0 (3):\penalty0
  610--636, 2017.

\bibitem[Ardia et~al.(2023)Ardia, Barras, Gagliardini, and Scaillet]{ardia2023}
Ardia, D.; Barras, L.; Gagliardini, P., and Scaillet, O.
\newblock Is it alpha or beta? a formal evaluation of hedge fund models.
\newblock \emph{Swiss Finance Institute Research Paper}, 2023.

\bibitem[Babu and Toreti(2016)]{babu2016}
Babu, G.J. and Toreti, A.
\newblock A goodness-of-fit test for heavy tailed distributions with unknown
  parameters and its application to simulated precipitation extremes in the
  euro-mediterranean region.
\newblock \emph{Journal of Statistical Planning and Inference}, 174:\penalty0
  11--19, 2016.

\bibitem[Bader et~al.(2018)Bader, Yan, and Zhang]{bader2018}
Bader, B.; Yan, J., and Zhang, X.
\newblock Automated threshold selection for extreme value analysis via ordered
  goodness-of-fit tests with adjustment for false discovery rate.
\newblock \emph{Annals of Applied Statistics}, 12\penalty0 (1):\penalty0
  310--329, 2018.

\bibitem[Bali et~al.(2007)Bali, Gokcan, and Liang]{bali2007}
Bali, T.G.; Gokcan, S., and Liang, B.
\newblock Value-at-risk and the cross-section of hedge fund returns.
\newblock \emph{Journal of Banking \& Finance}, 31\penalty0 (4):\penalty0
  1135--1166, 2007.

\bibitem[Bali et~al.(2014)Bali, Brown, and Caglayan]{bali2014}
Bali, T.G.; Brown, S.J., and Caglayan, M.O.
\newblock Macroeconomic risk and hedge fund returns.
\newblock \emph{Journal of Financial Economics}, 114\penalty0 (1):\penalty0
  1--19, 2014.

\bibitem[Balkema and de~Haan(1974)]{balkema}
Balkema, A.A. and de~Haan, L.
\newblock {Residual Life Time at Great Age}.
\newblock \emph{The Annals of Probability}, 2\penalty0 (5):\penalty0 792 --
  804, 1974.

\bibitem[Bee et~al.(2019)Bee, Dupuis, and Trapin]{bee2019}
Bee, M.; Dupuis, D.J., and Trapin, L.
\newblock {Realized Peaks over Threshold: A Time-Varying Extreme Value Approach
  with High-Frequency-Based Measures}.
\newblock \emph{Journal of Financial Econometrics}, 17\penalty0 (2):\penalty0
  254--283, 2019.

\bibitem[Beirlant and Goegebeur(2003)]{beirlant2003}
Beirlant, J. and Goegebeur, Y.
\newblock {Regression with response distributions of Pareto-type}.
\newblock \emph{Computational Statistics \& Data Analysis}, 42\penalty0
  (4):\penalty0 595--619, 2003.

\bibitem[Beirlant et~al.(2004)Beirlant, Joossens, and Segers]{beirlant2004b}
Beirlant, J.; Joossens, E., and Segers, J.
\newblock {``Generalized Pareto Fit to the Society of Actuaries’ Large Claims
  Database", Ana C. Cebri{\'a}n, Michel Denuit, and Philippe Lambert, July
  2003}.
\newblock \emph{North American Actuarial Journal}, 8\penalty0 (2):\penalty0
  108--111, 2004.

\bibitem[Biagini et~al.(2021)Biagini, Huber, Jaspersen, and
  Mazzon]{biagini2021}
Biagini, F.; Huber, T.; Jaspersen, J.G., and Mazzon, A.
\newblock Estimating extreme cancellation rates in life insurance.
\newblock \emph{Journal of Risk and Insurance}, 88\penalty0 (4):\penalty0
  971--1000, 2021.

\bibitem[Billio et~al.(2012)Billio, Getmansky, W.~Lo, and Pelizzon]{billio2012}
Billio, M.; Getmansky, M.; W.~Lo, A.W., and Pelizzon, L.
\newblock Econometric measures of connectedness and systemic risk in the
  finance and insurance sectors.
\newblock \emph{Journal of Financial Economics}, 104\penalty0 (3):\penalty0
  535--559, 2012.

\bibitem[Bollerslev et~al.(2015)Bollerslev, Todorov, and Xu]{bollerslev2015}
Bollerslev, T.; Todorov, V., and Xu, L.
\newblock Tail risk premia and return predictability.
\newblock \emph{Journal of Financial Economics}, 118\penalty0 (1):\penalty0
  113--134, 2015.

\bibitem[Castro-Camilo et~al.(2018)Castro-Camilo, de~Carvalho, and
  Wadsworth]{castro2018}
Castro-Camilo, D.; de~Carvalho, M., and Wadsworth, J.
\newblock {Time-varying extreme value dependence with application to leading
  European stock markets}.
\newblock \emph{The Annals of Applied Statistics}, 12\penalty0 (1):\penalty0
  283 -- 309, 2018.

\bibitem[Chavez-Demoulin and Davison(2005)]{chavez2005}
Chavez-Demoulin, V. and Davison, A.~C.
\newblock Generalized additive modelling of sample extremes.
\newblock \emph{Journal of the Royal Statistical Society: Series C (App.
  Stat.)}, 54\penalty0 (1):\penalty0 207--222, 2005.

\bibitem[Chavez-Demoulin et~al.(2016)Chavez-Demoulin, Embrechts, and
  Hofert]{chavez2016}
Chavez-Demoulin, V.; Embrechts, P., and Hofert, M.
\newblock An extreme value approach for modeling operational risk losses
  depending on covariates.
\newblock \emph{Journal of Risk and Insurance}, 83\penalty0 (3):\penalty0
  735--776, 2016.

\bibitem[Choulakian and Stephens(2001)]{choulakian2001}
Choulakian, V. and Stephens, M.~A.
\newblock {Goodness-of-fit tests for the generalized Pareto distribution}.
\newblock \emph{Technometrics}, 43\penalty0 (4):\penalty0 478--484, 2001.

\bibitem[Coles(2001)]{coles2001}
Coles, S.
\newblock \emph{An introduction to statistical modeling of extreme values}.
\newblock Springer, London, 2001.

\bibitem[Cuesta-Albertos et~al.(2008)Cuesta-Albertos, Matrán, and
  Mayo-Iscar]{cuesta2008}
Cuesta-Albertos, J.~A.; Matrán, C., and Mayo-Iscar, A.
\newblock Robust estimation in the normal mixture model based on robust
  clustering.
\newblock \emph{Journal of the Royal Statistical Society: Series B (Stat.
  Method.)}, 70\penalty0 (4):\penalty0 779--802, 2008.

\bibitem[Dacorogna et~al.(2023)Dacorogna, Debbabi, and Kratz]{dacorogna2023}
Dacorogna, M.; Debbabi, N., and Kratz, M.
\newblock Building up cyber resilience by better grasping cyber risk: A new
  algorithm for modelling cyber complaints filed at the {\it gendarmerie
  nationale}.
\newblock \emph{European Journal of Operation Research}, (under review), 2023.

\bibitem[Davison and Smith(1990)]{davisonsmith1990}
Davison, A.~C. and Smith, R.~L.
\newblock Models for exceedances over high thresholds.
\newblock \emph{Journal of the Royal Statistical Society: Series B (Stat.
  Method.)}, 52\penalty0 (3):\penalty0 393--425, 1990.

\bibitem[Dawid et~al.(2016)Dawid, Musio, and Ventura]{dawid2016}
Dawid, A.P.; Musio, M., and Ventura, A.
\newblock Minimum scoring rule inference.
\newblock \emph{Scandinavian Journal of Statistics}, 43\penalty0 (1), 2016.

\bibitem[de~Carvalho et~al.(2022)de~Carvalho, Pereira, Pereira, and
  de~Zea~Bermudez]{decarvalho2021}
de~Carvalho, M.; Pereira, S.; Pereira, P., and de~Zea~Bermudez, P.
\newblock An extreme value bayesian lasso for the conditional left and right
  tails.
\newblock \emph{Journal of Agricultural, Biological and Environmental
  Statistics}, 27:\penalty0 222--239, 2022.

\bibitem[Debbabi et~al.(2017)Debbabi, Kratz, and Mboup]{debbabi}
Debbabi, M.; Kratz, M., and Mboup, M.
\newblock A self-calibrating method for heavy tailed data modelling.
  application in neuroscience and finance.
\newblock \emph{ESSEC Working Paper \& arXiv1612.03974v2}, 2017.

\bibitem[Debbabi and Kratz(2014)]{debbabi2014}
Debbabi, N. and Kratz, M.
\newblock A new unsupervised threshold determination for hybrid models.
\newblock \emph{IEEE International Conference on Acoustics, Speech and Signal
  Processing (ICASSP)}, pages 3440--3444, 2014.

\bibitem[Diks et~al.(2011)Diks, Panchenko, and van Dijk]{diks2011}
Diks, C.; Panchenko, V., and van Dijk, D.
\newblock Likelihood-based scoring rules for comparing density forecasts in
  tails.
\newblock \emph{Journal of Econometrics}, 163\penalty0 (2):\penalty0 215--230,
  2011.

\bibitem[Dunn and G.K.(1996)]{dunn1996}
Dunn, P.K. and G.K., Smyth.
\newblock Randomized quantile residuals.
\newblock \emph{Journal of Computational and Graphical Statistics}, 5\penalty0
  (3):\penalty0 236--244, 1996.

\bibitem[Dupuis et~al.(2022)Dupuis, Engelke, and Trapin]{dupuis2022}
Dupuis, D.; Engelke, S., and Trapin, L.
\newblock Modeling panels of extremes.
\newblock \emph{Annals of Applied Statistics}, To appear, 2022.

\bibitem[Eastoe and Tawn(2009)]{eastoe}
Eastoe, E.F. and Tawn, J.A.
\newblock Modelling non-stationary extremes with application to surface level
  ozone.
\newblock \emph{Journal of the Royal Statistical Society. Series C (App.
  Stat.)}, 58\penalty0 (1):\penalty0 25--45, 2009.

\bibitem[Einmahl and He(2023)]{einmahl2023}
Einmahl, J.H.J. and He, Y.
\newblock Extreme value estimation for heterogeneous data.
\newblock \emph{Journal of Business \& Economic Statistics}, 41\penalty0
  (1):\penalty0 255--269, 2023.

\bibitem[Einmahl et~al.(2016)Einmahl, Kiriliouk, Krajina, and
  Segers]{einmahl2018}
Einmahl, J.H.J.; Kiriliouk, A.; Krajina, A., and Segers, J.
\newblock {An M-estimator of spatial tail dependence}.
\newblock \emph{Journal of the Royal Statistical Society. Series B (Stat.
  Method.)}, 78\penalty0 (1):\penalty0 275--298, 2016.

\bibitem[Embrechts et~al.(1997)Embrechts, Kl\"{u}ppelberg, and
  Mikosch]{embrechtsbook}
Embrechts, P.; Kl\"{u}ppelberg, C., and Mikosch, T.
\newblock \emph{Modelling Extremal Events for Insurance and Finance}.
\newblock Springer, 1997.

\bibitem[Fahrmeir et~al.(2013)Fahrmeir, Kneib, Lang, and Marx]{fahrmeir2013}
Fahrmeir, L.; Kneib, T.; Lang, S., and Marx, B.
\newblock \emph{{Regression : Models, Methods and Applications}}.
\newblock Springer, 2013.

\bibitem[Frazzini and Pedersen(2014)]{frazzini2014}
Frazzini, A. and Pedersen, L.H.
\newblock Betting against beta.
\newblock \emph{Journal of Financial Economics}, 111\penalty0 (1):\penalty0
  1--25, 2014.

\bibitem[Fung and Hsieh(2004)]{funghsieh}
Fung, W. and Hsieh, D.A.
\newblock Hedge {{Fund Benchmarks}}: {{A Risk-Based Approach}}.
\newblock \emph{Financial Analysts Journal}, 60\penalty0 (5):\penalty0 65--80,
  September 2004.

\bibitem[Gao et~al.(2018)Gao, Gao, and Song]{gao2018}
Gao, G.P.; Gao, P., and Song, Z.
\newblock {Do Hedge Funds Exploit Rare Disaster Concerns?}
\newblock \emph{The Review of Financial Studies}, 31\penalty0 (7):\penalty0
  2650--2692, 2018.

\bibitem[Gao et~al.(2019)Gao, Lu, and Song]{gao2019}
Gao, G.P.; Lu, X., and Song, Z.
\newblock Tail risk concerns everywhere.
\newblock \emph{Management Science}, 65\penalty0 (7):\penalty0 3111--3130,
  2019.

\bibitem[Getmansky et~al.(2015)Getmansky, Lee, and Lo]{getmansky2015}
Getmansky, M.; Lee, P.A., and Lo, A.W.
\newblock Hedge funds: A dynamic industry in transition.
\newblock \emph{Annual Review of Financial Economics}, 7\penalty0 (1):\penalty0
  483--577, 2015.

\bibitem[Hambuckers et~al.(2018)Hambuckers, Groll, and Kneib]{hambuckers2018}
Hambuckers, J.; Groll, A., and Kneib, T.
\newblock {Understanding the economic determinants of the severity of
  operational losses: A regularized generalized Pareto regression approach}.
\newblock \emph{Journal of Applied Econometrics}, 33\penalty0 (6):\penalty0
  898--935, 2018.

\bibitem[He et~al.(2022)He, Peng, Zhang, and Zhao]{he2022}
He, Y.; Peng, L.; Zhang, D., and Zhao, Z.
\newblock {Risk Analysis via Generalized Pareto Distributions}.
\newblock \emph{Journal of Business \& Economic Statistics}, 40\penalty0
  (2):\penalty0 852--867, 2022.

\bibitem[Hothorn et~al.(2014)Hothorn, Kneib, and Bühlmann]{hothorn2014}
Hothorn, T.; Kneib, T., and Bühlmann, P.
\newblock Conditional transformation models.
\newblock \emph{Journal of the Royal Statistical Society Series B (Stat.
  Method.)}, 76:\penalty0 3--27, 2014.

\bibitem[Hu and Zidek(2002)]{hu2002}
Hu, F. and Zidek, J.V.
\newblock The weighted likelihood.
\newblock \emph{Canadian Journal of Statistics}, 30\penalty0 (3):\penalty0
  347--371, 2002.

\bibitem[Huang et~al.(2012)Huang, Liu, {Ghon Rhee}, and Wu]{huang2012}
Huang, W.; Liu, Q.; {Ghon Rhee}, S., and Wu, F.
\newblock Extreme downside risk and expected stock returns.
\newblock \emph{Journal of Banking \& Finance}, 36\penalty0 (5):\penalty0
  1492--1502, 2012.

\bibitem[H\"{u}ser and Davison(2014)]{huser2014}
H\"{u}ser, R. and Davison, A.~C.
\newblock Space–time modelling of extreme events.
\newblock \emph{Journal of the Royal Statistical Society: Series B (Stat.
  Method.)}, 76\penalty0 (2):\penalty0 439--461, 2014.

\bibitem[Jensen et~al.(2021)Jensen, Kelly, and Pedersen]{kelly2021}
Jensen, T.; Kelly, B., and Pedersen, L.
\newblock Is there a replication crisis in finance?
\newblock \emph{The Journal of Finance}, forthcoming, 2021.

\bibitem[Joenv\"{a}\"{a}r\"{a} et~al.(2021)Joenv\"{a}\"{a}r\"{a}, Kauppila,
  Kosowski, and Tolonen]{joenv2021}
Joenv\"{a}\"{a}r\"{a}, J.; Kauppila, M.; Kosowski, R., and Tolonen, P.
\newblock Hedge fund performance: Are stylized facts sensitive to which
  database one uses?
\newblock \emph{Critical Finance Review}, 10:\penalty0 1--70, 2021.

\bibitem[Jun-Haeng~Heo et~al.(2013)Jun-Haeng~Heo, Shin, Woosung~Nam, Om, and
  Jeong]{heo2013}
Jun-Haeng~Heo, J-H.; Shin, H.; Woosung~Nam, W.; Om, J., and Jeong, C.
\newblock Approximation of modified anderson–darling test statistics for
  extreme value distributions with unknown shape parameter.
\newblock \emph{Journal of Hydrology}, 499:\penalty0 41--49, 2013.

\bibitem[Kelly and Jiang(2014)]{kelly2014}
Kelly, B. and Jiang, H.
\newblock {Tail Risk and Asset Prices}.
\newblock \emph{The Review of Financial Studies}, 27\penalty0 (10):\penalty0
  2841--2871, 06 2014.

\bibitem[Kneib et~al.(2021)Kneib, Silbersdorff, and S\"{a}fken]{kneib2021}
Kneib, T.; Silbersdorff, A., and S\"{a}fken, B.
\newblock Rage against the mean -- a review of distributional regression
  approaches.
\newblock \emph{Econometrics and Statistics}, 2021.

\bibitem[Koenker and Bassett(1978)]{koenker1978}
Koenker, R. and Bassett, G.
\newblock Regression quantiles.
\newblock \emph{Econometrica}, 46\penalty0 (1):\penalty0 33--50, 1978.

\bibitem[McNeil et~al.(2015)McNeil, Frey, and Embrechts]{mcneil2015}
McNeil, A.J.; Frey, R., and Embrechts, P.
\newblock \emph{Quantitative Risk Management: Concepts, Techniques and Tools -
  Revised Edition}.
\newblock Princeton University Press, 2015.

\bibitem[Mhalla et~al.(2022)Mhalla, Hambuckers, and Lambert]{mhalla2022}
Mhalla, L.; Hambuckers, J., and Lambert, M.
\newblock Extremal connectedness of hedge funds.
\newblock \emph{Journal of Applied Econometrics}, 37\penalty0 (5):\penalty0
  987--1009, 2022.

\bibitem[Moskowitz et~al.(2012)Moskowitz, Ooi, and Pedersen]{mosko2012}
Moskowitz, T.J.; Ooi, Y.H., and Pedersen, L.H.
\newblock Time series momentum.
\newblock \emph{Journal of Financial Economics}, 104\penalty0 (2):\penalty0
  228--250, 2012.

\bibitem[Naveau et~al.(2016)Naveau, Huser, Ribereau, and Hannart]{naveau2016}
Naveau, P.; Huser, R.; Ribereau, P., and Hannart, A.
\newblock Modeling jointly low, moderate, and heavy rainfall intensities
  without a threshold selection.
\newblock \emph{Water Resources Research}, 52\penalty0 (4):\penalty0
  2753--2769, 2016.

\bibitem[Opitz et~al.(2018)Opitz, H\"{u}ser, Bakka, and Rue]{opitz2018}
Opitz, T.; H\"{u}ser, R.; Bakka, H., and Rue, H.
\newblock {INLA goes extreme: Bayesian tail regression for the estimation of
  high spatio-temporal quantiles}.
\newblock \emph{Extremes}, 21:\penalty0 441--462, 2018.

\bibitem[Pastor and Stambaugh(2003)]{pastor2003}
Pastor, L. and Stambaugh, R.F.
\newblock Liquidity risk and expected stock returns.
\newblock \emph{Journal of Political Economy}, 111\penalty0 (3):\penalty0
  642--685, 2003.

\bibitem[Patton and Ramadorai(2013)]{patton2013}
Patton, A.J. and Ramadorai, T.
\newblock On the {{High-Frequency Dynamics}} of {{Hedge Fund Risk Exposures}}.
\newblock \emph{The Journal of Finance}, 68\penalty0 (2):\penalty0 597--635,
  2013.

\bibitem[Pickands(1975)]{pickands}
Pickands, J.
\newblock {Statistical Inference Using Extreme Order Statistics}.
\newblock \emph{The Annals of Statistics}, 3\penalty0 (1):\penalty0 119 -- 131,
  1975.

\bibitem[Reynkens et~al.(2017)Reynkens, Verbelen, Beirlant, and
  Antonio]{reynkens2017}
Reynkens, T.; Verbelen, R.; Beirlant, J., and Antonio, K.
\newblock Modelling censored losses using splicing: A global fit strategy with
  mixed erlang and extreme value distributions.
\newblock \emph{Insurance: Mathematics and Economics}, 77:\penalty0 65--77,
  2017.

\bibitem[Rigby and Stasinopoulos(2005)]{rigby2005}
Rigby, R. and Stasinopoulos, D.
\newblock {Generalized additive models for location, scale and shape}.
\newblock \emph{Journal of the Royal Statistical Society. Series C (App.
  Stat.)}, 54\penalty0 (3):\penalty0 507--554, 2005.

\bibitem[Stulz(2007)]{stulz2007}
Stulz, R.M.
\newblock Hedge funds: Past, present, and future.
\newblock \emph{Journal of Economic Perspectives}, 21\penalty0 (2):\penalty0
  175--194, 2007.

\bibitem[Van~der Vaart(2000)]{van2000asymptotic}
Van~der Vaart, A.W.
\newblock \emph{Asymptotic statistics}, volume~3.
\newblock Cambridge university press, 2000.

\bibitem[Vandewalle et~al.(2007)Vandewalle, Beirlant, Christmann, and
  Hubert]{vandewalle2007}
Vandewalle, B.; Beirlant, J.; Christmann, A., and Hubert, M.
\newblock {A robust estimator for the tail index of Pareto-type distributions}.
\newblock \emph{Computational Statistics \& Data Analysis}, 51\penalty0
  (12):\penalty0 6252--6268, 2007.

\bibitem[Wan and Zidek(2005)]{wang2005}
Wan, X.G. and Zidek, J.V.
\newblock {Selecting likelihood weights by cross-validation}.
\newblock \emph{The Annals of Statistics}, 33\penalty0 (2):\penalty0 463 --
  500, 2005.

\bibitem[Zou(2006)]{zou2006}
Zou, H.
\newblock {The Adaptive Lasso and Its Oracle Properties}.
\newblock \emph{{Journal of the American Statistical Association}},
  101\penalty0 (476):\penalty0 1418--1429, 2006.

\end{thebibliography}
\end{singlespace}

\newpage
\appendix
\begin{center}
    {\large\bf APPENDIX}
\end{center}

In Section~\ref{app:simu}, we present the results of a second simulation study using increasing contamination rates in DGP II from the main manuscript. In Section~\ref{app:mean_filter}, we detail the mean estimation and filtering procedure used to process the data before conducting the tail risk analysis. In Section~\ref{sec:empirical_suite}, we provide additional estimates of the tail risk model splitting the funds according to their use of financial leverage or not, and using only a subset of the covariates. 
In Section~\ref{sec:theoretical_suite}, we discuss the conditions presented in Section~\ref{sec:theory} in a simplified setting, considering a Pareto distribution.

\section{Additional simulation results}\label{app:simu}

In this section, we present additional simulation results to complement Section~\ref{sec:simu}. We use the simulation setting of DGP II outlined in Section~\ref{subsec:simu_corspec}, with a single change: instead of contaminating the 10\% smallest observations, we use 3 other contamination rates $c$, namely 7.5\%, 20\%, and 30\%. Parameters of the G-E-GPD and GPD distributions remain unchanged. 

\begin{figure}[htbp]
    \centering
    \begin{tabular}{ccc}
\includegraphics[scale=.32]{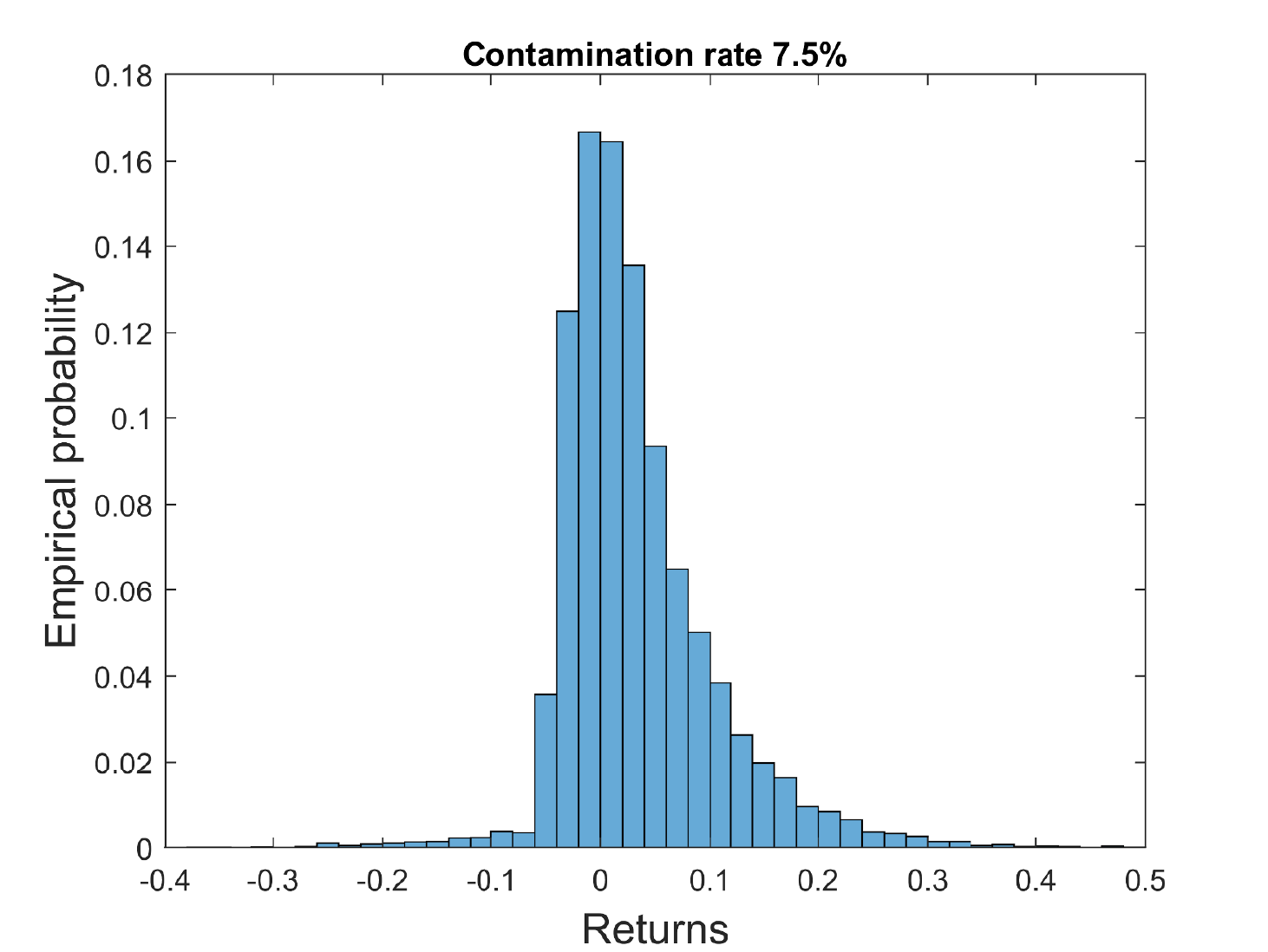} & \includegraphics[scale=.32]{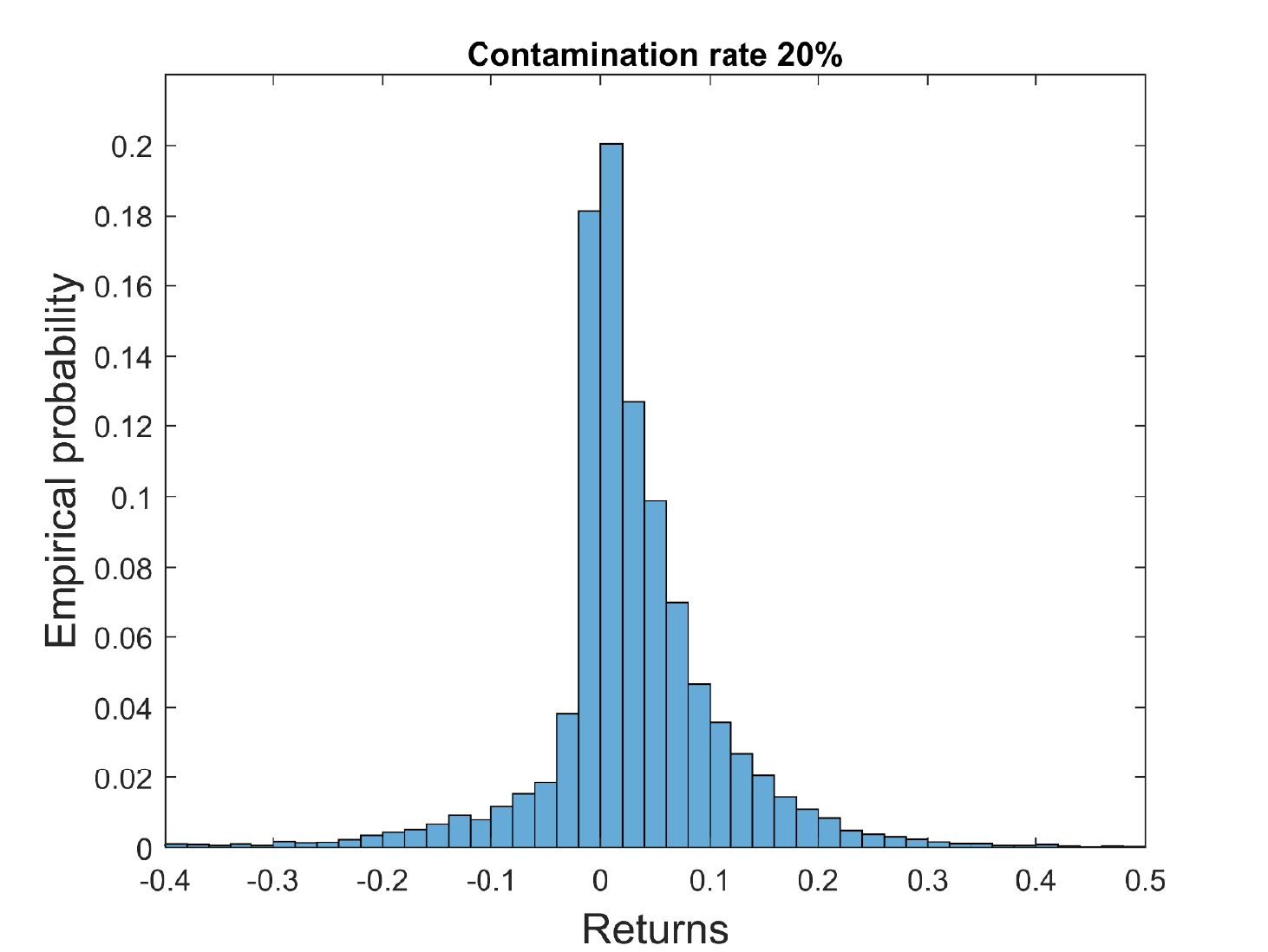} &\includegraphics[scale=.32]{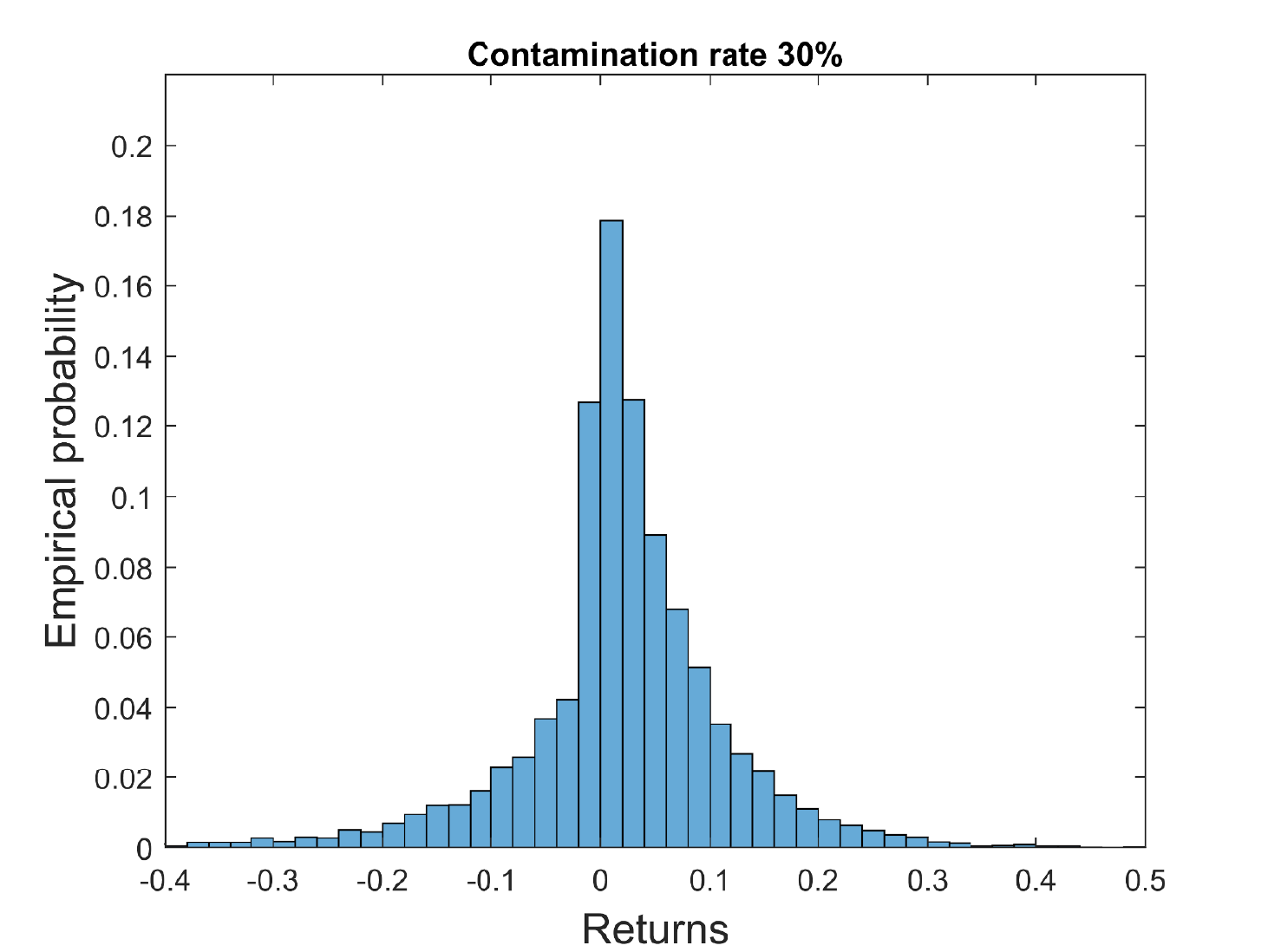}\\
(i) $c=0.075$ & (ii) $c=0.20$ & (iii) $c=0.30$
\end{tabular}
    \caption{\footnotesize Simulated data obtained with different contamination thresholds $c$.}
    \label{fig:histo_appendix}
\end{figure}

\begin{figure}[htbp]
    \centering
    \begin{tabular}{ccc}
\includegraphics[scale=.32]{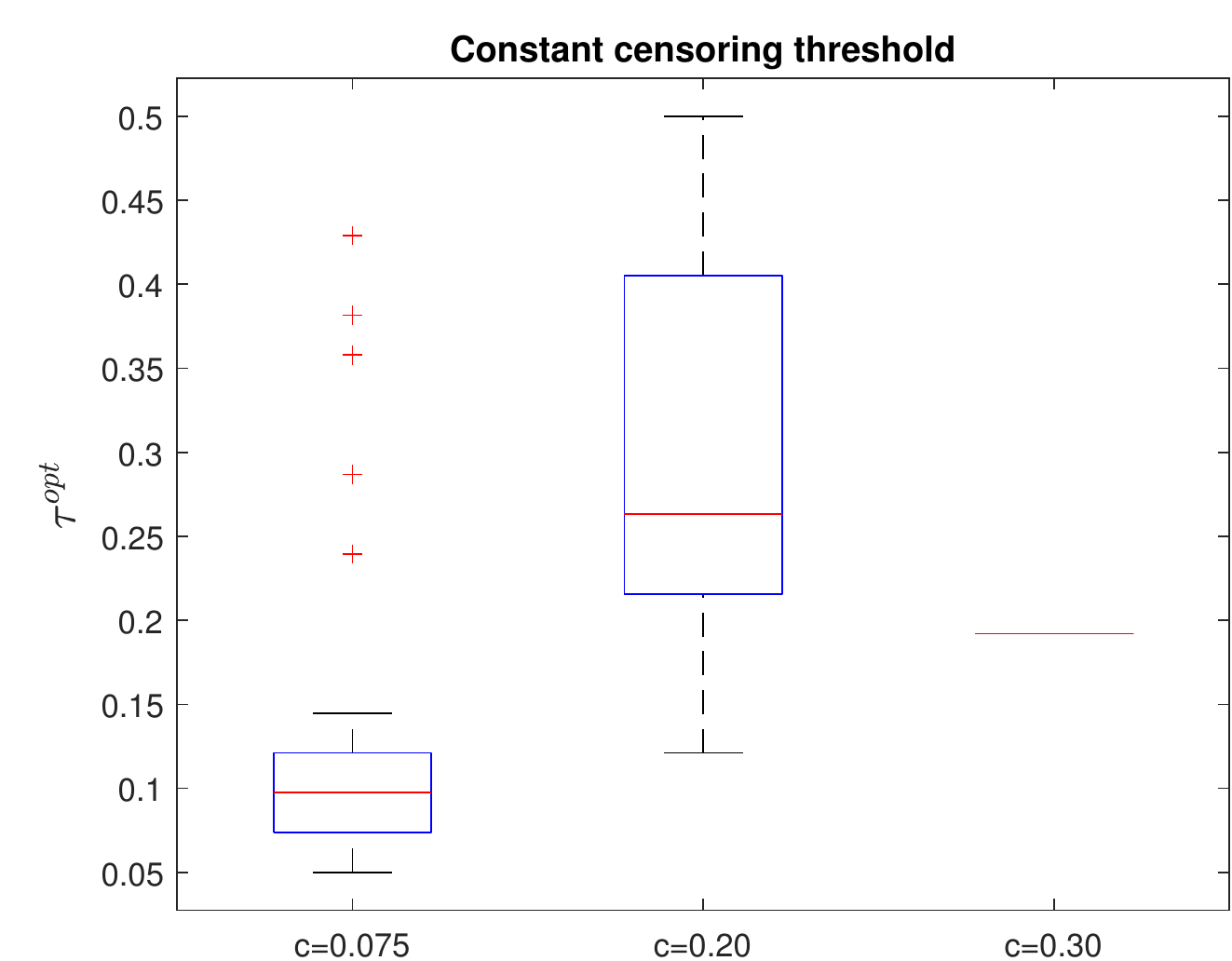} & \includegraphics[scale=.32]{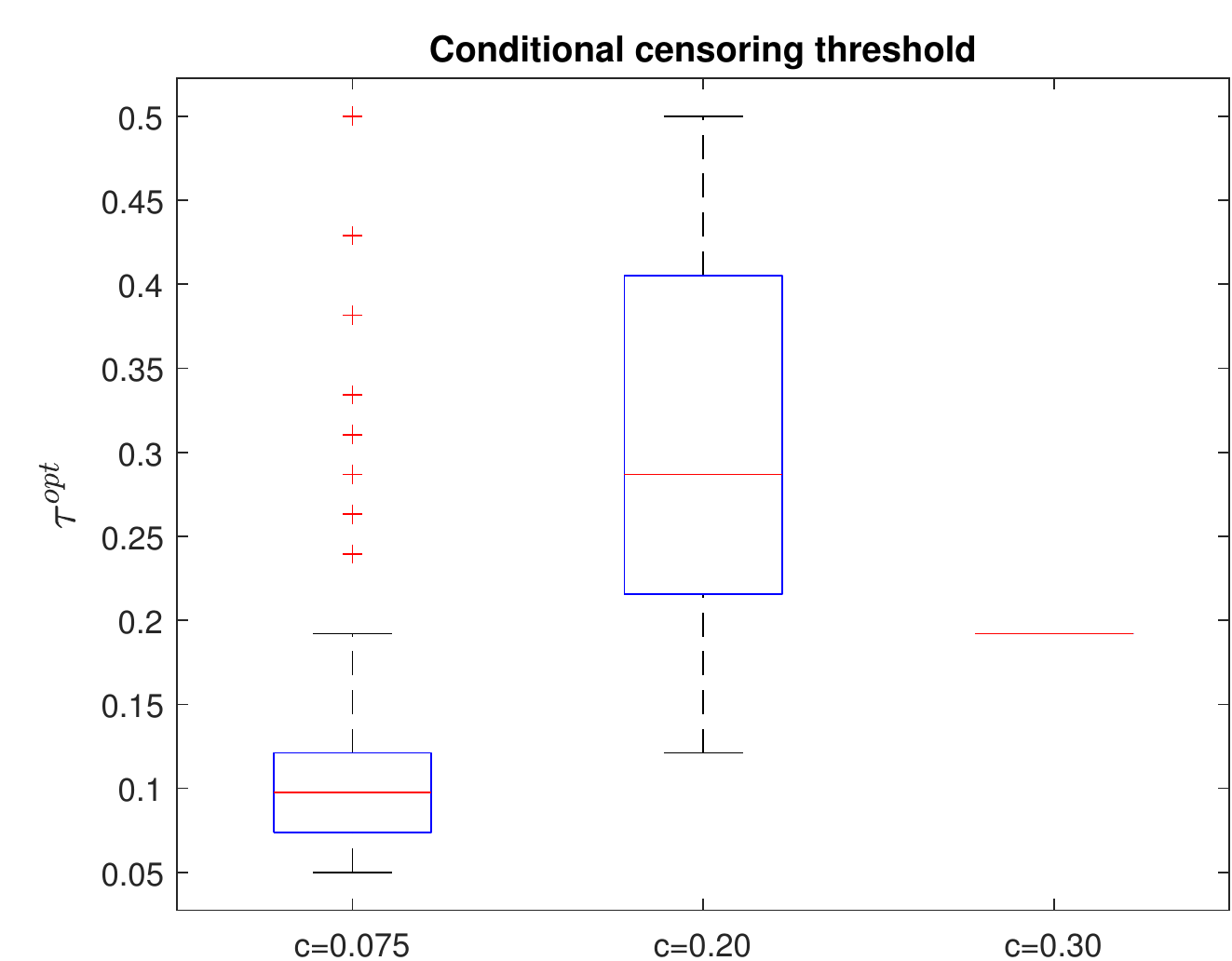} &\includegraphics[scale=.32]{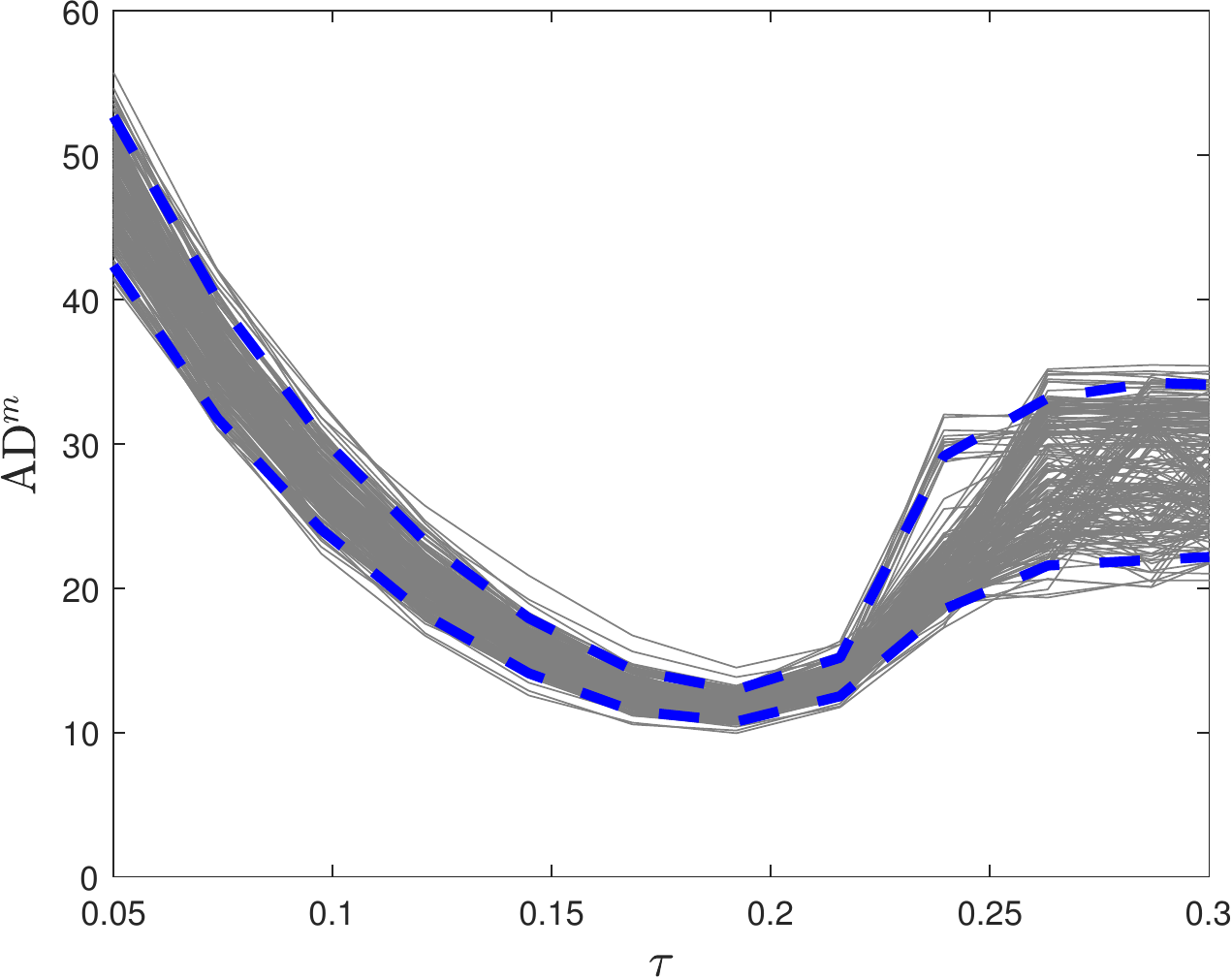}\\
(i) & (ii)& (iii) 
\end{tabular}
    \caption{\footnotesize Selected censoring thresholds using either an unconditional quantile (panel (i)) or a conditional quantile (panel (ii)) in the weight function. Panel (iii): AD$^{m}$ curves for $c=30\%$. Blue: 5\% and 95\% empirical quantile of AD$^{m}$ for fixed values of $\tau$.}
    \label{fig:histo_tau}
\end{figure}

In Figure \ref{fig:histo_appendix}, we display representative histograms of the simulated data. The data are severely skewed for the contamination rates 7.5\% and 20\%. In Figure \ref{fig:histo_tau}, we portray the selected $\tau^{opt}$ for the different DGPs, and an example of AD$^{m}$ curves for $c=0.30$ (see equation (12) and following for exact definitions of these two quantities). For the smallest contamination rate ($c=0.075$), we observe small $\tau$ values around 10\%, although it can sometimes be much larger. The parameter $\tau^{opt}$ takes values between 0.2 and 0.4 for $c=0.20$. Interestingly, for the largest contamination rate ($c=0.30$), we always select $\tau^{opt}=0.20$.
  
\begin{figure}[!h]
\centering
\begin{tabular}{ccc}
\includegraphics[scale = .4]{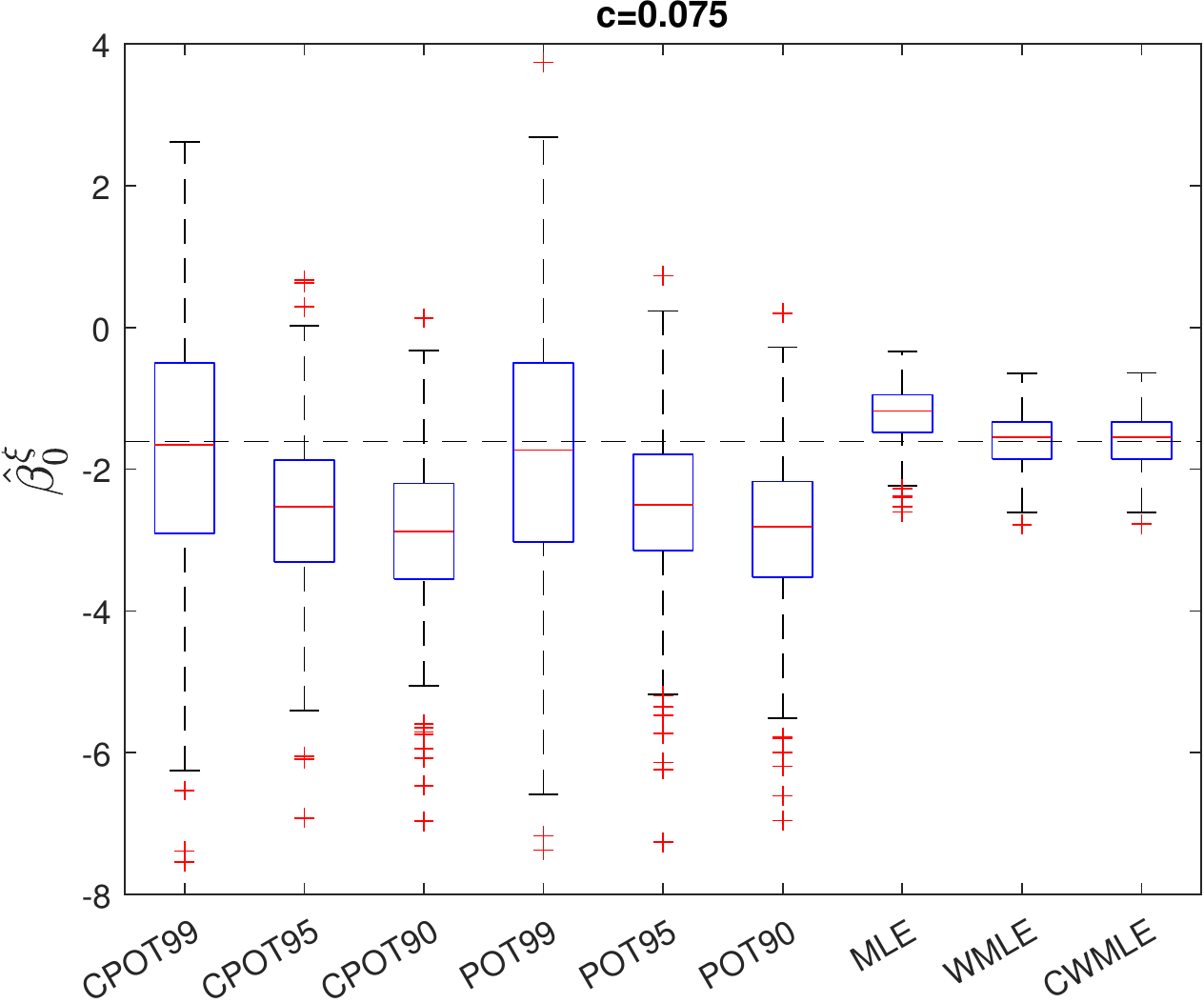}  &    \includegraphics[scale = .4]{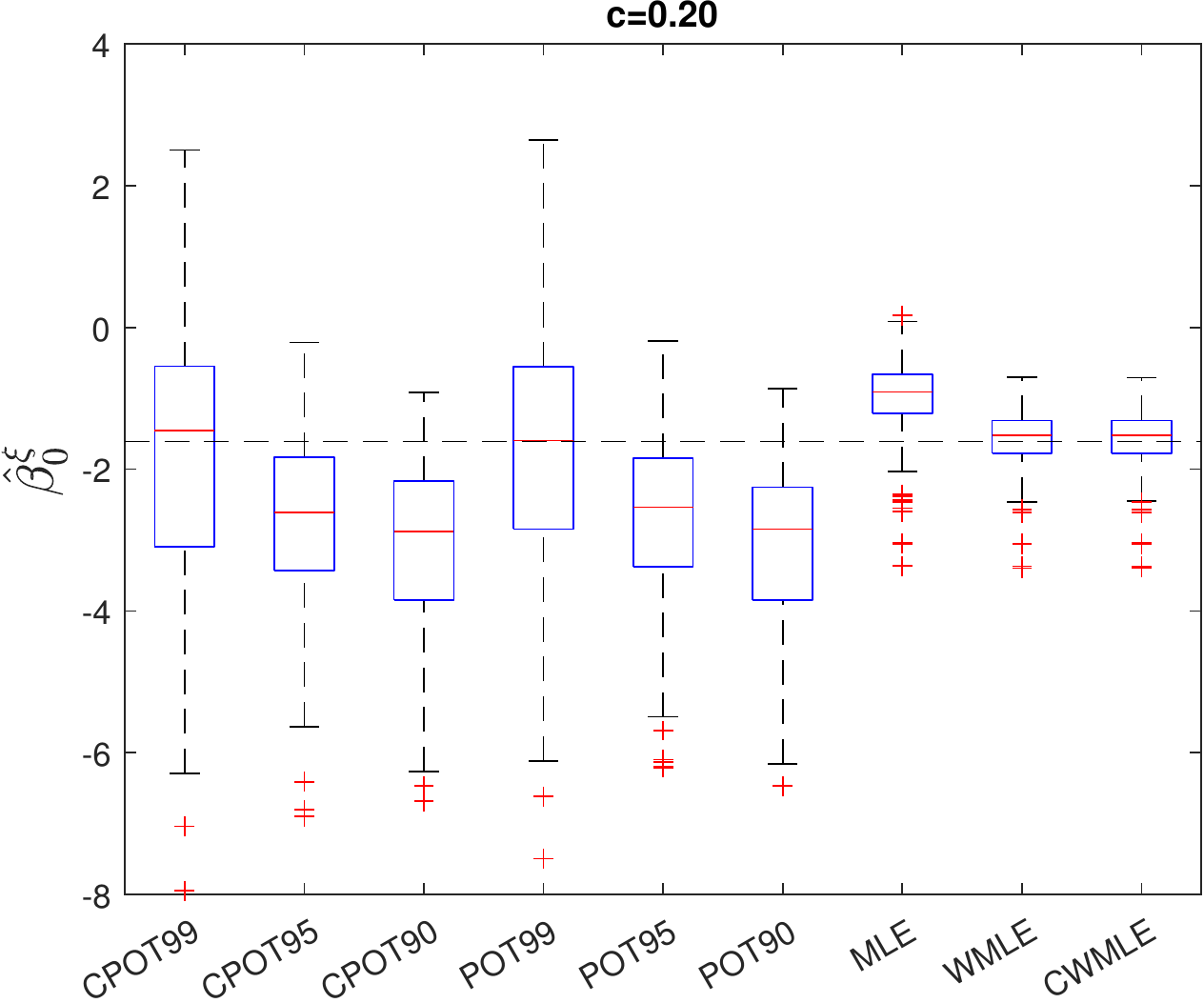} &   \includegraphics[scale = .4]{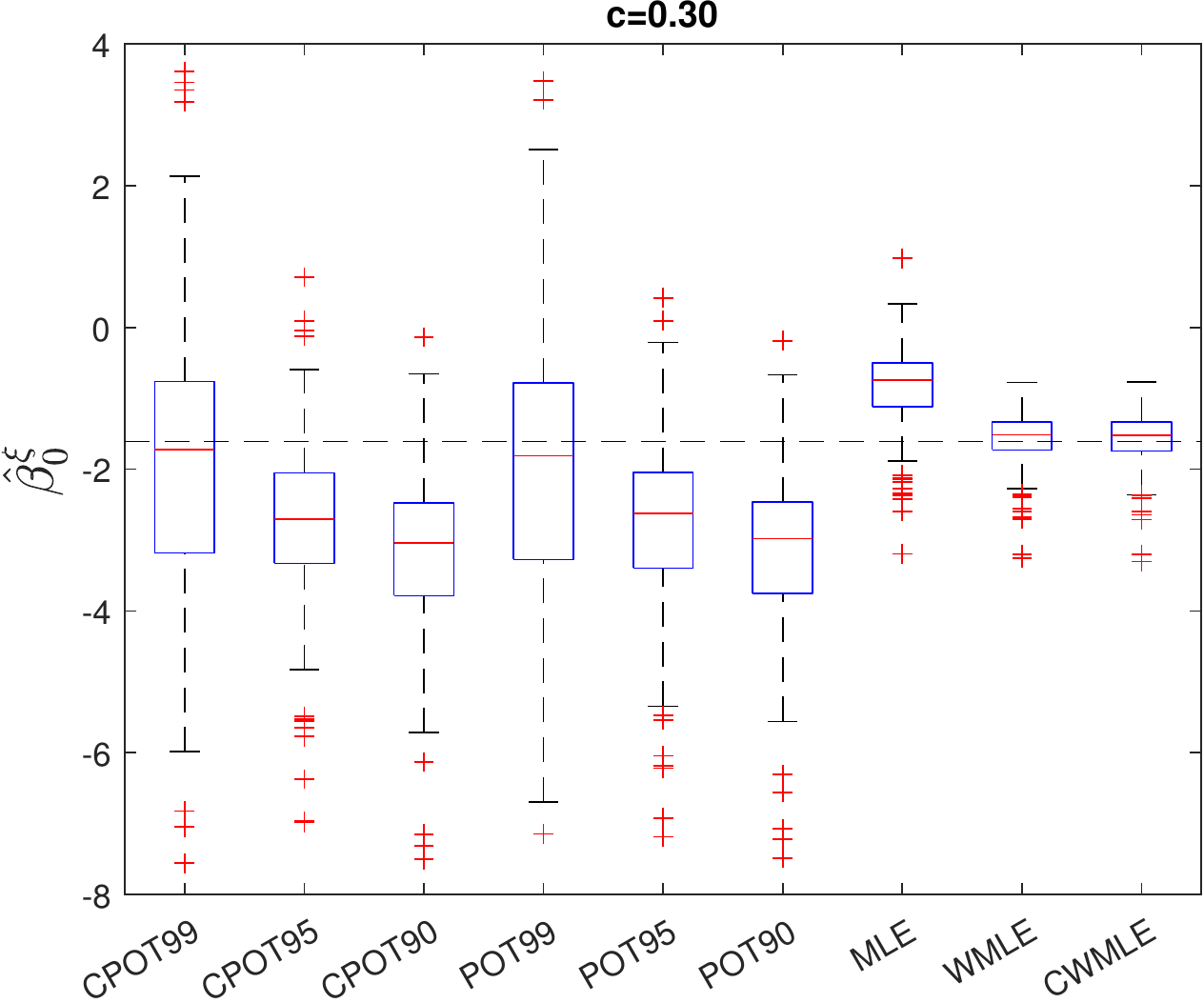}\\       \includegraphics[scale = .4]{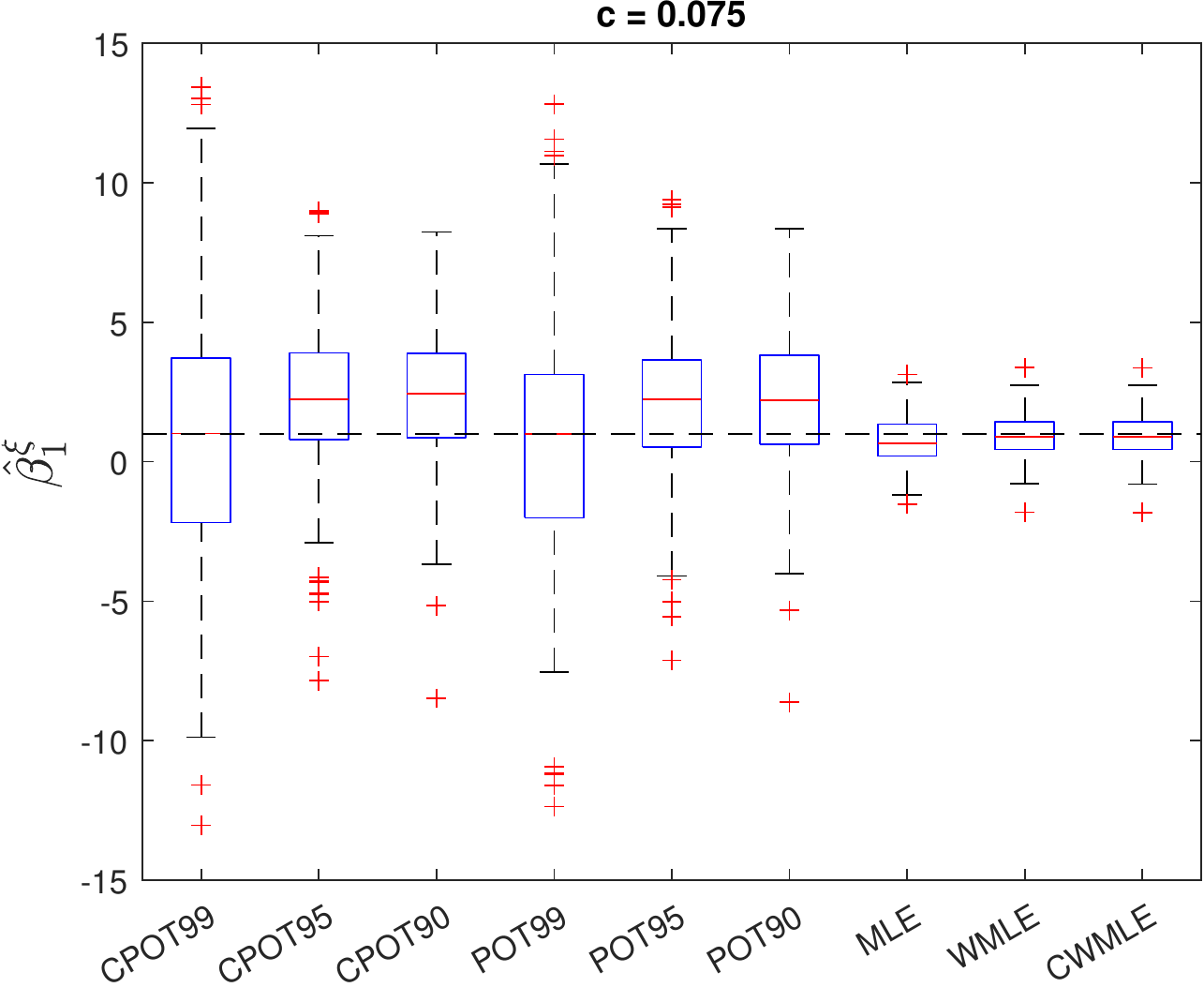} &   \includegraphics[scale = .4]{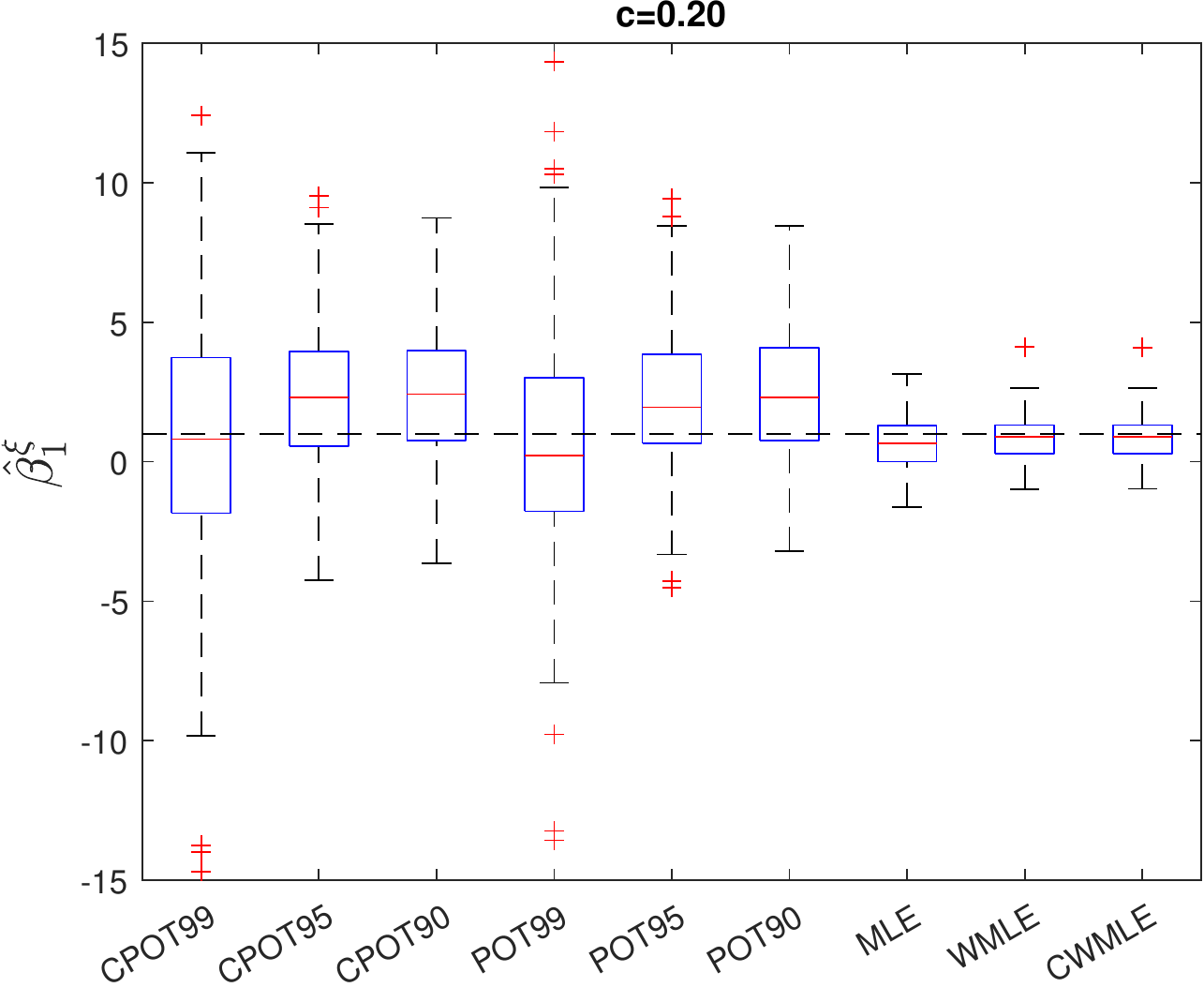}
  &              \includegraphics[scale = .4]{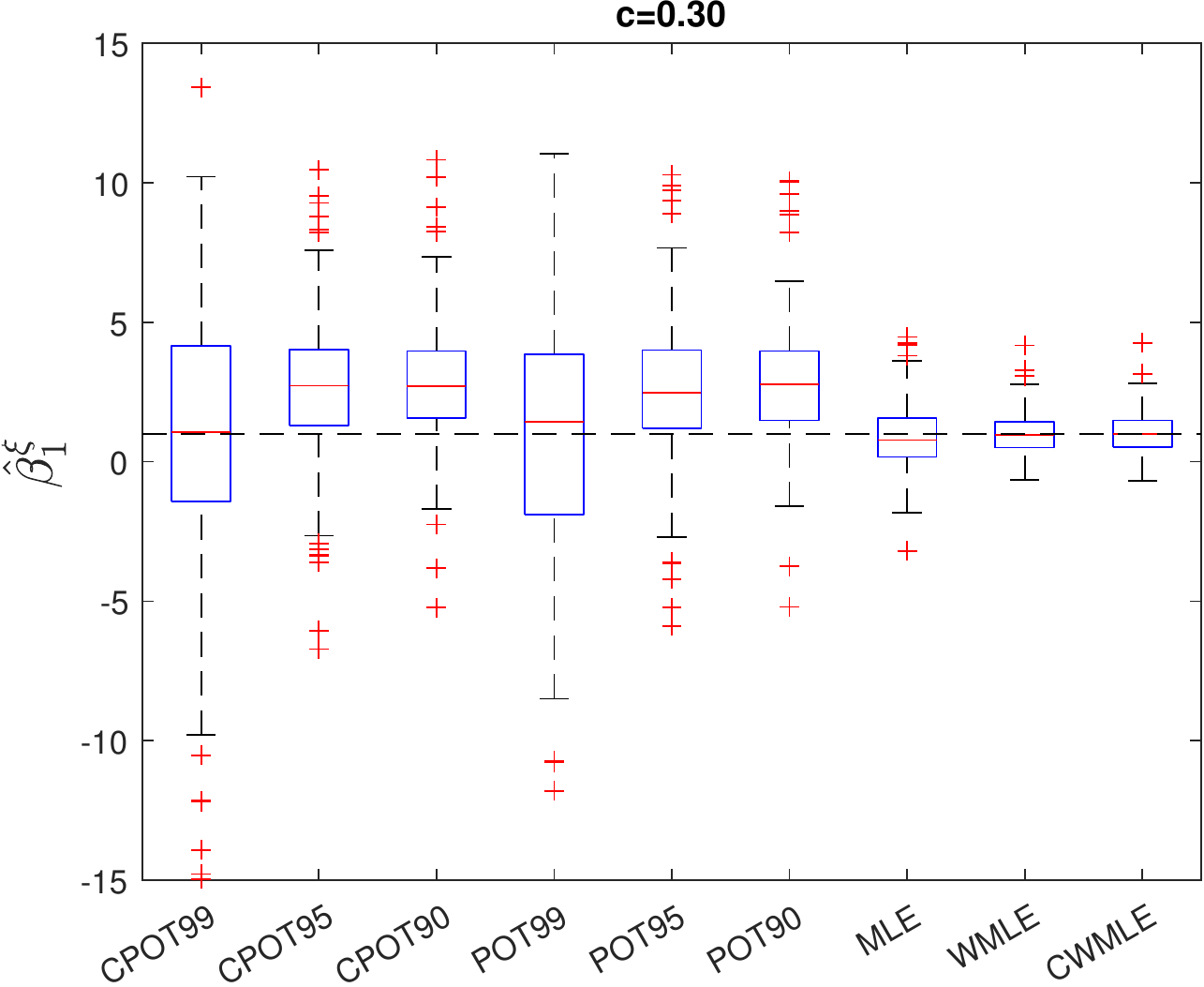}\\
(i) & (ii) & (iii)
\end{tabular}
    \caption{\footnotesize Estimated parameters $\hat{\beta}_{0}^{\xi}$ (top row) and $\hat{\beta}_{1}^{\xi}$ (bottom row) obtained with the G-E-GPD method and with various POT approaches for DGP I to DGP III (from left to right). Dashed: values of the true parameters.}
    \label{fig:estimates_appendix}
\end{figure}

\begin{figure}[htbp]
    \centering
    \begin{tabular}{ccc}
        \includegraphics[scale=.37]{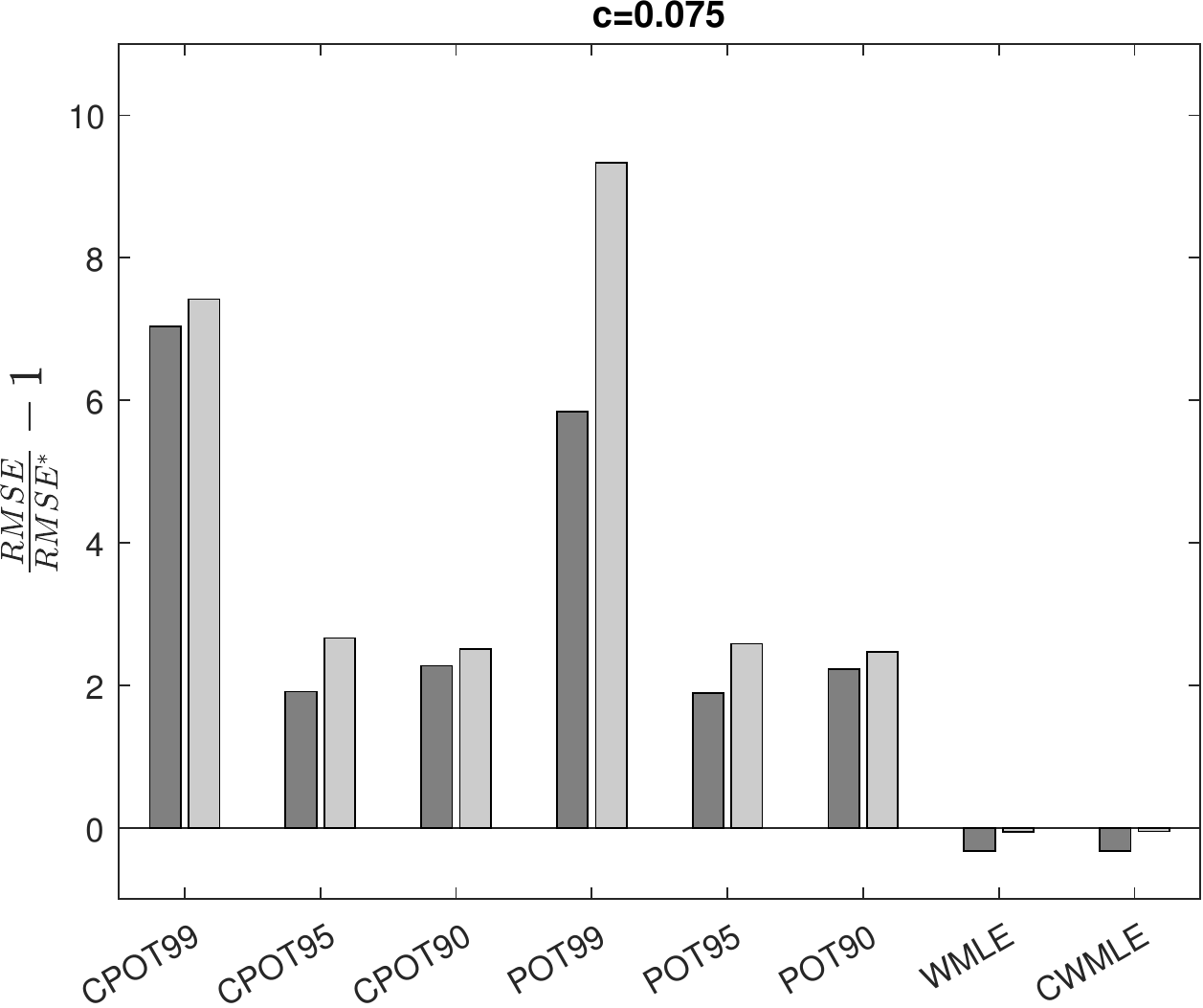} &\includegraphics[scale=.37]{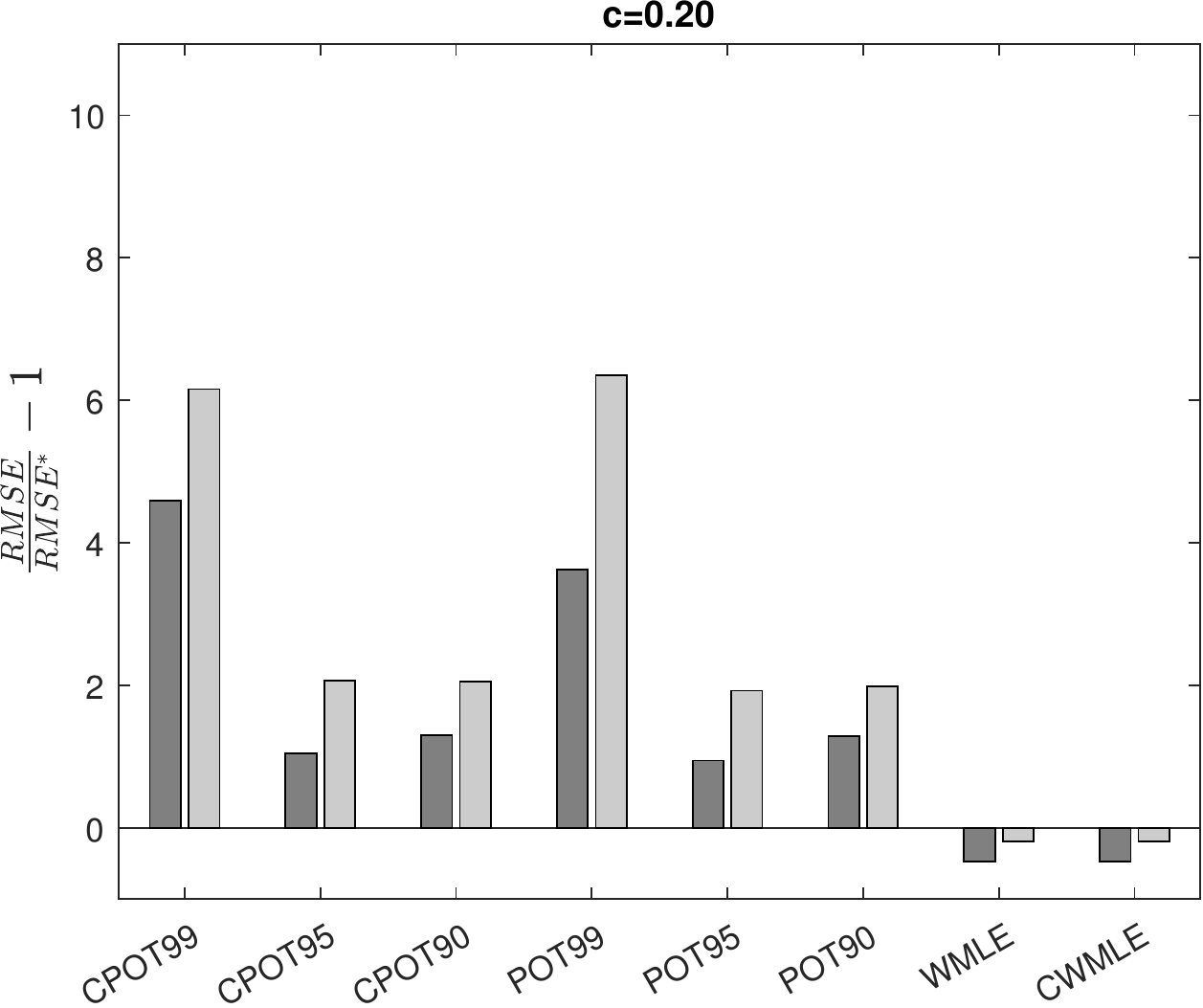} &\includegraphics[scale=.37]{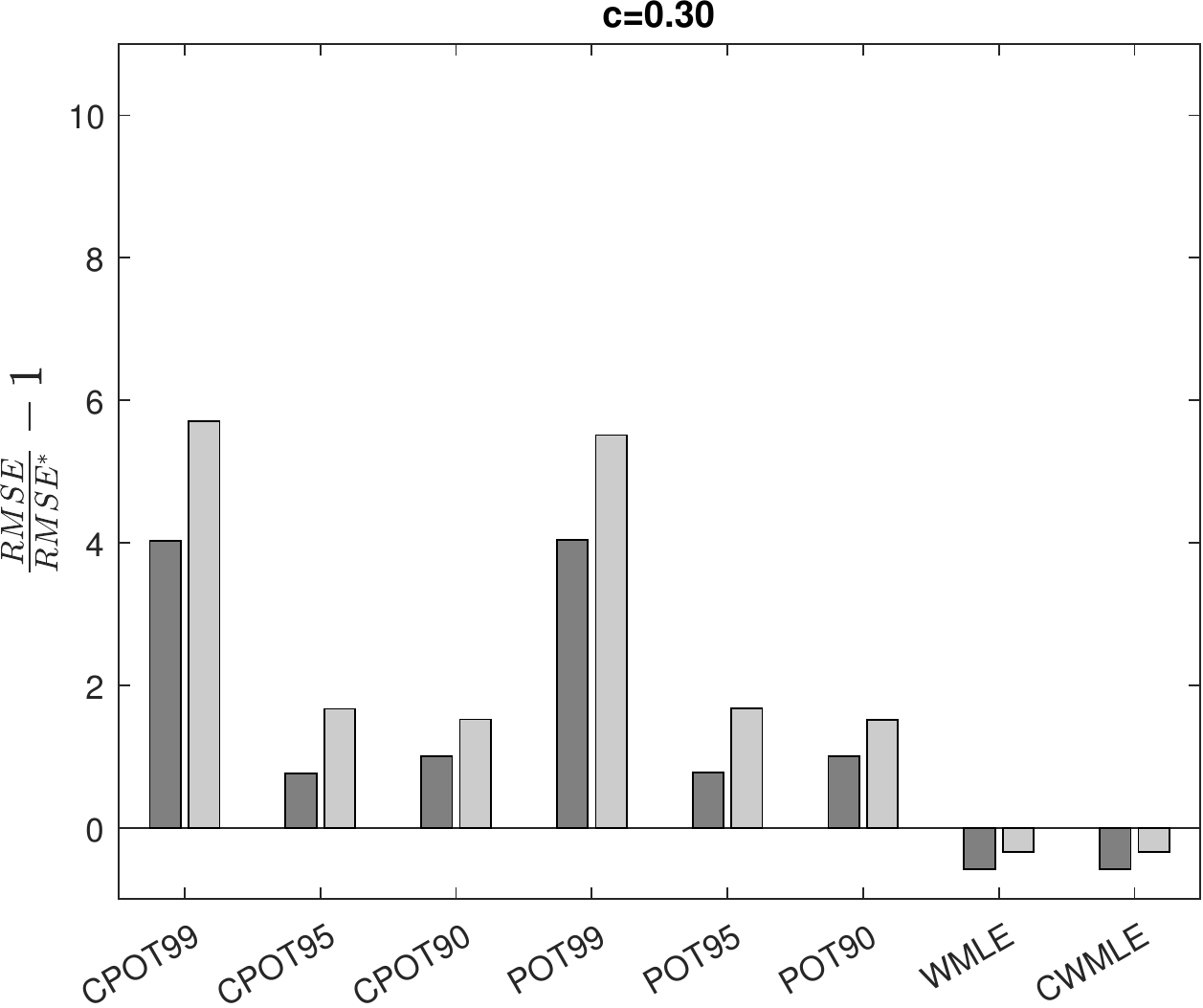}  \\
        (i) & (ii) & (iii)
    \end{tabular}
    \caption{\footnotesize $RMSE/RMSE^{*}-1$ for $\hat{\beta}_{0}^{\xi}$ (dark grey) and $\hat{\beta}_{1}^{\xi}$ (light grey), for the different contamination rates.}
    \label{fig:rmse_appendix}
\end{figure}

 In Figures \ref{fig:estimates_appendix} and \ref{fig:rmse_appendix}, we report the estimated regression coefficient for $\xi$ and the corresponding RMSE, scaled by the RMSE of the MLE. Estimates obtained from the G-E-GPD model with a censoring estimation procedure (referred as WMLE) are virtually unbiased and much less dispersed than the POT-based estimates or the classical MLE. In terms of RMSE, WMLE and CWMLE are much better than MLE. Decreases in relative RMSE are stronger with increasing contamination rates. Gains in RMSE with respect to the MLE varies between $5\%$ and 33\% for $\beta^{\xi}_{1}$, and between 32\% and 58\% for $\beta^{\xi}_{0}$. 

\begin{figure}[t]
    \centering
    \begin{tabular}{ccc}
\includegraphics[scale=.6]{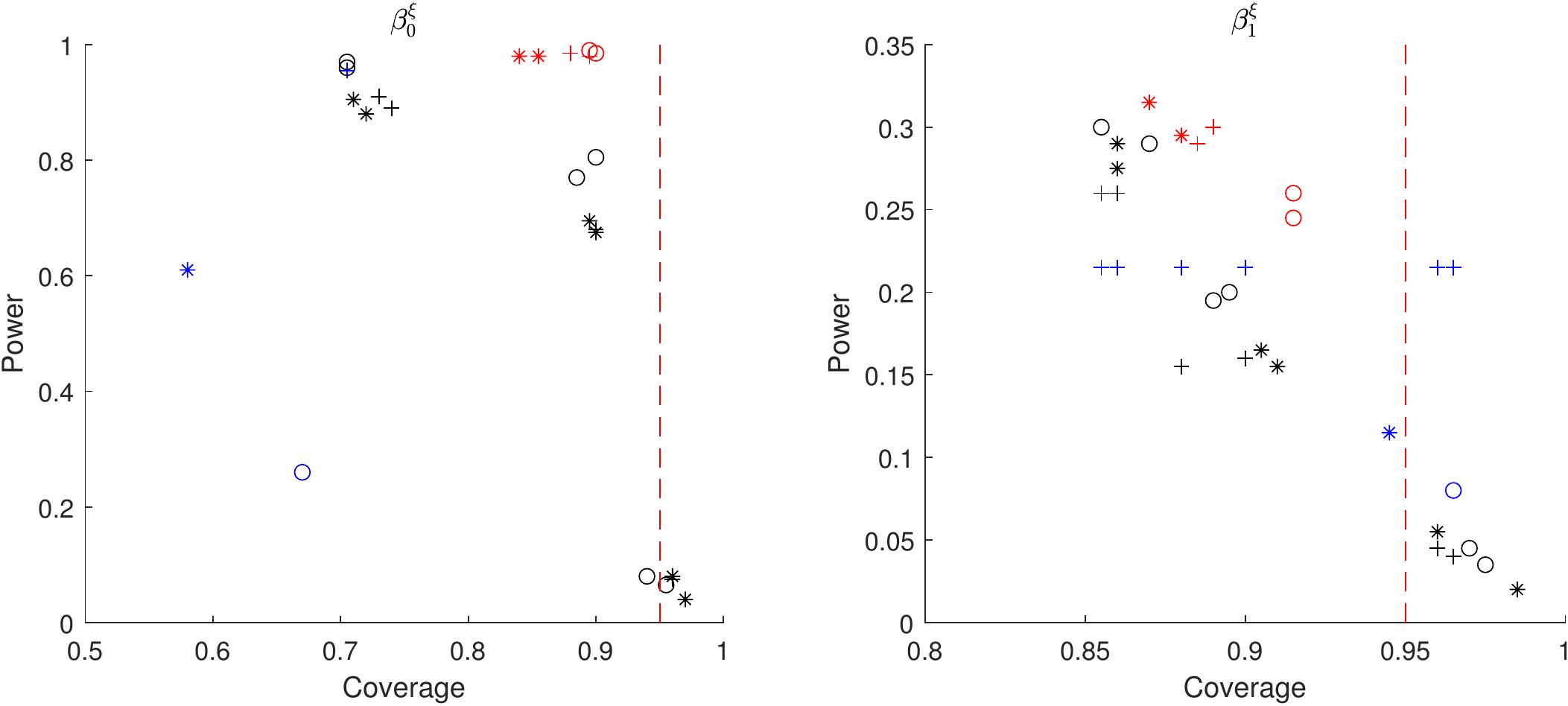} 
\end{tabular}
    \caption{\footnotesize Power of rejecting the null hypothesis $H_{0}:\beta^{\xi}_{j}=0$, for $j=0$ (left) and $j=1$ (right), as a function of the empirical coverage of a 95\% confidence interval. Black: POT-based method. Blue: MLE. Red: (C)WMLE. $+$: $c=0.075$, $*$: $c=0.20$, $\circ$: $c=0.3$. Dashed red: nominal coverage of the confidence intervals.}
    \label{fig:power_coverage}
\end{figure}

In Figure \ref{fig:power_coverage}, we display the power of Wald-type tests \citep{fahrmeir2013} against the null hypothesis $H_{0}:\beta^{\xi}_{j}=0$,
 as a function of the coverage rate of 95\% confidence intervals. WMLE systematically provides a good trade-off, with its coverage rate being close to the nominal 95\% (dashed red curve) and its power being equivalent to or dominating the power of the other methods. 
 
\begin{figure}[H]
\centering
\begin{tabular}{ccc}
\includegraphics[scale = .4]{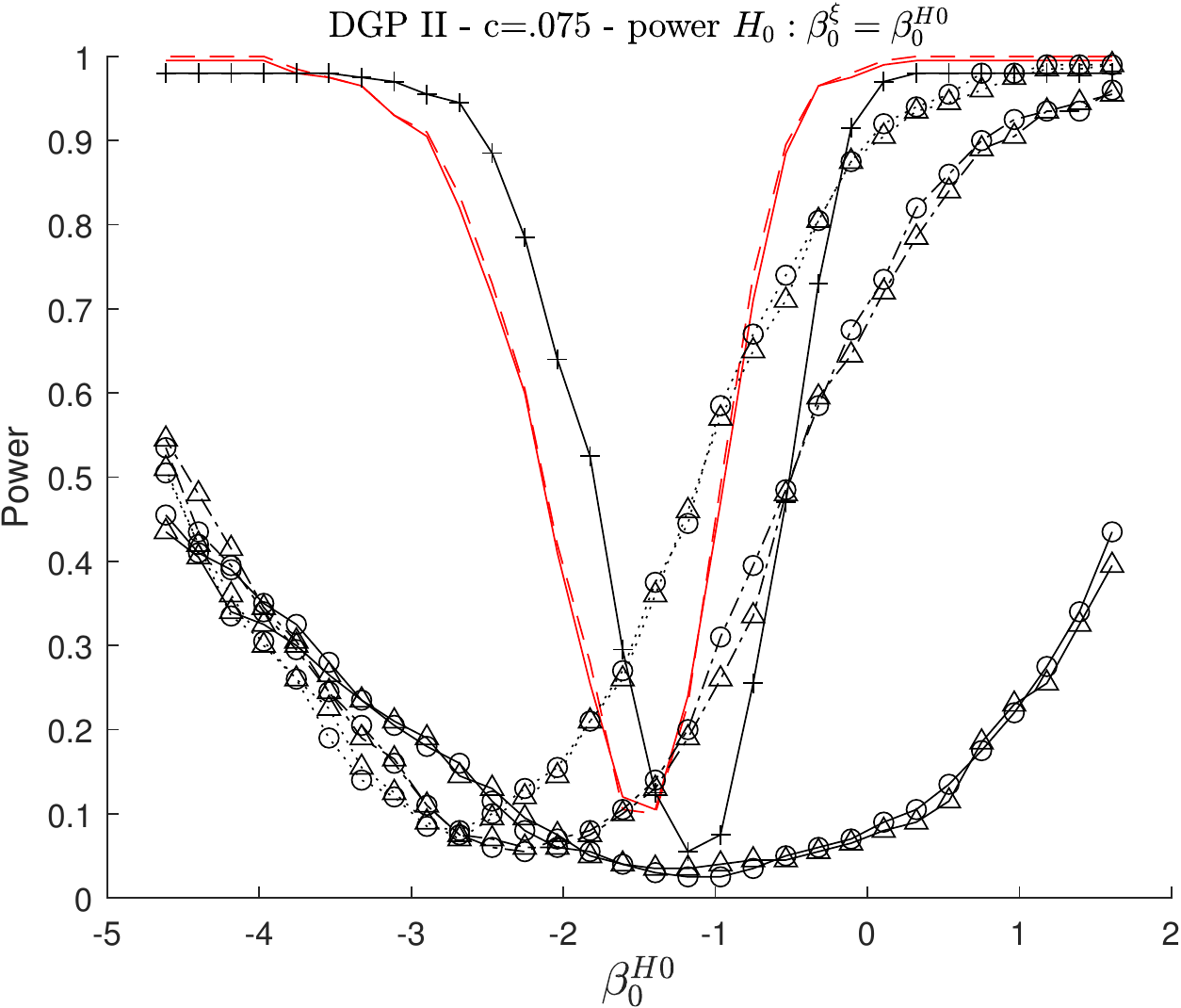}  &   \includegraphics[scale = .4]{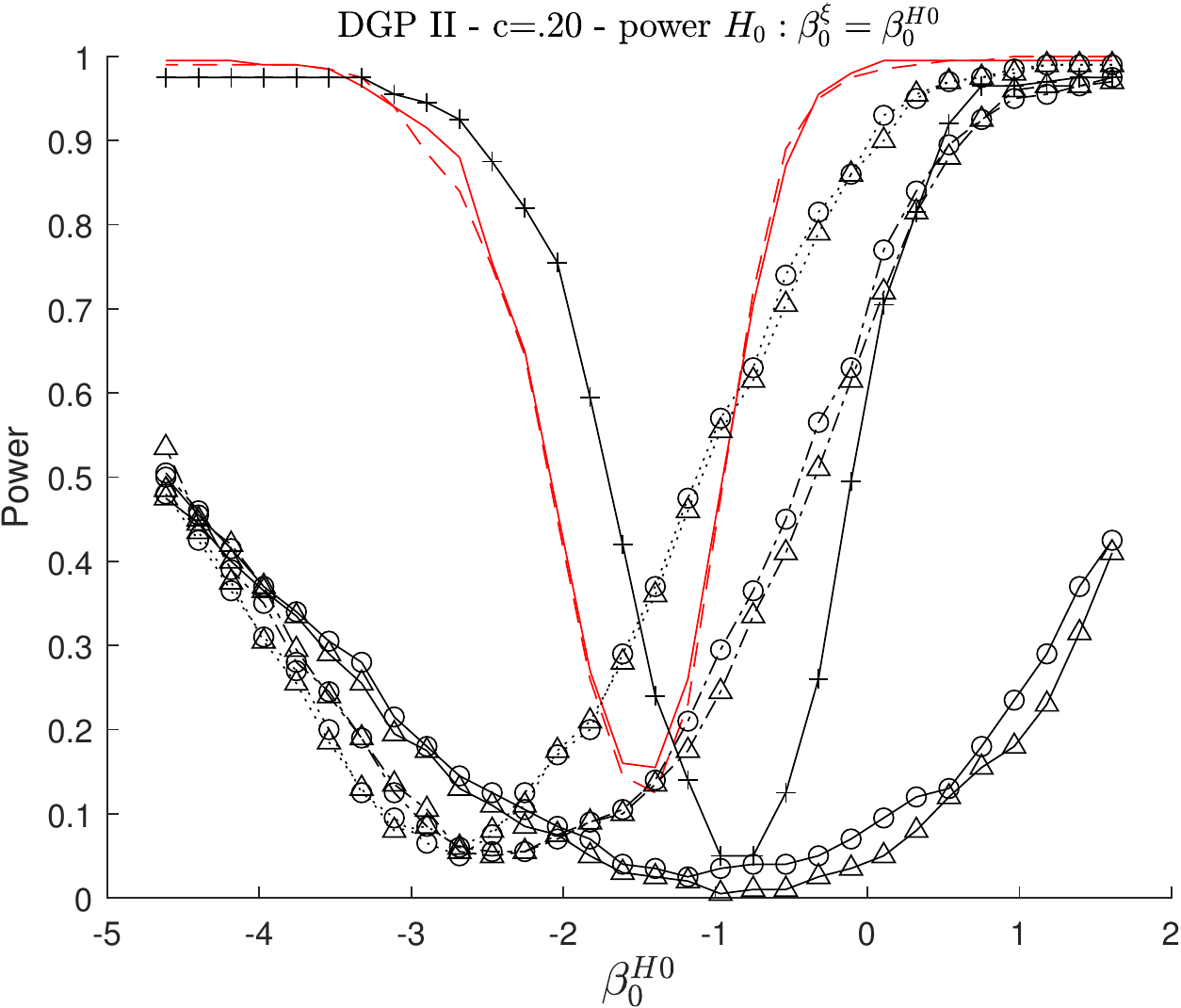}    &\includegraphics[scale = .45]{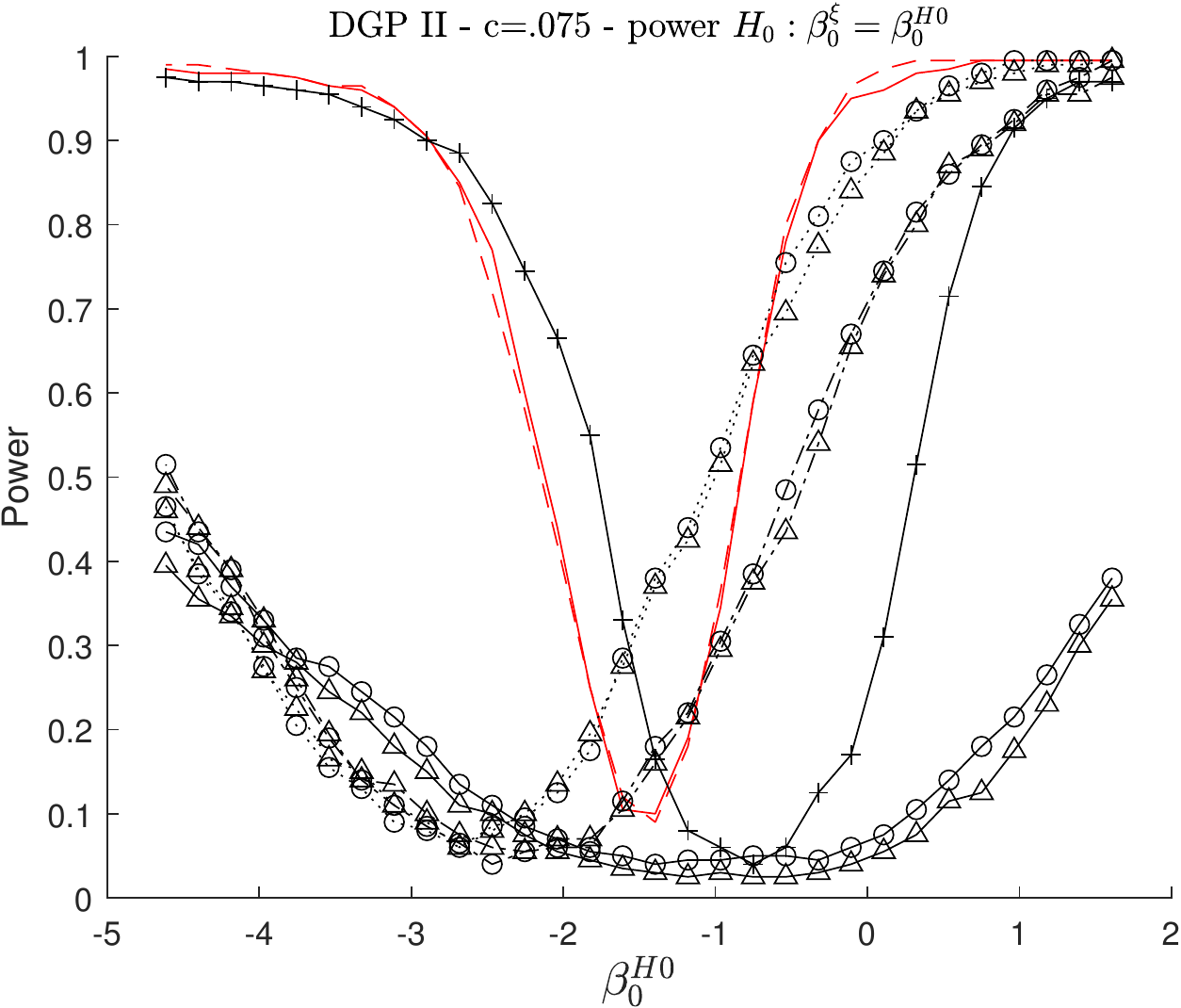}\\       \includegraphics[scale = .4]{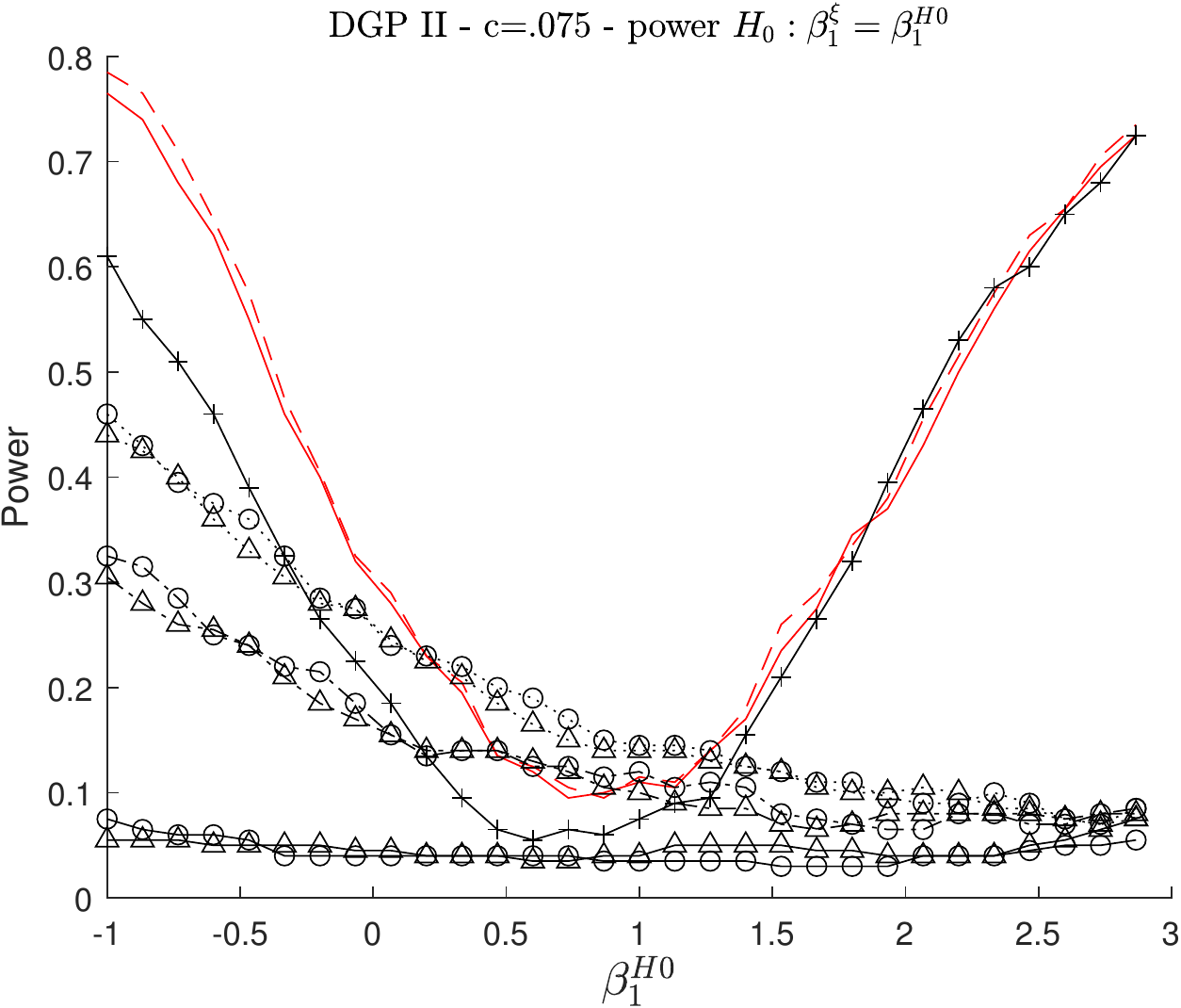}
  &              \includegraphics[scale = .4]{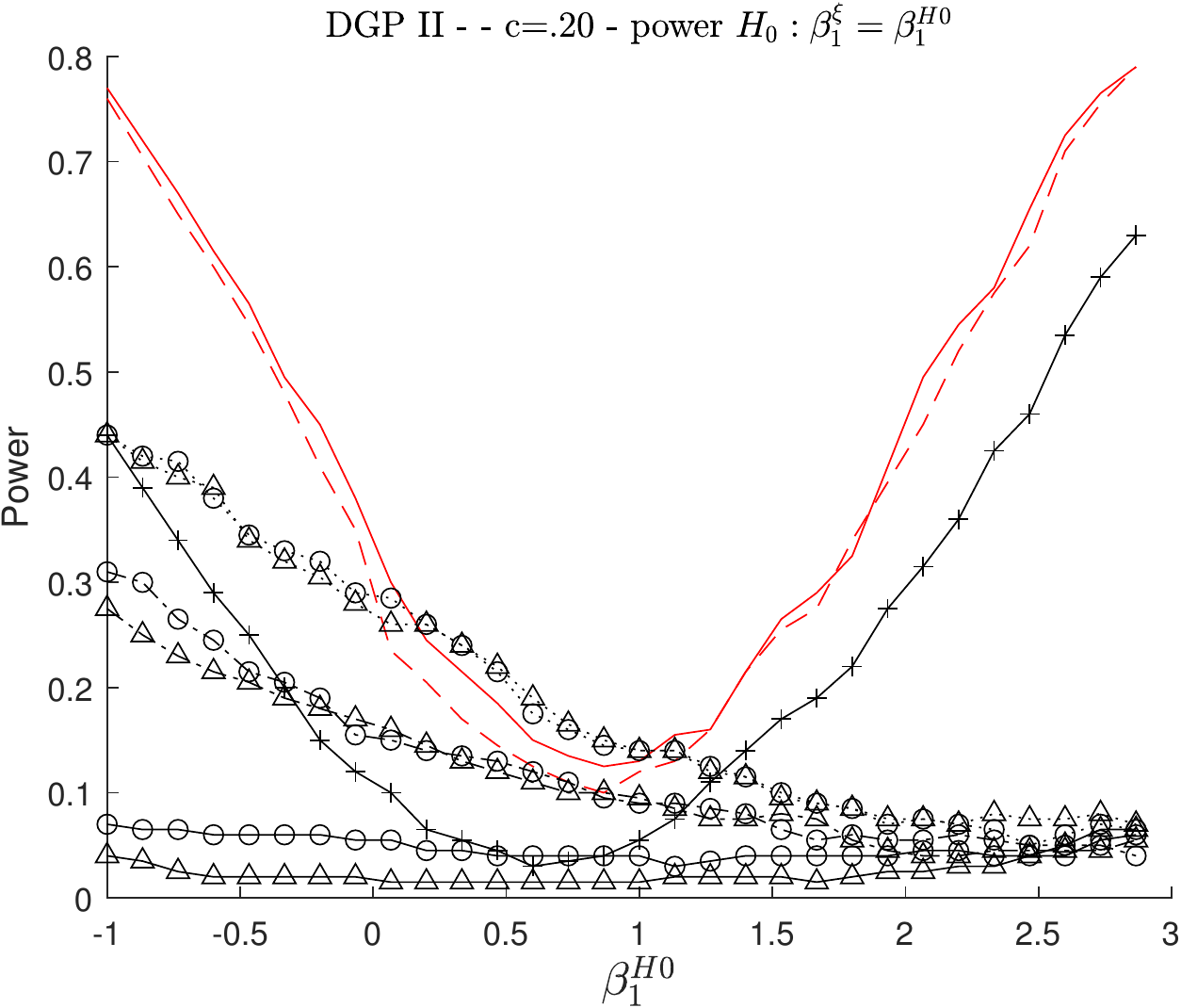}  &              \includegraphics[scale = .4]{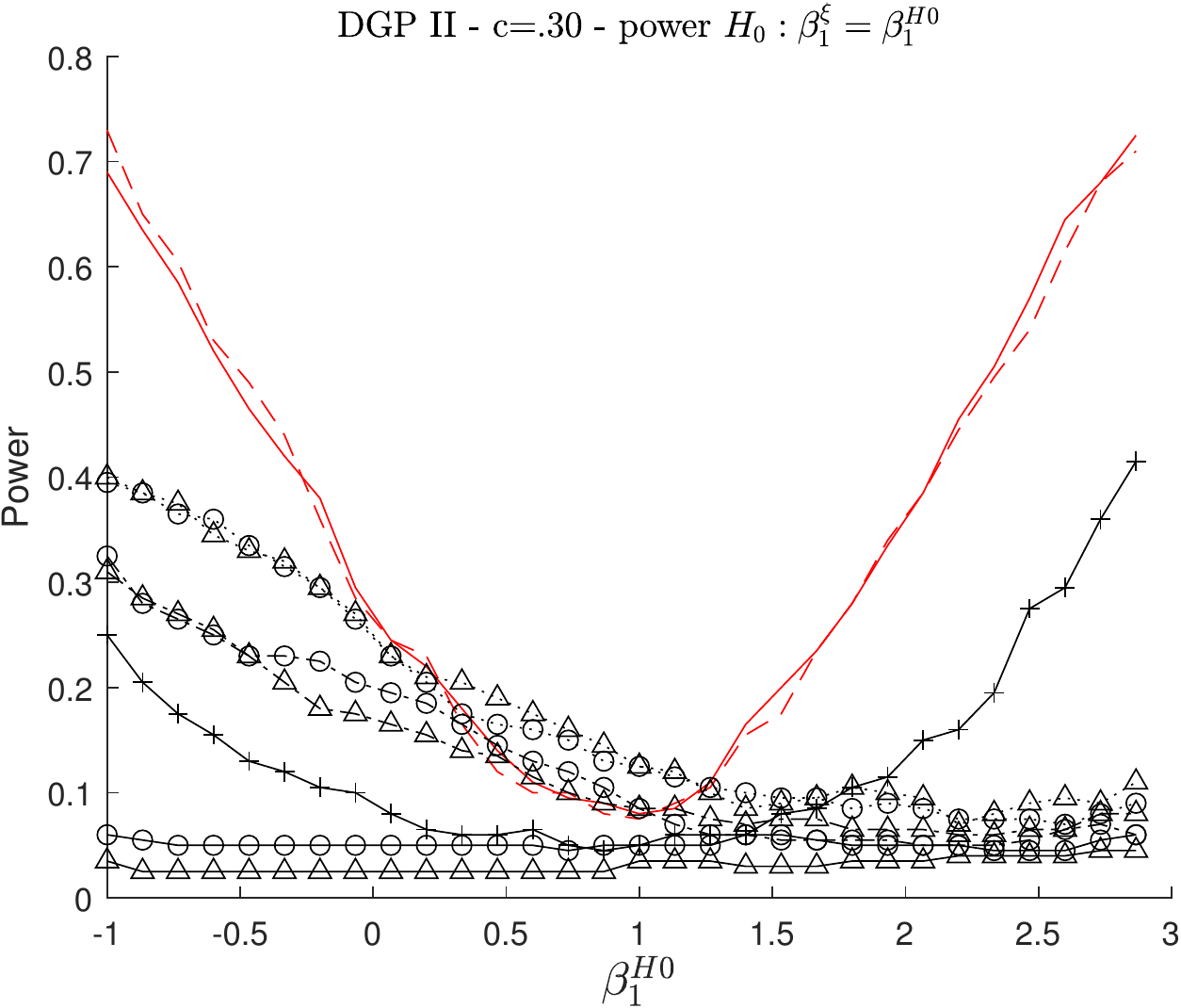}
\end{tabular}
    \caption{\footnotesize Power curves for the null hypotheses $H_{0}:\beta_{0}^{\xi}=\beta_{0}^{H0}$ (top) and $H_{0}:\beta_{1}^{\xi}=\beta_{1}^{H0}$ (bottom), for different values of $\beta_{0}^{H0}$ and $\beta_{1}^{H0}$. From left to right: DGP II with $c=0.075$, $c=0.20$ and $c=0.30$. For $\beta_{0}^{H0}=\log(.2)$ and $\beta_{1}^{H0}=1$, we expect a rejection rate equal to the 5\% test level. Red solid and dashed: \texttt{WMLE} and \texttt{CWMLE}. $+$: \texttt{MLE}. $\triangle$: POT estimates based on unconditional thresholds. $o$: POT estimates based on conditional thresholds. Thresholds are taken as quantiles at the 99\% (black solid), 95\% (dashed dotted), and 90\% (dotted) levels in the POT approaches.}
    \label{fig:power_curves_app}
\end{figure}

\begin{figure}[htbp]
    \centering
    \begin{tabular}{ccc}
        \includegraphics[scale=.37]{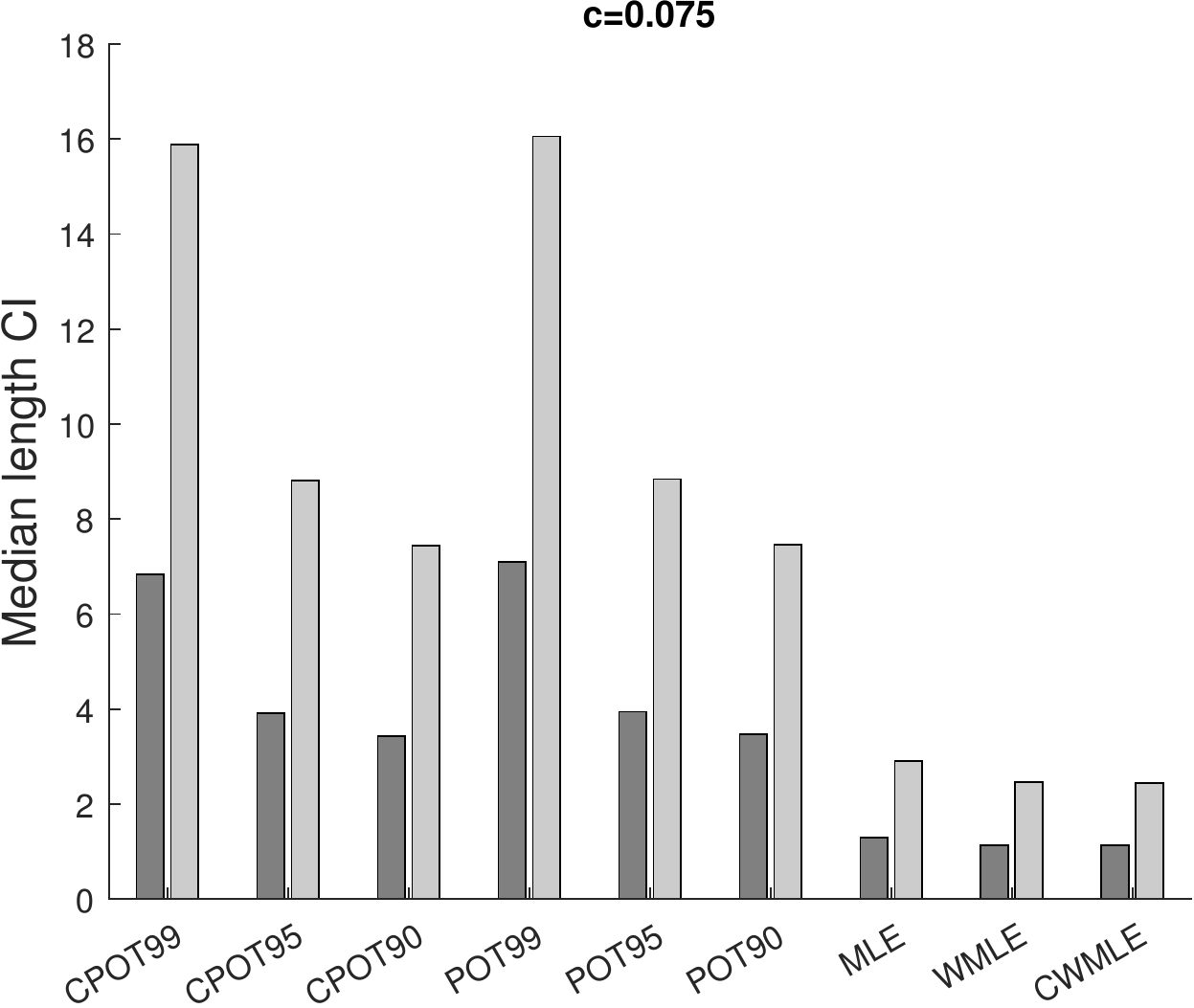} &\includegraphics[scale=.37]{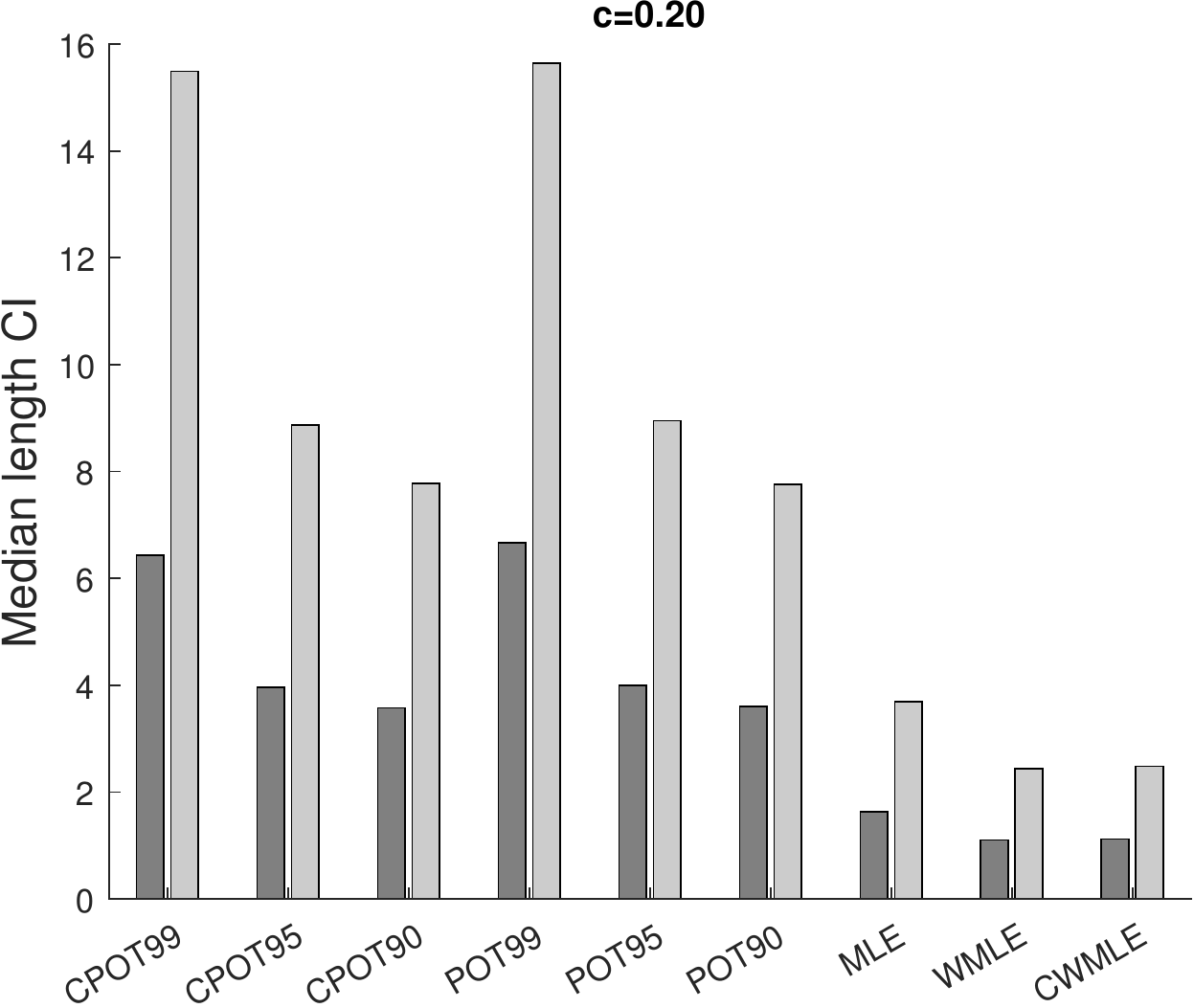} &\includegraphics[scale=.37]{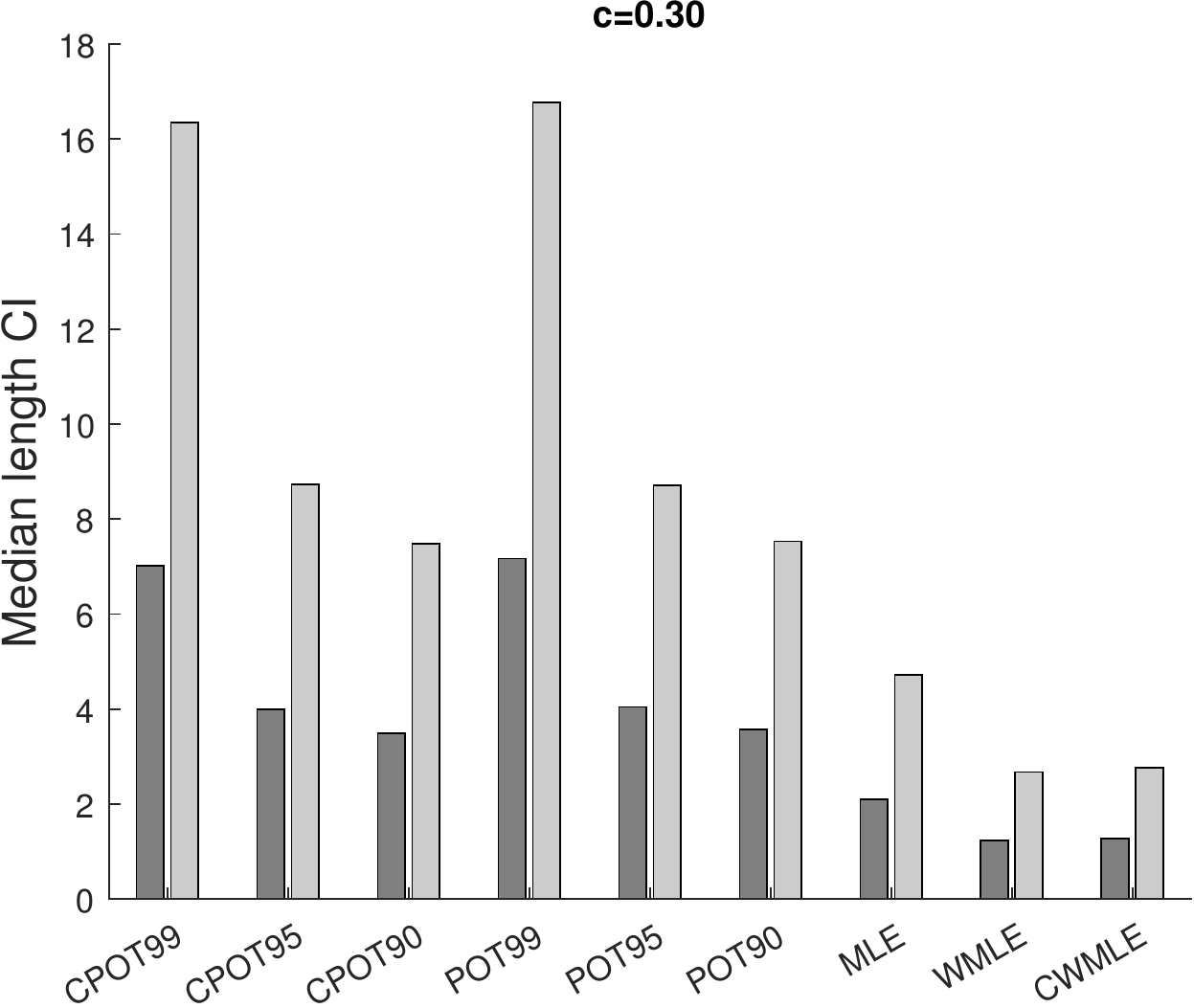} \\
        (i) & (ii) & (iii)
    \end{tabular}
    \caption{\footnotesize Median length of the 95\% confidence intervals (CI) obtained with the different methods, for the different contamination rates. Dark gray: length of the CI for $\beta^{\xi}_{0}$. Light gray: length of the CI for $\beta^{\xi}_{1}$.}
    \label{fig:length_appendix}
\end{figure}

 Figure \ref{fig:power_curves_app} displays the full power curves for null hypotheses with increasing degrees of deviation from the true null. One may observe a comparable performance across DGPs for the POT-based approaches and WMLE, with a clear superiority of WMLE. On the contrary, MLE not being a robust method, its performance decreases dramatically with increasing contamination rates. Finally, the G-E-GPD estimated with the WML approach also dominates in terms of median lengths of the confidence intervals (Figure \ref{fig:length_appendix}).

In conclusion, these results confirm the superiority of the G-E-GPD approach with the censoring estimation method over the POT-based methods, for these new DGPs. This performance is particularly remarkable, since increasing contamination rates favors the POT approaches.\\

\section{Estimation of the conditional mean model of \cite{patton2013} - A brief description}\label{app:mean_filter}

In Section~\ref{sec:empirical}, we conduct our analysis on hedge fund returns from which we removed the conditional mean, following the idea of \cite{kelly2014}. To estimate this conditional mean component, we use the asset pricing model of \citet{patton2013}. \citet{patton2013} propose refining the classical seven-factor asset pricing model of \cite{funghsieh} to account for intra-month variations in exposure to risk factors. Starting from the model of \cite{funghsieh}, we have
\begin{align}
    r_{it} = \alpha_i + \sum\limits_{j}^{p}\beta_{ijt} f_{jt} + y_{it},
\end{align}
where $r_{it}$ denotes monthly returns of fund $i$ at time $t$. \citet{patton2013} assume in addition that intra-month returns (e.g. daily returns) are given by:
\begin{align}\label{eq:patton_daily}
    r_{id}^\star = \alpha_i + \sum\limits_{j=1}^{p}\beta_{ijd} f^{\star}_{jd} + \varepsilon^\star_{id}
\end{align}
where $r_{id}$ is the daily return of fund $i$ on day $d$, for $d \in \mathcal{M}(t)$, $\mathcal{M}(t)$ denoting the set of days belonging to month $t$, $f^{\star}_{jd}$ is the $j$-th risk factor measured on day $t$ and $\beta_{ijd}$ describes the time-variant exposure to this factor for fund $i$. The parameter $\beta_{ijd}$ is made time-variant by conditioning it to a covariate $Z$ measured at the monthly and daily frequencies, and common to all funds. We assume the following dynamics for $\beta_{ijd}$: 
\begin{align}
    \beta_{ijd} & = \beta_{0ij} + \gamma_{ij} Z_{t-1} + \delta_{ij} Z_{d-1}^{\star}.
    \label{eq:patton_gilin}
\end{align}
Here, $Z_{t-1}$ is a monthly covariate and $Z_{d-1}^{\star}$ is measured at a daily frequency\footnote{This equation can be extended to the case of $K$ covariates driving $\beta_{ijd}$, but for simplicity of exposition, we consider here that only one variable is involved in changes of intra-month risk exposure.}. Substituting \eqref{eq:patton_gilin} in \eqref{eq:patton_daily}, we obtain
\begin{align}
    r_{id}^\star = \alpha_i + \sum\limits_{j=1}^{p}\beta_{0ij}f^{\star}_{jd} +\sum\limits_{j=1}^{p}\gamma_{ij} Z_{t-1}f^{\star}_{jd}+\sum\limits_{j=1}^{p}\delta_{ij} Z_{d-1}^{\star}f^{\star}_{jd}+ \varepsilon^\star_{id}
\end{align}
Returns on individual hedge funds being only available at a monthly frequency, we need to express monthly returns from the daily returns. Denoting by $n_t$ the number of observations in the set $\mathcal{M}(t)$, we assume that:
\begin{equation}
   r_{it} = \sum_{d \in \mathcal{M}(t)} r_{id}^\star
\end{equation}
such that
\begin{equation}
    r_{it} = \:  \alpha_i n_t + \sum\limits_{j}^{p}\beta_{0ij}\sum_{d \in \mathcal{M}(t)}f^\star_{jd}  + \sum\limits_{j}^{p}\gamma_{ij} \bar{f}_{jt} Z_{t-1} + \sum\limits_{j}^{p}\delta_{ij}\sum_{d \in \mathcal{M}(t)}f^\star_{jd} Z^{\star}_{d-1} + \sum_{d \in \mathcal{M}(t)} \epsilon^\star_{id}.
    \label{eq:patton_general}
\end{equation}
Following \citet{patton2013}, we consider up to two variables as conditioning covariates $\mathbf{Z}_{t}=\left(Z_{t1},Z_{t2}\right)$ and $\mathbf{Z}^{*}_{d}=\left(Z^{*}_{d1},Z^{*}_{d2}\right)$:
\begin{enumerate}
    \item Market liquidity, measured by the TED spread (3-month LIBOR rate minus the 3-month T-Bill rate) 
    \item Market volatility, proxied with the VIX.
\end{enumerate}
Regarding the risk factors ($f_{jt}$ and $f_{jd}^{*}$), we use the four factors, out of the seven provided by \citet{funghsieh}, that are available on a daily basis:
\begin{enumerate}
    \item The equity risk factor (SP500), measured by the excess return of the S\&P500.
    \item The small-minus-big factor (SMB), measured by the difference between the Russell 2000 index and the S\&P500 index returns. 
    \item The bond market factor (TCM10Y) measured by the change in the 10-year treasury constant maturity yield. 
    \item The credit spread factor (BAAMTSY) measured by the change in the Moody's Baa yield minus the 10-year treasury constant maturity yield.
\end{enumerate}
To obtain a parsimonious model, we perform both model estimation and selection simultaneously using an adaptive LASSO approach \citep{zou2006}. The penalization parameter is chosen using the Bayesian information criterion (BIC). Final estimates of the coefficients are obtained from the post-LASSO estimator (i.e. the OLS estimates using only the active set identified by the LASSO at the model selection step). Our tail risk analysis is then conducted on the estimated residuals of this model, i.e. on 
\begin{equation}
    \hat{y}_{it} = r_{it} - \hat{\alpha}_i + \sum\limits_{j=1}^{p}\hat{\beta}_{ijt} f_{jt}.
\end{equation}

\section{Additional empirical results}\label{sec:empirical_suite}

In this section, we present additional results to the main analysis, as well as the results of two additional regression models, complementing Section~\ref{sec:empirical}. 

\subsection{Additional results for the main analysis}

In Section~\ref{sec:empirical}, we discuss the bias of the different estimators. Table~\ref{tab:average_coef} reports the average bias for each parameter and each DGP. Although all methods exhibit a certain degree of bias in DGP III, estimates obtained from \texttt{WMLE} and \texttt{CWMLE} tend to be closer to the true coefficient $\beta_{1}^{\xi}$ than the MLE and the POT-based approaches.
\begin{table}[htbp]
    \centering
    \begin{footnotesize}
    \begin{tabular}{ccccccccccc}
    \hline
    \hline
        DGP & Crit. &  \texttt{CPOT99} &\texttt{CPOT95} &\texttt{CPOT90} &\texttt{POT99} &\texttt{POT95} & \texttt{POT90} &\texttt{MLE} & \texttt{WMLE}&\texttt{CWMLE}\\
         \hline
         \hline
         DGP I & $\beta_{0}^{\xi}$& -0.901 &   -1.302 &   -1.454 &   -1.008 &  -1.309 &  -1.458 &   -0.077 & -0.081 & -0.076\\
         & $\beta_{1}^{\xi}$&  -0.516 &  1.546 &   1.542 & -0.259 & 1.536 &  1.533 & 0.014 & 0.010 &   0.011\\
         \hline
         DGP II & $\beta_{0}^{\xi}$& -1.026 &  -1.161  &   -1.507  &   -0.876  &   -1.163  &   -1.488  &    0.347  &   -0.051  &   -0.049\\
         & $\beta_{1}^{\xi}$&  -0.343 &    1.388 & 1.640 &  -0.568 &    1.302 &   1.572 &   -0.192 &   -0.034 &   -0.039\\
\hline
         DGP III & $\beta_{0}^{\xi}$& -1.115 &  -0.437 &   -0.619 &   -0.984 &   -0.483 &   -0.673 &    0.772 &    0.467 &  0.479
\\
         & $\beta_{1}^{\xi}$&-0.271 &    0.404 &    0.631 &   -0.888 &    0.488 &    0.760 &    0.037 &   -0.004  & -0.051\\
         \hline
         \hline
    \end{tabular}
    \end{footnotesize}
    \caption{\footnotesize Estimated bias computed as $\dfrac{1}{B}\sum\limits_{b=1}^{B}\left(\hat{\beta}^{\xi}_{0}-\beta^{\xi}_{0}\right)$ and $\dfrac{1}{B}\sum\limits_{b=1}^{B}\left(\hat{\beta}^{\xi}_{1}-\beta^{\xi}_{1}\right)$, from $B=200$ simulated samples for each DGP. True values of the parameters for all DGPs are $\beta_{0}^{\xi}=\log(.2)$ and $\beta_{1}^{\xi}=1$ }
    \label{tab:average_coef}
\end{table}

\subsection{Additional analyses}

In Table \ref{tab:leverage}, we reproduce our main regression analysis from Section~\ref{sec:empirical}, however splitting the sample of funds between leveraged (first column) and non-leveraged ones (second column). Results are similar across groups. Then, we also re-estimate the regression model on the complete sample, however removing variables not found to be significant in the main regression analysis (third column). Estimates remain unchanged. This model is then used in Subsection 4.2 to conduct the performance analysis of hedge funds. 

\begin{table}[h]
    \centering
   \begin{footnotesize}
    \begin{tabular}{ccccccc}
    \hline
    \hline
    Covariate & Leverage & Not Leveraged & Significant MA \\
    \hline
    \hline
$\beta^{\xi}_{0}$   &  -1.29$^{***}$	&	-1.27$^{***}$	&	-1.31$^{***}$	\\
	 & 	$[-1.42,-1.16]$	&   $[-1.54,-1.01]$	& $[-1.45,-1.18]$\\
  $\beta^{\xi}(\texttt{Crisis})$ & 0.04	&	0.04	&	-	\\
  &$[-0.50,0.58]$	&   $[-0.82, 0.89]$	& -\\
$\beta^{\xi}(\texttt{FSI})$		& 	-0.67$^{***}$	&	-0.67$^{***}$	&	-0.67$^{***}$	\\
	&	$[-0.80,-0.54]$	&   $[-0.87,-0.47]$	&	$[-0.78,-0.56]$\\
$\beta^{\xi}(\texttt{VIX})$		&	-0.19$^{**}$	&	-0.19	&	-0.19$^{***}$	\\
&		$[-0.34,-0.03]$	&   $[-0.47,0.09]$	&	$[-0.31,-0.07]$ \\
$\beta^{\xi}(\Delta\texttt{MSCI})$		&	0.03	&	0.03	&	-	\\
	&	$[-0.05,0.11]$	&   $[-0.06,0.12]$	&	-\\
$\beta^{\xi}(\texttt{MOM})$		& 0.22$^{***}$	&	0.21$^{***}$	&	0.21$^{***}$	\\
&	$[0.15,0.28]$	&    $[0.14,0.28]$	&	 $[0.17,0.26]$\\
$\beta^{\xi}(\texttt{Liq})$		&  -0.03	&	-0.03	&	-	\\
&	$[-0.08,0.03]$	&   $[-0.11,0.05]$	&	-	\\
$\beta^{\xi}(\texttt{CredSpr})$		& -0.47$^{***}$	&	-0.46$^{***}$	&	-0.47$^{***}$	\\
		&	$[-0.55,-0.38]$	&   $[-0.62,-0.31]$	& $[-0.54,-0.39]$\\
		\hline
		\hline
$\beta^{\sigma}_{0}$	&	-3.88$^{***}$	&	-3.86$^{***}$	&	-3.90$^{***}$	\\
& $[-3.89,-3.87]$	&   $[-3.87,-3.85]$	& $[-3.91,-3.88]$\\
$\beta^{\sigma}(\texttt{VIX})$		&0.07$^{***}$	&	0.07$^{***}$	&	0.08$^{***}$	\\
	&	$[0.06,0.08]$	&    $[0.06,0.08]$	& $[0.07,0.09]$\\
$\beta^{s}_{0}$		&	-3.52$^{***}$	&	-3.52$^{***}$	&	-3.56$^{***}$	\\							&	$[-3.54,-3.51]$	&   $[-3.55,-3.50]$	&	$[-3.58,-3.55]$\\							
$\beta^{s}(\texttt{VIX})$		&	0.02$^{***}$	&	0.02$^{**}$	&	0.02$^{***}$	\\		
	&  $[0.01,0.03]$	&    $[0.00,0.03]$	&	$[0.01,0.03]$\\								
$\beta^{m}_{0}$	&	-0.002$^{***}$		&		-0.002$^{***}$		&-0.002$^{***}$ 		\\						
&	$[-0.002,-0.002]$	&	$[-0.002,-0.002]$ &	$[-0.002,-0.002]$	\\	
\hline
\hline
Sample size & 102,598& 86,414 & 189,014\\
$\tau^{opt}$ & 0.253 & 0.234 & 0.253\\
\hline
\hline
    \end{tabular}

    \end{footnotesize}
    \caption{\footnotesize Estimated regression effects for the different estimation methods, for the funds recorded as leveraged (first column), not leveraged (second columns), or using only the covariates found to be significant in the main analysis (third column). Confidence intervals at the 95\% level are in brackets below the estimates obtained with the $\texttt{WMLE}$ approach. $^{***}$ and $^{**}$ indicate coefficients significant at the 1\% and 5\% test levels, respectively.}
    \label{tab:leverage}
\end{table}

\newpage

\section{A discussion on WML estimation}\label{sec:theoretical_suite}

In Section~\ref{sec:theory}, we introduce a set of conditions for the weighted maximum likelihood (WML) estimator to be consistent and asymptotically normal. These conditions being tedious to check with our hybrid regression model (and the asymptotic covariance matrix not being explicit), we discuss them here on a simpler (unconditional) example that exhibits a heavy tail. For illustration, we consider the case of a Pareto distribution (with survival distribution $y^{-1/\xi_0}$ for all $y>1$). 
The (explicit) ML estimator (or WML estimator with $\tau=0$) is known to be consistent, unbiased, and asymptotically normal with variance $\xi_0^2$. 

If $\tau \neq 0$, the WML estimator $\hat{\xi}^w(\tau)$ becomes the solution of 
$$
\sum \limits_{i=1}^n \psi_{\xi}(y_i)=0, \quad\text{where} \;
\psi_{\xi}(y):=-\frac1\xi \mathds{1}_{\{y > q(\tau) \}} +\frac1{\xi^2} \log(y) \mathds{1}_{\{y > q(\tau) \}} -\frac1{\xi^2} \log(q(\tau)) \frac{q(\tau)^{-\frac1\xi}}{1-q(\tau)^{-\frac1\xi}} \mathds{1}_{\{y \leq q(\tau) \}},
$$
leading to the fixed-point relationship:

$$ \hat{\xi}^w(\tau)= \frac{ \frac{1}{n} \sum \limits_{i=1}^n \log(y_i) \mathds{1}_{\{y_i>q(\tau) \}} - \frac{\log(q(\tau)) q(\tau)^{-1/\hat{\xi}^w(\tau)} }{1-q(\tau)^{-1/\hat{\xi}^w(\tau)}} \frac{1}{n} \sum \limits_{i=1}^n \mathds{1}_{\{y_i \leq q(\tau) \}}}{ \frac{1}{n} \sum \limits_{i=1}^n \mathds{1}_{\{y_i>q(\tau)\}}}. $$
Assuming, for convenience, $q(\tau)$ to be known (in practice, it would have to be estimated, making the theoretical study more involved) and equal to $(1-\tau)^{-\xi_0}$ under a correct specification, then straightforward calculations lead to  
$$ \left\{
    \begin{array}{ccl}
        \frac{1}{n} \sum \limits_{i=1}^n \log(y_i) \mathds{1}_{\{y_i>q(\tau) \}}  & = & \mathbb{E} \left[ \log(y_i) \mathds{1}_{\{y_i>q(\tau) \}}  \right] (1+o_{\mathbb{P}}(1))=\xi_0 (1-\tau) (1-\log(1-\tau)) (1+o_{\mathbb{P}}(1)), \\
        \frac{1}{n} \sum \limits_{i=1}^n \mathds{1}_{\{y_i>q(\tau) \}} & = & \mathbb{P} [y_i>q(\tau)](1+o_{\mathbb{P}}(1))=(1-\tau)(1+o_{\mathbb{P}}(1)), \\
        \frac{1}{n} \sum \limits_{i=1}^n \mathds{1}_{\{y_i \leq q(\tau) \}} & =  & \mathbb{P} [y_i \leq q(\tau)](1+o_{\mathbb{P}}(1))=\tau (1+o_{\mathbb{P}}(1)).
    \end{array}
\right. $$
Notice that $|\mathbb{E}[\psi_{\xi}(y)]|>0$ if $\xi \neq \xi_0$, and condition $(i)$ for the consistency (see the end of Section 2.3) may be obtained, at least on a large interval of $\mathbb{R}_+$, by the law of large numbers. The consistency of $\hat{\xi}^w(\tau)$ may thus be deduced from the following equation:
$$ 
\frac{\hat{\xi}^w(\tau)}{\xi_0} = \frac{(1-\tau) (1-\log(1-\tau)) + \frac{\log(1-\tau)(1-\tau)^{\xi_0/\hat{\xi}^w(\tau)}}{1-(1-\tau)^{\xi_0/\hat{\xi}^w(\tau)}} \tau }{1-\tau} (1+o_{\mathbb{P}}(1)).
$$
Moreover, for $0<\varepsilon<\xi_0$, the derivative $\frac{\partial}{\partial \xi} \psi_{\xi}(y_i)$ is clearly bounded on $\xi \in (\xi_0-\varepsilon,\xi_0+\varepsilon)$, then Condition (I) holds with 
\begin{multline*}
    \overline{\psi}(y)= \frac{1}{(\xi_0-\varepsilon)^2} \mathds{1}_{\{y>q(\tau)\}} + \frac{2}{(\xi_0-\varepsilon)^3} \log(y) \mathds{1}_{\{y>q(\tau)\}} + \frac{2}{(\xi_0-\varepsilon)^3} \log(q(\tau))   \\ +\frac{1}{(\xi_0-\varepsilon)^4} \frac{\log(q(\tau))^2 q(\tau)^{-1/(\xi_0+\varepsilon)}}{\left( 1-q(\tau)^{-1/(\xi_0+\varepsilon)} \right)^2} - \frac{2}{(\xi_0-\varepsilon)^3} \frac{\log(1-\tau)}{\tau}.
\end{multline*}
Note also that $\mathbb{E}[\overline{\psi}(y)^2]<\infty$ since $\mathbb{E}[\log(y)^2 \mathds{1}_{\{y>q(\tau)\}}]=\frac{\log(q(\tau))^2+2\xi_0 \log(q(\tau))+2\xi_0^2}{q(\tau)^{1/\xi_0}}<\infty$. \\
In addition, it is straightforward to check that Condition (II) is satisfied with
$$ V_{\xi_0}= (\tau-1)\, \frac{1+\frac{\log(1-\tau)^2}{\tau}}{\xi_0^2}. $$\\[-3ex]
$\hat{\xi}^w(\tau)$ being computed as the solution of $\sum \limits_{i=1}^n \psi_{\xi}(y_i)=0$, Condition (III) is fulfilled, hence the asymptotic distribution \\[-2ex]
$$ 
\sqrt{n} \left( \hat{\xi}^w(\tau)- \xi_0 \right) \rightarrow \mathcal{N} \left( 0, \frac{\xi_0^2}{1-\tau+\frac{\log(1-\tau)^2 (1-\tau)}{\tau}} \right),\; \text{as}\, n \rightarrow \infty. 
$$
Note that the case $\tau=0$ gives back the classical asymptotic distribution of the MLE.

To conclude, consistency and rate of convergence were obtained for the WML estimator of the Pareto parameter in this simplified setting. Considering our hybrid (and conditional) model G-E-GPD would make the theoretical study much more challenging, where we have to deal with many parameters and constraints. However, the asymptotic variance may be computed and estimated accurately (numerically). 
\nocite{}

\end{document}